# Unveiling Structural Bottlenecks of Dynamic Disorder in a Density-Tunable Glass Former: From Strong to Fragile Regimes

Shubham Kumar[1] and Shinji Saito[1,2,*]

[1]Institute for Molecular Science, Myodaiji, Okazaki, Aichi 444-8585, Japan

[2]The Graduate University for Advanced Studies (SOKENDAI), Myodaiji, Okazaki, Aichi 444-8585, Japan

*Author for correspondence: shinji@ims.ac.jp

## Abstract

Fragility characterizes how rapidly a glass-forming liquid slows down upon supercooling, but whether strong and fragile behaviors arise from the same microscopic relaxation mechanism remains unclear. Here, we address this question using a density-tunable soft-repulsive binary mixture spanning distinct fragility regimes and analyze particle jump dynamics within the framework of dynamic disorder. Across these regimes, we show that increasing fragility leads to progressively broader cage-lifetime distributions and increasingly non-exponential survival probabilities, revealing non-Poisson cage-to-jump statistics governed by fluctuating jump rates and slowly evolving structural variables. To characterize their structural origin, we first identify the neighbor ranks most strongly coupled to jump motion using Kullback–Leibler (KL) divergence and Pearson correlation analyses. We then introduce a structural slowness parameter (SSP) that combines these neighbor-distance fluctuations into a reduced slow coordinate for constructing the slow-fluctuation survival probability. A comparison with the actual survival probability shows that localized neighbor-distance fluctuations control the jump rate in the strong regime, whereas extended neighbor rearrangements become relevant in the intermediate and fragile regimes, increasing the effective dimensionality of the slow-variable space. In the fragile regime, distance-based descriptors alone become insufficient at the lowest temperature, where the Voronoi free volume captures additional cage-volume fluctuations in the rate-controlling slow variable. Point-to-set correlations (PTS) grow with fragility, but the spatial extent of the slow variables exceeds the PTS length. These results show that fragility changes the structural bottleneck for microscopic rate fluctuations, linking dynamic disorder and multidimensional slow variables.

## I. INTRODUCTION

Glass-forming liquids exhibit a dramatic slowdown of dynamics upon cooling, often without substantial changes in conventional structural measures, yet the microscopic mechanisms responsible for this phenomenon remain elusive.[1–5] While the growth of structural relaxation times is ubiquitous across glass formers, the nature of this slowdown varies widely across systems.[6–8] In particular, liquids classified as *strong* display nearly Arrhenius temperature dependence of relaxation times, whereas *fragile* liquids exhibit strongly super-Arrhenius behavior.[9,10] This diversity raises a fundamental question: do strong and fragile glass formers slow down through the same microscopic mechanisms, or does the nature of relaxation itself change across fragility regimes?[1,11]

Many theoretical and computational approaches implicitly treat relaxation in strong and fragile glass formers as arising from qualitatively similar microscopic processes, attributing differences in fragility primarily to variations in cooperativity,[2,12,13] dynamical heterogeneity,[14–19] or effective activation barriers[7,20] rather than distinct relaxation mechanisms. Within this view, fragile behavior is often interpreted as an amplification of cooperativity or heterogeneity already present in strong liquids.[13,15–21] In this context, Tah *et al.* showed that the same machine-learning-based softness descriptor predicts particle rearrangement propensity across strong and fragile regimes.[21] However, direct microscopic evidence for a common underlying relaxation mechanism across fragility regimes remains limited.[16,17,21,22] In particular, it is unclear whether fragility reflects changes in the processes that *control* the occurrence of relaxation events, changes in how these events are *accommodated* by the surrounding medium, or a more fundamental reorganization of the slow degrees of freedom underlying glassy dynamics.

In this context, progress toward resolving this issue has been hindered by the complexity of most glass-forming materials, where changes in fragility are accompanied by variations in interaction topology, bonding directionality, molecular anisotropy, chemical composition, or the softness of interparticle repulsion.[16,19,22–31] As a result, isolating the dynamical origins of fragility from material-specific structural effects remains challenging. Model systems that permit continuous tuning of fragility while maintaining simple interactions and avoiding additional structural complexity are therefore essential for disentangling universal mechanisms of glassy slowdown from system-dependent features.[7,32] To address this problem, we exploit a well-established class of soft-repulsive binary mixtures in which fragility can be tuned independently of interaction topology and chemical composition. By varying density, these systems exhibit strong or fragile relaxation behavior over the corresponding temperature ranges, without introducing directional bonding and network formation.[21,32,33] This permits a controlled examination of how microscopic relaxation dynamics evolve across different fragility regimes.

At the microscopic level, structural relaxation in glass-forming liquids proceeds through intermittent, localized rearrangements, commonly referred to as particle jump events.[34–43] These events correspond to elementary cage-breaking processes that enable long-time relaxation. It is now well established that jump dynamics are intrinsically heterogeneous and become increasingly intermittent upon supercooling, with cage-time distributions deviating strongly from Poisson statistics.[38,39,43–45] At a coarse-grained level, such statistics are commonly described within continuous-time random walk (CTRW) representations of structural relaxation, in which relaxation emerges from a sequence of stochastic jump events separated by waiting-time distributions.[38,42,46–49] However, as supercooling deepens, the assumption of statistically independent jump events becomes inadequate, reflecting the influence of slowly evolving local environments on the underlying jump rates.[43–45]

The involvement of slowly evolving local environments implies that the probability of a jump is not governed by a single, time-independent rate. Instead, relaxation reflects a distribution of microscopic jump rates that persist over timescales comparable to, or longer than, the cage lifetime.[50–54] This behavior is commonly described within the framework of *dynamic disorder*, in which temporal correlations of the local environment give rise to *non-Markovian jump dynamics*.[50,54–56] From this perspective, deviations from Poisson statistics reflect the presence of slow degrees of freedom that regulate jump rates over extended timescales.[43–45,51–53,57,58] Identifying the relevant slow variables and determining how their influence on jump dynamics evolves across different fragility regimes is, therefore, central to a microscopic description of glassy slowdown.

In this study, we address this question by examining how jump dynamics and dynamic disorder evolve across different fragility regimes within a single class of glass-forming liquids. Using a density-tunable soft-repulsive binary mixture,[21,32,33] we analyze relaxation in terms of particle-resolved jump events and their associated cage-time statistics over a broad range of temperatures. This approach allows us to directly compare the nature of rate heterogeneity, temporal correlations, and non-Markovian effects across different fragility regimes, while avoiding changes in interaction topology or chemical composition. Thus, we establish a unified microscopic framework for understanding how dynamic disorder evolves with fragility.

Our results reveal that the nature of the slow dynamical variables governing relaxation differs systematically from a strong regime to the intermediate and fragile regimes. To quantify this dependence, we introduce a *structural slowness parameter* (SSP) that captures the modulation of jump probabilities by local environment, thereby providing a quantitative measure of the slow degrees of freedom that regulate relaxation. We find that in the strong regime, jump dynamics are primarily controlled by a relatively small number of localized structural constraints, whereas increasingly extended neighbor rearrangements become relevant in the intermediate and fragile regimes, enhancing intermittency and rate heterogeneity. This evolution reflects a systematic change in the structural bottleneck that controls jump-rate fluctuations, rather than a simple amplification of local cooperativity. We further compare the spatial extent of these slow variables with point-to-set correlations to assess their relationship with static amorphous order. These findings establish a direct microscopic connection between fragility and the nature of the slow variables underlying structural relaxation in glass-forming liquids.

The organization of the rest of this paper is as follows. In Sec. II, we discuss the theoretical and computational details employed in this study. In Sec. III, we present a detailed discussion of the results obtained from our simulations. In this section, we first characterize the relaxation dynamics across different fragility regimes and then analyze the structural evolution during jump motion. We next examine jump statistics, cage-lifetime distributions, survival probabilities, and dynamic disorder to characterize the transition in the nature of jump dynamics across fragility regimes. Subsequently, we identify the microscopic origin of dynamic disorder by systematically comparing the strong, intermediate, and fragile regimes using the SSP and its slow-fluctuation survival probability. We further compare the spatial extent of the rate-controlling slow variables with the point-to-set (PTS) correlation length. In Sec. IV, we summarize the present study and highlight potential directions for future research.

## II. THEORETICAL AND COMPUTATIONAL DETAILS

### A. Models and simulation details

We investigated a three-dimensional equimolar (50:50) binary mixture composed of particles A and B that interact through purely repulsive pair potentials. The model was designed to provide a controlled route for tuning glass-forming behavior from strong to fragile regimes through density variation while suppressing crystallization and phase separation.[21,32,33] The interaction between particles $\alpha$ and $\beta$ ($\alpha, \beta \in \{A, B\}$) was described by a harmonic soft-sphere potential of the form

$$\upsilon_{\alpha\beta}(r) = \begin{cases} \varepsilon_{\alpha\beta}\left(1 - \dfrac{r}{\sigma_{\alpha\beta}}\right)^2, & r < \sigma_{\alpha\beta}, \\ 0, & r \geq \sigma_{\alpha\beta}, \end{cases} \tag{1}$$

where $r$ denotes the interparticle separation, and $\sigma_{\alpha\beta}$ and $\varepsilon_{\alpha\beta}$ represent the interaction length and energy scales, respectively. The interaction parameters were chosen as $\sigma_{AA} = 1.0$, $\sigma_{AB} = 1.2$, and $\sigma_{BB} = 1.4$, with $\varepsilon_{\alpha\beta} = 1.0$ for all pairs, and particle masses were set to $m = 1$. All quantities were expressed in reduced units, with length measured in units of $\sigma_{AA}$, energy in units of $\varepsilon$, temperature in units of $\varepsilon/k_B$, and time in units of $\sqrt{m\sigma_{AA}^2/\varepsilon}$. All molecular dynamics simulations were performed using the Large-scale Atomic/Molecular Massively Parallel Simulator (LAMMPS) package.[59] All simulations were carried out using a system size of $N = 10000$ particles with periodic boundary conditions applied in all three spatial directions. The equations of motion were integrated using the velocity-Verlet algorithm[60] with a time step $\delta t = 0.01$ in reduced units.

Simulations were performed at number densities $\rho = 0.65$, 0.69, 0.72, 0.78, and 0.82, which spanned strong to fragile dynamical regimes. The temperature range explored at each density extended from the normal liquid regime to deeply supercooled conditions approaching dynamical arrest; the simulation state points are summarized in Table S1. Configurations at lower temperatures were prepared by gradual cooling from equilibrated higher-temperature states to minimize aging effects.

At each density and temperature, the system was first equilibrated in the canonical (NVT) ensemble using a Nosé–Hoover thermostat[61] with a relaxation time chosen to ensure stable temperature control while minimizing perturbations to the intrinsic dynamics. Following equilibration, production trajectories were generated in the microcanonical (NVE) ensemble for data collection. The equilibration runs at each state point were performed for durations at least three to five times longer than the characteristic structural relaxation time $\tau_\alpha$, and the production runs were extended to at least five times longer to ensure adequate statistical sampling. Here, the structural relaxation time $\tau_\alpha$ was obtained from the decay of the self-overlap function $Q(t)$, as described in Sec. III A. Trajectory configurations were saved at intervals of 10 $\delta t$. We also examined the dependence of our results on the simulation ensemble by performing selected analyses in the NVT ensemble using the Nosé–Hoover thermostat and found no significant differences in the statistical properties of particle dynamics.

### B. Identification of jump events

To identify jump motion associated with structural relaxation, we use the hop function $h(t)$, which quantifies particle displacements between two neighboring time intervals.[36,37,43–45,62,63] For each particle $i$, the hop function $h_i(t)$ is defined by comparing its positions averaged over two adjacent time windows surrounding time $t$,

$$h_i(t) = \sqrt{\left\langle \left(\mathbf{r}_i(t) - \langle \mathbf{r}_i(t) \rangle_B\right)^2 \right\rangle_A \left\langle \left(\mathbf{r}_i(t) - \langle \mathbf{r}_i(t) \rangle_A\right)^2 \right\rangle_B}, \tag{2}$$

where $\mathbf{r}_i(t)$ denotes the position vector of particle $i$ at time $t$. The symbols $\langle \cdot \rangle_A$ and $\langle \cdot \rangle_B$ denote time averages over two intervals of duration $\Delta t/2$ defined as $A = [t - \Delta t/2, t]$ and $B = [t, t + \Delta t/2]$, respectively. These averages are given by $\langle X(t) \rangle_A = \frac{1}{\Delta t/2} \int_{t-\Delta t/2}^{t} X(\tau) d\tau$ and $\langle X(t) \rangle_B = \frac{1}{\Delta t/2} \int_{t}^{t+\Delta t/2} X(\tau) d\tau$. The averaging window $\Delta t$ is selected from the plateau regime of the mean-squared displacement (MSD), $\langle \delta r^2(t) \rangle$ [Fig. S3], by locating the minimum of its logarithmic derivative $d \log \langle \delta r_\alpha^2(t) \rangle / d \log t$ [Fig. S4], corresponding to the timescale of maximal cage confinement.[36–39,43,44] Moderate variations of $\Delta t$ within this range do not affect the jump statistics.[62] The specific values of $\Delta t$ used for the representative densities are summarized in Tables S2-S4.

To distinguish localized cage motion from jump events, we analyze the cumulative distribution of hop amplitudes, which shows a crossover between small displacements associated with oscillations within a cage and a tail corresponding to jumps between neighboring cages.[62,64] The threshold $h^*$ obtained from this crossover is nearly independent of density and temperature [Figs. S5 and S6], with $h^* = 0.2$ for A particles and $h^* = 0.1$ for B particles. Using a consistent threshold allows direct comparison of jump dynamics across systems with different fragilities.

### C. Survival probability and rate fluctuations

Following the identification of jump events using the threshold $h^*$, we analyze the temporal statistics of the cage state prior to a jump. The residence-time distribution $\psi_{cage}(t)$ describes the probability density that a particle remains in the cage state ($h < h^*$) for a duration $t$ before undergoing a jump. The probability that a particle remains confined beyond time $t$, denoted as the residence probability $C_R(t)$, is obtained as[58]

$$C_R(t) = \int_t^\infty \psi_{cage}(\tau)\, d\tau. \tag{3}$$

The survival probability is then defined as[54,58]

$$C_S(t) = \frac{1}{\langle t \rangle} \int_t^\infty C_R(\tau)\, d\tau, \tag{4}$$

where $\langle t \rangle$ is the mean residence time. The survival probability, therefore, characterizes the persistence of the cage state preceding a jump.

Deviations from Poisson statistics in the cage-state residence times are quantified using the randomness parameter $R$, defined as[65,66]

$$R = \frac{\langle t^2 \rangle - \langle t \rangle^2}{\langle t \rangle^2}, \tag{5}$$

where $\langle t^n \rangle = \int t^n \psi_{cage}(t) dt$ denotes the $n$th moment of the residence-time distribution. For a Poisson process governed by a single rate, the residence-time distribution follows $\psi_{cage}(t) = k e^{-kt}$, for which $R = 1$. Deviations from unity reflect fluctuations in the jump rates, providing a measure of dynamic disorder.

The survival probability can be formally expressed in terms of a time-dependent jump rate as[50,54]

$$C_S(t) = \left\langle \exp\left( -\int_0^t k(\tau)\, d\tau \right) \right\rangle, \tag{6}$$

where $k(t)$ denotes an instantaneous jump rate that fluctuates due to variations in the local environment, and the angular brackets denote an average. Depending on the timescale of rate fluctuations, two limiting cases can be considered.

In the fast-fluctuation limit, rate fluctuations are rapid compared to the residence time, and the dynamics are characterized by an effective average rate $k_{fast}$. The survival probability reduces to a single exponential,[43,50,54,58]

$$C_{fast}(t) = \exp\left(-k_{fast} t\right), \tag{7}$$

with $k_{fast} = N_J / \left(N_{tot} \Delta t\right)$, where $N_{tot}$ is the total number of sampled time steps, $N_J$ is the number of steps at which $h(t) \geq h^*$, and $\Delta t$ denotes the sampling interval.

In the slow-fluctuation limit, rate fluctuations persist over times much longer than the average cage residence time, so the dynamics can be described in terms of quasi-static substates with different jump rates. The survival probability then takes a multi-exponential form determined by the distribution of rates,[50]

$$C_{slow}(t) = \left\langle \exp\left(-kt\right) \right\rangle_k . \tag{8}$$

By discretizing the rate distribution into $N$ substates with rates $k_i$ and populations $p_i$, this expression can be written as

$$C_{slow}(t) = \sum_{i=1}^{N} p_i \exp\left(-k_i t\right), \tag{9}$$

To determine $p_i$ and $k_i$, the trajectory is partitioned into substates defined by the value of a given slow variable. Let $N_{tot,i}$ denote the number of sampled time steps belonging to substate $i$ out of the total $N_{tot}$, and $N_{J,i}$ the number of sampled steps at which a jump is detected while the system is in substate $i$. The population and rate associated with substate $i$ are therefore $p_i = N_{tot,i}/N_{tot}$ and $k_i = N_{J,i}/(N_{tot,i}\Delta t)$, respectively.[43,54,58] The corresponding survival probability is then given by

$$C_{slow}(t) = \sum_{i=1}^{N} \frac{N_{tot,i}}{N_{tot}} \exp\left( -\frac{N_{J,i}}{N_{tot,i}\Delta t} t \right). \tag{10}$$

This expression provides the basis for evaluating slow-fluctuation contributions using different structural descriptors as candidate slow variables.

## D. Identification of candidate slow variables

The slow-fluctuation framework described in Sec. II C requires structural variables that evolve on timescales comparable to jump dynamics and modulate jump probabilities. In principle, multiple structural quantities may serve as candidate slow variables.[43–45] Since jump events involve rearrangements of the local environment surrounding a particle, structural quantities based on nearest-neighbor distances provide natural candidates.[43–45] We therefore evaluate nearest-neighbor distances of particles undergoing jumps as a function of the hop amplitude and use statistical measures, including the Kullback–Leibler (KL) divergence[67] and the Pearson correlation coefficient[68], to identify the neighbor ranks relevant for describing jump-rate fluctuations.

### (i) Kullback-Leibler (KL) divergence analysis

We consider the probability distributions $P^{\mathrm{eq}}(r_k)$ and $P(r_k,h^*)$ of nearest-neighbor distances for particles undergoing jumps, obtained from equilibrium configurations and configurations near the hop threshold $h^*$, respectively. Here, $r_k$ denotes the distance between a jumping particle and its $k$-th nearest neighbor. The difference between these distributions is quantified using the Kullback–Leibler (KL) divergence,[67]

$$D_{eq}(k) = \int dr_k \; P^{eq}(r_k) \log \frac{P^{eq}(r_k)}{P(r_k, h^*)}. \tag{11}$$

The KL divergence, therefore, provides a quantitative measure of how strongly the local structural environment associated with the $k$-th neighbor differs between equilibrium configurations and configurations at the hop threshold $h^*$. Neighbor ranks exhibiting large values of $D_{eq}(k)$ are therefore considered relevant for describing structural changes preceding a jump and for constructing candidate slow variables.

### (ii) Pearson correlation analysis

In addition to the KL divergence analysis, we compute the Pearson correlation coefficient[68] during cage-to-jump transitions to identify suitable neighbor ranks for constructing candidate slow variables. For each jump event identified by the threshold $h^*$, the time $t$ at which particle $i$ first enters the jump state is determined. An event-dependent time interval $W_i$ is then defined by tracing backward from $t$ to the nearest earlier time at which the hop amplitude reaches its equilibrium cage value. The increments associated with the transition are defined as

$$\Delta h_i = h_i(t) - h_i(t - W_i), \tag{12}$$

$$\Delta r_{k,i} = r_{k,i}(t) - r_{k,i}(t - W_i), \tag{13}$$

where $r_{k,i}(t)$ denotes the distance between particle $i$ and its $k$-th nearest neighbor at time $t$. We consider only the onset of each identified jump event in this analysis and exclude successive configurations after the jump, since the particle has already entered the jump state.

The Pearson correlation coefficient for neighbor rank $k$ is computed over the ensemble of jump events as

$$\rho_P(k) = \frac{\langle \Delta h \, \Delta r_k \rangle_{\mathrm{jump}}}{\sqrt{\langle (\Delta h)^2 \rangle_{\mathrm{jump}} \langle (\Delta r_k)^2 \rangle_{\mathrm{jump}}}}. \tag{14}$$

This analysis quantifies the coupling between variations in the hop amplitude and the structural changes associated with different neighbor ranks during the cage-to-jump transition. Neighbor ranks exhibiting large values of $\rho_P(k)$ are therefore considered as candidates for constructing slow variables.

### E. Structural descriptors for slow-variable construction

The analyses described in Sec. II D provide a procedure for selecting neighbor ranks relevant to characterizing structural variations associated with jump dynamics. Using these structural quantities directly to define substates would lead to a multidimensional description involving several neighbor distances, which becomes impractical for evaluating the survival probability in the slow-fluctuation limit. Moreover, structural rearrangements accompanying jump motion involve heterogeneous displacements of surrounding particles across coordination shells, so that different neighbor ranks may respond differently during the transition. Consequently, simple averaging over multiple structural components does not provide a well-defined representation of the local structural state. These considerations motivate the construction of reduced structural descriptors that combine the contributions of relevant neighbor ranks into a small number of variables that can serve as slow variables.

Based on this approach, we construct a structural slowness parameter (SSP) using the neighbor ranks identified in Sec. II D. As an additional descriptor, we also consider the Voronoi free volume associated with each particle.

### (i) Structural slowness parameter (SSP)

Using the selected set of neighbor ranks $K$, we construct a composite structural descriptor based on nearest-neighbor distances. For each rank $k \in K$, the equilibrium mean distance is defined as

$$\mu_{k,eq} = \left\langle r_k(t) \right\rangle_{eq}, \tag{15}$$

and the corresponding equilibrium variance is

$$\sigma_{k,eq}^2 = \left\langle \left[ r_k(t) - \mu_{k,eq} \right]^2 \right\rangle_{eq}, \tag{16}$$

where $r_k(t)$ denotes the distance between a particle and its $k$-th nearest neighbor at time $t$. To characterize structural variations associated with configurations near the hopping threshold, the corresponding mean distance $\mu_{k,hop}$ is obtained from configurations with hop amplitudes in a small interval around $h^*$. To account for the different sensitivities of individual neighbor ranks to structural changes associated with jump motion, we introduce rank-dependent weights. The weight associated with neighbor rank $k$ is then defined as

$$w_k = \frac{\mu_{k,hop} - \mu_{k,eq}}{\sigma_{k,eq}^2}. \tag{17}$$

These weights assign larger contributions to neighbor ranks whose distance distributions change more strongly near the hop threshold. The structural slowness parameter (SSP) is then defined as

$$S(t) = \sum_{k \in K} w_k \left[ r_k(t) - \mu_{k,eq} \right]. \tag{18}$$

The SSP provides a reduced structural coordinate that combines the contributions of several neighbor ranks associated with cage-to-jump transitions and serves as a candidate slow variable for defining substates in the slow-fluctuation analysis.

### (ii) Voronoi free volume

As an additional structural descriptor, we consider the Voronoi free volume associated with each particle, which provides a geometric measure of local packing fluctuations.[69] The Voronoi volume $V(t)$ is obtained from a radical Voronoi tessellation of the instantaneous configuration using the Voro++ library,[70] which constructs a Voronoi polyhedron surrounding each particle. The free volume is defined as

$$v(t) = V(t) - v_p, \tag{19}$$

where $v_p = (\pi/6)\sigma_\alpha^3$ denotes the intrinsic volume of a spherical particle of species $\alpha$, with $\sigma_\alpha$ being the corresponding particle diameter. The Voronoi free volume is examined as an additional structural descriptor within the same framework.

For each structural descriptor considered above, we analyze the equilibrium distribution and the distribution near $h^*$. These distributions are used to define substates of the structural variable for evaluating the survival probability in the slow-fluctuation limit described in Sec. II C.

## F. Static point-to-set correlations

To examine the correspondence between the spatial extent of static amorphous correlations and the slow variables governing relaxation, we evaluate the static PTS correlation length $\xi_{PTS}$. The PTS length quantifies the spatial extent over which frozen boundary conditions constrain particle configurations.[33,71–79]

The PTS length is determined using a spherical-cavity protocol in which particles located at distances greater than $R$ from the cavity center are frozen at their initial positions, while particles inside the cavity evolve dynamically.[33,75,77,78] For each cavity size, the system is equilibrated in the NVT ensemble at the target temperature, followed by production runs in the NVE ensemble for data collection. The system is discretized into cubic cells of volume $\upsilon = 0.36\sigma_A^3$, where $\sigma_A$ is the diameter of A particles, and cells containing frozen particles are excluded from the calculation. Overlap measurements are restricted to the interior region of the cavity to minimize boundary effects. The static overlap is evaluated within a central cubic region of the cavity consisting of $5^3$ grid cells and is defined as[75,78]

$$q(R,t) = \frac{\sum_i \langle n_i(t) n_i(0) \rangle}{\sum_i \langle n_i(0) \rangle}, \tag{20}$$

where $n_i(t) \in \{0, 1\}$ denotes the occupation number of the $i$-th cubic cell at time $t$.

The long-time overlap $q_c(R) \equiv \lim_{t\to\infty} q(R,t)$ is obtained from the plateau value of $q(R,t)$ after relaxation within the cavity. Simulation times are chosen to be 100-1000 times the structural relaxation time $\tau_\alpha$ (defined from the decay of the self-overlap function; see Sec. III A), depending on temperature and cavity size, to ensure convergence. For each value of $R$, the results are averaged over 25 independent cavity realizations. To extract the PTS length, the excess overlap $\tilde{q}(R) = q_c(R) - q_{rand}$ (with $q_{rand} = \rho\upsilon$, where $\rho$ is the number density and $\upsilon$ is the cell volume) is fitted to a compressed exponential form,[75,78]

$$\tilde{q}(R) = A \exp\left[ -\left( \frac{d-a}{\xi_{PTS}} \right)^{\eta} \right], \tag{21}$$

where $A$ and $\eta$ are fitting parameters, and $a$ = 1 accounts for the minimal cavity size of the order of the particle diameter.

# III. RESULTS AND DISCUSSION

## A. Relaxation dynamics across fragility regimes

We first calculated the self-overlap function across all densities to characterize the relaxation dynamics. The self-overlap function is defined as

$$Q(t)=\frac{1}{N}\sum_{i=1}^{N} w\left(\left|\mathbf{r}_i(t)-\mathbf{r}_i(0)\right|\right), \tag{22}$$

where $\mathbf{r}_i(t)$ denotes the position of particle $i$ at time $t$, and the window function satisfies $w(x) = 1$ for $x < a$ and $w(x) = 0$ otherwise. The cutoff distance was chosen as $a = 0.3$ to suppress decorrelation arising from vibrational motion within cages.[33,76] Temperature-dependent $Q(t)$ curves for representative densities are shown in Fig. S1. Their structural relaxation times $\tau_\alpha$ were extracted from $Q(t)$ as the times at which $Q(t) = e^{-1}$. To quantify the temperature dependence of the relaxation dynamics across densities, the structural relaxation times were fitted using the Vogel–Fulcher–Tammann (VFT) relation[80–82]

$$\tau_\alpha = \tau_0 \exp\left[\frac{1}{K_{VFT}\left(T/T_{VFT}-1\right)}\right], \tag{23}$$

where $\tau_0$, $T_{VFT}$, and $K_{VFT}$ are fitting parameters, with $T_{VFT}$ denoting the divergence temperature and $K_{VFT}$ characterizing the kinetic fragility. To facilitate comparison across densities, the glass transition temperature $T_g$ was defined as the temperature at which $\tau_\alpha = 5\times10^6$, consistent with previous studies of this model.[21,32,33]

Figure 1 depicts the Angell plot[9,10] of $\tau_\alpha$ as a function of the rescaled inverse temperature $T_g/T$ for densities ranging from $\rho = 0.65$ to 0.82. At the lowest density studied ($\rho = 0.65$), the relaxation times exhibit an approximately Arrhenius temperature dependence over the investigated range, indicating strong behavior. In contrast, at the highest density ($\rho = 0.82$), $\tau_\alpha$ displays a pronounced super-Arrhenius growth upon cooling, characteristic of fragile dynamics. Intermediate densities show a systematic evolution between these limiting behaviors. It is noted that at the lowest temperatures examined for each density, the structural relaxation times attain comparable values of $\tau_\alpha$, ensuring that the systems are compared at similar relaxation timescales across densities. The fragility indices $K_{VFT}$, obtained from the VFT fits, are shown in the inset of Fig. 1 and increase monotonically with density. These results are consistent with previous studies of this model system.[21,32,33] Angell plots obtained separately for particles A and B, using the species-resolved relaxation times $\tau_{\alpha,A}$ and $\tau_{\alpha,B}$, are shown in Fig. S2, together with the corresponding kinetic fragilities. At higher densities ($\rho \geq 0.78$), particle A exhibits slightly larger $K_{VFT}$ values than particle B, whereas at lower densities the fragilities of the two species become comparable. In the following sections, we focus on three representative densities, $\rho = 0.65$, 0.72, and 0.82, referred to as the strong, intermediate, and fragile regimes hereafter.

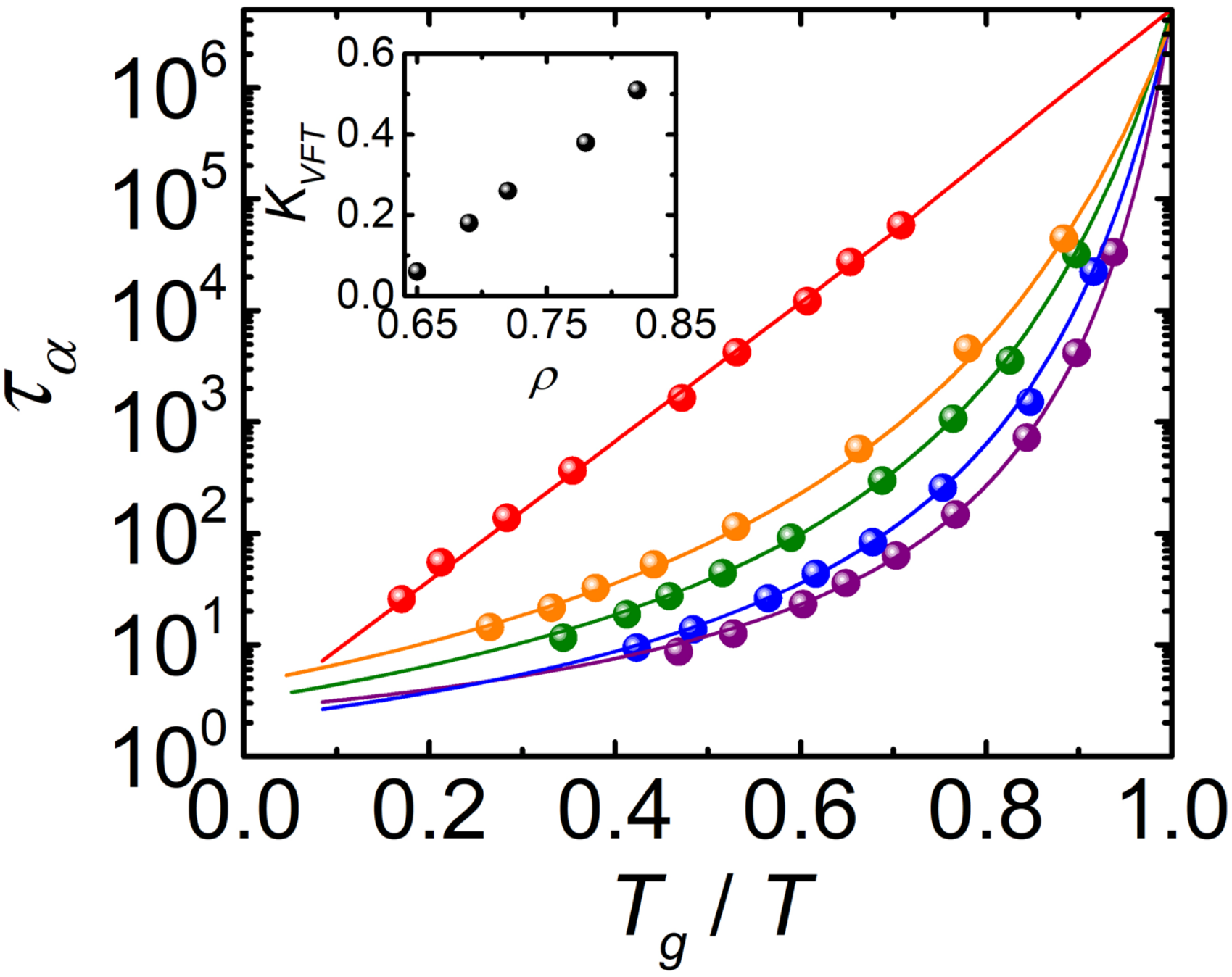

**FIG. 1.** Angell plot of the structural relaxation time $\tau_\alpha$ as a function of the rescaled inverse temperature $T_g/T$ for the soft-repulsive binary mixture at different densities. Symbols denote simulation data, and solid lines represent fits to the VFT relation. The densities $\rho = 0.65$, 0.69, 0.72, 0.78, and 0.82 are shown in red, orange, green, blue, and purple, respectively. The inset shows the kinetic fragility $K_{VFT}$ obtained from the VFT fits as a function of density. Increasing density leads to progressively stronger super-Arrhenius growth of $\tau_\alpha$, with $\rho = 0.65$ exhibiting strong behavior and $\rho = 0.82$ exhibiting fragile behavior. Statistical uncertainties are smaller than the symbol size.

## B. Structural evolution during jump motion

As discussed in Sec. I, structural relaxation in glass-forming liquids proceeds through intermittent particle jump events in which particles escape transient cages formed by their neighbors. These cage-breaking rearrangements constitute the elementary processes responsible for long-time relaxation in both strong and fragile liquids.[37,38,43–45] Since the present system exhibits different fragilities depending on density, it allows us to examine how these jump events occur across different fragility regimes. To identify such events, we analyzed the hop function $h(t)$, defined in Sec. II.B, which measures particle displacements between neighboring time intervals.

Figure 2 shows the representative time evolution of the hop function $h(t)$ for particles A and B at densities $\rho = 0.65$, 0.72, and 0.82, obtained at the lowest temperature studied for each density and plotted as a function of the scaled time $t/\tau_\alpha$. In all cases, $h(t)$ exhibits long intervals of small fluctuations associated with vibrational motion within a cage, interrupted by sudden increases corresponding to particle jumps between cages. These representative trajectories suggest that jump events become more intermittent in the fragile regime than in the strong regime; this behavior is quantified in Sec. III C through residence-time distributions and survival probabilities. Consistent with the size difference between the two species, particle B generally exhibits smaller values of $h$ compared to particle A. To distinguish localized cage motion from the jump state, we used the threshold value $h^*$, as discussed in Sec. II.B, with $h_A^* = 0.2$ and $h_B^* = 0.1$. Configurations with $h < h^*$ correspond to the cage state, whereas those with $h > h^*$ correspond to the jump state.[43–45] At higher temperatures, where particle motion becomes more continuous, the hop construction primarily serves as an operational criterion for identifying large displacements.[38,39]

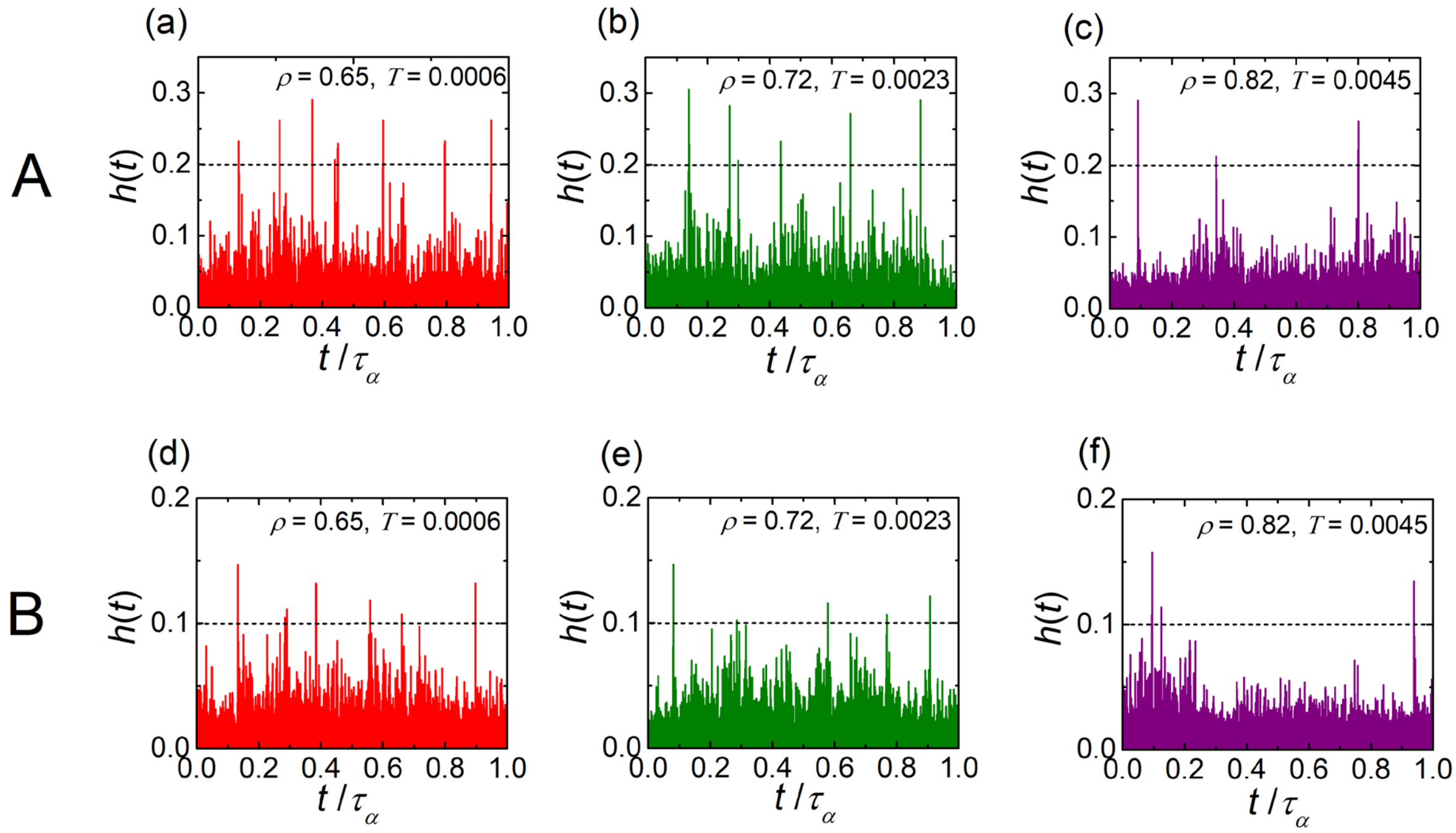

**FIG. 2.** Time series of the hop function $h(t)$ for particle A [(a)–(c)] and B [(d)–(f)] at the lowest temperature studied for each representative density. The densities and temperatures are $\rho$ = 0.65, $T$ = 0.0006 [(a), (d)]; $\rho$ = 0.72, $T$ = 0.0023 [(b), (e)]; and $\rho$ = 0.82, $T$ = 0.0045 [(c), (f)], corresponding to the strong, intermediate, and fragile regimes, respectively. The time axis is scaled by $\tau_\alpha$, obtained from the self-overlap function calculated over all particles. Dashed horizontal lines denote the hop thresholds $h_A^* = 0.2$ for particle A and $h_B^* = 0.1$ for particle B.

We next characterized the microscopic changes associated with jump motion by treating the hop amplitude $h$ as a reaction coordinate and analyzed thermodynamic and structural quantities as functions of $h$ for particles A and B across the densities and temperatures studied. The free-energy profiles along the hop coordinate $\Delta F(h)/(k_BT)$ are shown in Fig. S7 and discussed in the Supplementary Material. These profiles show that large-hop configurations become increasingly suppressed upon cooling, with a more pronounced temperature dependence in the fragile regime.

To examine the structural evolution accompanying jump motion, we analyzed the hop-resolved radial distribution functions $g_A(r, h)$ and $g_B(r, h)$ between a jumping particle of species A or B and its surrounding neighbors, without distinguishing the neighbor species, over different ranges of the hop amplitude $h$. The corresponding results at the lowest temperatures for the representative densities $\rho = 0.65$, 0.72, and 0.82 are shown in Fig. S8. For both particle types, the first coordination shell exhibits two closely spaced peaks. The peak at smaller $r$ primarily reflects contributions from the smaller A neighbors, whereas the peak at slightly larger $r$ is associated with B neighbors. At $\rho = 0.65$, increasing $h$ leads to a slight increase in the height of the peak at smaller $r$, together with a decrease in the height of the peak at larger $r$. A similar qualitative trend is observed at $\rho = 0.72$. In contrast, at $\rho = 0.82$, the heights of both peaks decrease as $h$ increases. In all cases, the minimum separating the first and second coordination shells becomes progressively shallower as $h$ increases. These trends indicate that jump motion is accompanied by a density-dependent reorganization of the local cage structure around the jumping particle. At $\rho = 0.65$ and 0.72, the increase of the peak at smaller $r$ together with the reduction of the peak at larger $r$, suggests a redistribution of the relative contributions of A and B neighbors within the first coordination shell, whereas at $\rho = 0.82$, both peaks decrease simultaneously, reflecting a weakening of the first-shell structure during the jump.

To quantify the evolution of the first coordination shell along the jump coordinate, we calculated the hop-dependent coordination numbers $CN_A(h)$ and $CN_B(h)$ by integrating the radial distribution functions up to the minimum separating the first and second coordination shells. The corresponding results for particles A and B over the investigated temperature range for the representative densities are shown in Fig. 3. At $\rho = 0.65$, both $CN_A(h)$ and $CN_B(h)$ increase systematically with $h$ at all temperatures considered, with each coordination number increasing by approximately one near the hop threshold $h^*$ at the lowest temperature. For $\rho =$ 0.72, the coordination numbers exhibit only a weak dependence on $h$ across the temperature range studied. In contrast, for $\rho = 0.82$, both coordination numbers decrease systematically with $h$ at all temperatures considered, with each coordination number decreasing by approximately one near $h^*$ at the lowest temperature. These results indicate that the population of neighbors within the first coordination shell evolves differently across densities. At $\rho = 0.65$, the increase in coordination number likely reflects the relatively open local packing, which allows neighbors near the first-shell boundary to enter the first shell during the jump, leading to a more highly coordinated local environment rather than an opening of the cage. At the intermediate density ($\rho = 0.72$), the weak dependence of the coordination number on $h$ suggests that jump motion primarily involves a redistribution of neighbors within the first shell. At $\rho = 0.82$, the decrease in coordination number indicates a partial loss of first-shell neighbors, suggesting an opening of the local cage during the jump. The decrease observed in the fragile regime is qualitatively similar to that reported for fragile water[43] and the Kob–Andersen Lennard-Jones (KALJ) system.[45] However, a decrease in coordination number is also observed in a strong glass former such as silica,[44] indicating that the sign of the coordination change is not determined by fragility alone but also depends on the local structural organization and interaction potential.

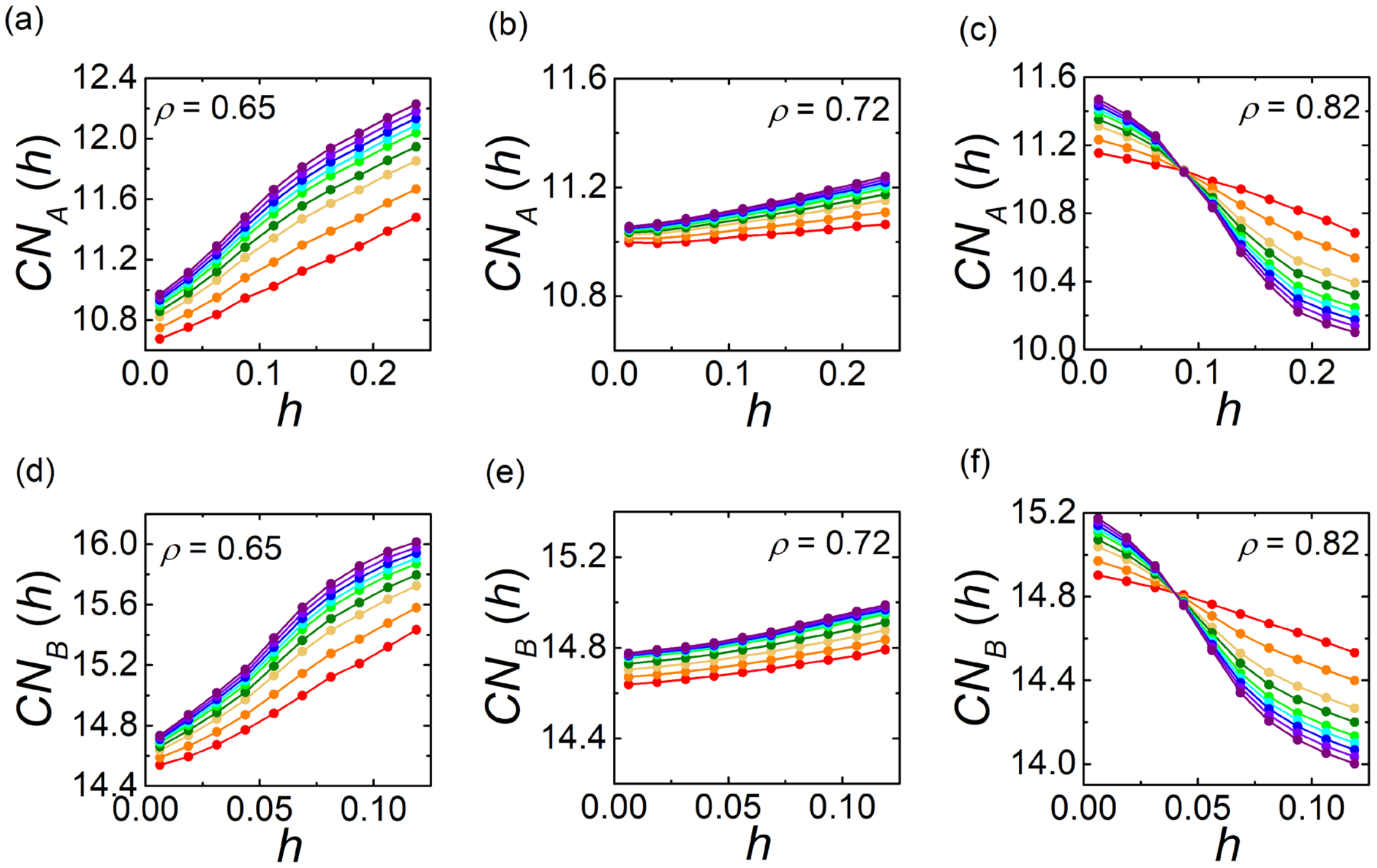

**FIG. 3.** Hop-dependent coordination numbers $CN_A(h)$ and $CN_B(h)$ for jumping particles A and B, respectively, at the representative densities. The coordination numbers were obtained by integrating the corresponding hop-resolved radial distribution functions up to the minimum separating the first and second coordination shells. Panels (a)–(c) show $CN_A(h)$, and panels (d)–(f) show $CN_B(h)$, for $\rho = 0.65$, 0.72, and 0.82, corresponding to the strong, intermediate, and fragile regimes, respectively. Colors indicate decreasing temperature from red to purple; the corresponding temperatures are listed in Table S1.

To further examine how jump motion affects the surrounding particles, we analyzed the displacements of particles located in successive coordination shells around a jumping particle. For each jump event, we calculated the average hop amplitude $\bar{h}_n$ of particles belonging to the $n$th coordination shell ($n$ = 1–4).[43,45] To compare these displacements with the equilibrium motion of the system, the shell-resolved hop amplitudes $\bar{h}_n$ were normalized by the equilibrium average hop amplitude $h_{eq}$. The temperature dependence of $\bar{h}_n/h_{eq}$ for the representative densities considered here is shown in Figs. S9–S11. At all densities, $\bar{h}_n/h_{eq}$ increases upon cooling. The magnitude of $\bar{h}_n/h_{eq}$ is largest in the first coordination shell and decreases systematically with increasing shell index, indicating that the displacement associated with a jump weakens with distance from the jumping particle.

To examine how this shell-resolved displacement varies across fragility regimes, we compare the values of $\bar{h}_n/h_{eq}$ for particles A and B at the lowest temperatures studied for the representative densities (Fig. 4). In the strong regime ($\rho$ = 0.65), $\bar{h}_n/h_{eq}$ is largest in the first coordination shell and remains noticeable in the second shell, but becomes very small in the third shell and essentially vanishes in the fourth shell. In the intermediate regime ($\rho$ = 0.72), the displacement extends farther and remains visible up to the third coordination shell. In contrast, in the fragile regime ($\rho$ = 0.82), $\bar{h}_n/h_{eq}$ remains finite up to the fourth coordination shell for both particle types. Qualitatively similar propagation of jump-induced displacements up to the fourth coordination shell has been reported in supercooled water in its fragile regime and the KALJ system.[43,45] The present results show that the spatial extent of the displacements associated with a jump depends strongly on fragility. In the strong regime, these displacements remain noticeable only up to the second coordination shell, whereas in the fragile regime, they persist up to the fourth coordination shell. This difference indicates that jump motion in the fragile regime involves more extended rearrangements of the surrounding particles, consistent with the more cooperative nature of relaxation in fragile liquids compared to strong liquids.[16–18,33,83,84]

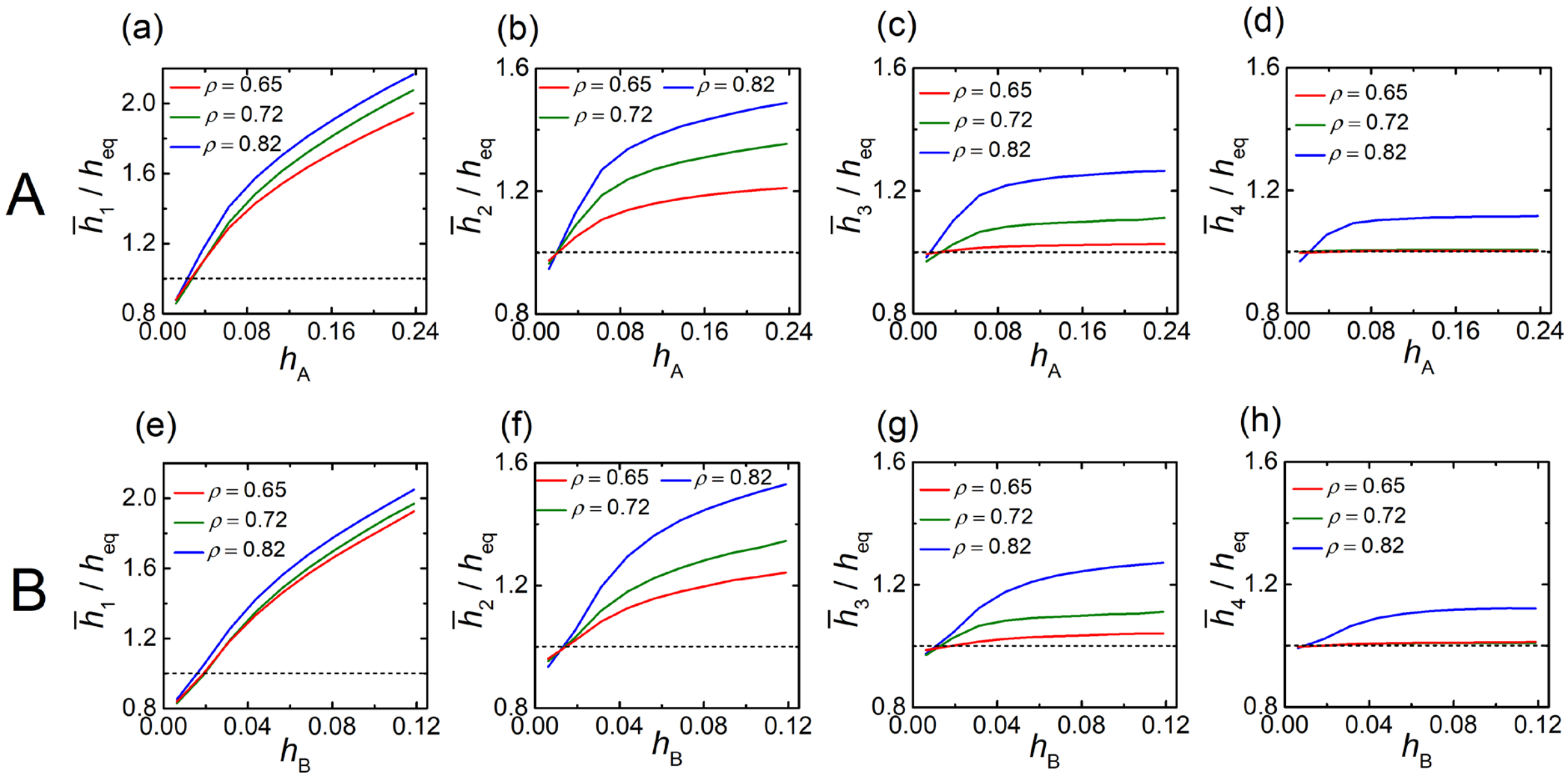

**FIG. 4.** Normalized average hop amplitude $\bar{h}_n/h_{eq}$ of particles in the $n$th coordination shell around a jumping particle, plotted as a function of the hop amplitude of the jumping particle. Results are shown for jumping particles A [(a)–(d)] and B [(e)–(h)]. The subscripts $n$ = 1–4 correspond to the first to fourth coordination shells. Data are shown at the lowest temperatures studied for the representative densities: $\rho$ = 0.65, $T$ = 0.0006; $\rho$ = 0.72, $T$ = 0.0023; and $\rho$ = 0.82, $T$ = 0.0045. Dashed horizontal lines denote $\bar{h}_n/h_{eq} = 1$, corresponding to the equilibrium average hop amplitude. Larger values of $\bar{h}_n/h_{eq}$ in outer shells at higher density indicate more extended displacements in the fragile regime.

## C. Transition in the nature of jump dynamics across fragility regimes

To examine how jump dynamics evolve upon cooling across fragility regimes, we analyzed the residence time distributions of the cage and jump states, identified using the threshold hop value $h^*$. Figures 5(a)–5(c) show the temperature-dependent residence-time distributions of the jump state, $\psi_{jump}(t)$, for particle A at $\rho$ = 0.65, 0.72, and 0.82. At all three densities, $\psi_{jump}(t)$ remains short and exhibits only a weak temperature dependence. Upon cooling, a slight broadening of the distribution is observed at each density, consistent with earlier observations of jump dynamics in glass-forming liquids.[42–44,85] In addition, $\psi_{jump}(t)$ becomes progressively broader as the density decreases from $\rho$ = 0.82 to 0.65. This behavior suggests that, at lower densities, weaker packing constraints may allow the surrounding particles to reorganize more gradually during a jump, leading to a wider range of jump durations.

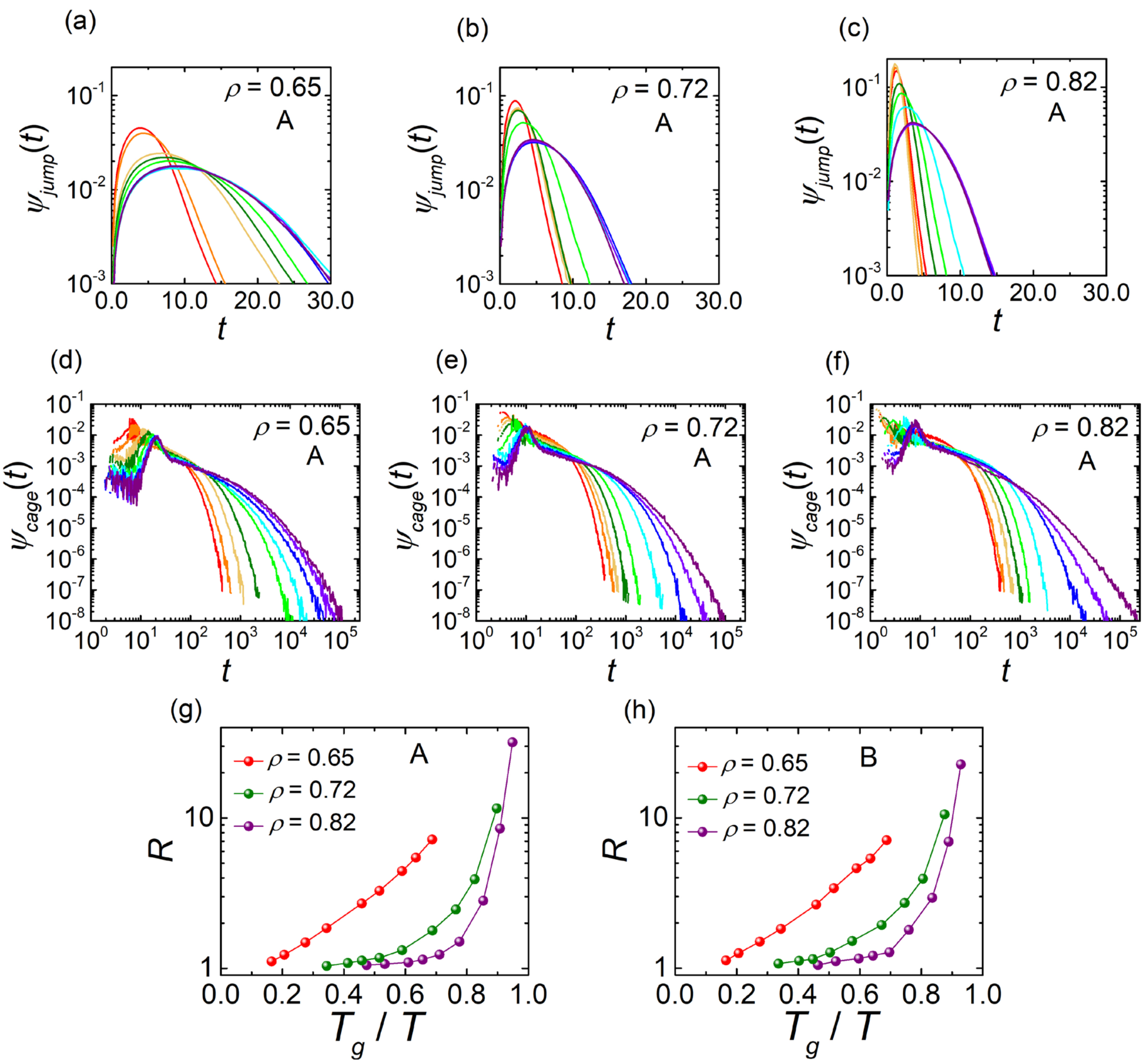

**FIG. 5.** Residence-time distributions and randomness parameter $R$ characterizing the transition from Poisson to non-Poisson statistics. The jump-state residence-time distributions $\psi_{jump}(t)$ for particle A are shown in (a)–(c), and the corresponding cage-state residence-time distributions $\psi_{cage}(t)$ are shown in (d)–(f), for $\rho = 0.65$, 0.72, and 0.82, respectively. Colors indicate decreasing temperature from red to purple in each plot; the corresponding temperature values are listed in Table S1. The jump-state distributions remain short and weakly temperature dependent, whereas the cage-state residence-time distributions broaden strongly and develop long tails upon cooling. The temperature dependence of $R$, shown in (g) and (h) for particles A and B, respectively, quantifies deviations from Poisson statistics in the cage-state residence times. Larger $R$ values at higher densities indicate more pronounced deviations from Poisson statistics, particularly in the fragile regime.

Figures 5(d)–5(f) show the residence-time distributions of the cage state, $\psi_{cage}(t)$, for particle A at the representative densities. In contrast to the jump-time distributions, $\psi_{cage}(t)$ exhibits a strong temperature dependence at all densities. At high temperatures, the distributions decay approximately exponentially, $\psi_{cage}(t) \sim \exp(-t/\tau)$, indicating that cage lifetimes follow nearly Poisson statistics with an approximately constant jump rate. Upon cooling, $\psi_{cage}(t)$ progressively deviates from exponential behavior and develops a pronounced long-time tail, reflecting non-Poisson statistics characterized by a broad distribution of cage lifetimes and increasingly intermittent dynamics.[39,43–45]

Further, the evolution of $\psi_{cage}(t)$ upon cooling varies systematically across the fragility regimes. In the strong regime ($\rho = 0.65$), the distribution remains relatively narrow over the investigated temperature range, although a clear long-time tail develops upon cooling. In the intermediate regime ($\rho = 0.72$), the long-time tail becomes more pronounced, reflecting a broader distribution of cage lifetimes. In contrast, in the fragile regime ($\rho = 0.82$), $\psi_{cage}(t)$ exhibits a strongly extended tail at low temperatures, spanning a wide range of timescales. These results indicate that cage lifetimes become increasingly heterogeneous with fragility, leading to more intermittent cage-to-jump dynamics. For particle B, similar trends are observed in both the jump- and cage-time distributions, and the corresponding results are shown in Fig. S12.

To quantify the deviation from Poisson statistics, we analyzed the randomness parameter $R$ (Eq. 5), which measures temporal heterogeneity in the cage-time distributions. Figures 5(g) and 5(h) show the temperature dependence of $R$, plotted as a function of $T_g/T$, for particles A and B at representative densities. $R$ exhibits a clear separation across densities, closely analogous to the fragility-dependent behavior observed in the Angell plots of structural relaxation times, indicating that the onset and growth of temporal heterogeneity depend systematically on fragility. At high temperatures $\left(T_g/T \ll 1\right)$, $R$ remains close to unity for all densities, consistent with exponential cage-time distributions and Poisson statistics, where cage lifetimes are characterized by a single timescale. Upon cooling, $R$ increases systematically, indicating the emergence of non-Poisson dynamics associated with a broad distribution of cage lifetimes.

The manner in which $R$ increases upon cooling differs across fragility regimes. In the strong regime ($\rho = 0.65$), $R$ increases progressively over a broad temperature range, indicating that deviations from Poisson statistics develop gradually upon cooling. This behavior is consistent with the nearly Arrhenius temperature dependence of $\tau_\alpha$, where the cage-to-jump statistics change continuously with temperature. In contrast, in the fragile regime ($\rho = 0.82$), $R$ remains close to unity over a wider temperature range and then increases rapidly at lower temperatures. This sharp growth indicates a sudden enhancement of temporal heterogeneity, consistent with the super-Arrhenius growth of $\tau_\alpha$. The intermediate regime ($\rho = 0.72$) exhibits behavior between these two limits. At the lowest temperatures investigated for each density, $R$ in the fragile regime significantly exceeds that in the strong regime, reflecting enhanced temporal heterogeneity and a broader distribution of cage lifetimes.

To further examine the deviations from Poisson behavior identified above, we analyzed the survival probability of the cage state, $C_S(t)$, which quantifies the probability that a particle remains in the cage state up to time $t$, and compared it with its fast fluctuation limit, $C_{fast}(t)$.[43–45,54] Figures 6(a)–6(i) show representative results for particle A at three temperatures for each density; the corresponding results for particle B are shown in Fig. S13. At high temperatures, $C_S(t)$ closely follows $C_{fast}(t)$, exhibiting an approximately exponential decay at all densities. This agreement indicates that the jump rate can be approximated as the average rate given by $k_{fast} = N_J / \left( N_{tot} \Delta t \right)$.[43,44] Upon cooling, $C_S(t)$ deviates systematically from $C_{fast}(t)$, showing a slower, non-exponential decay. This deviation indicates that the relaxation cannot be described by a single constant rate, and the approximation $k(t) \sim k_{fast}$ is no longer valid.[43–45,54] This behavior is observed at all densities, although the extent of the deviation varies across densities. To quantify how this non-exponential behavior varies across densities, we fit $C_S(t)$ to a stretched exponential form,[43–45,86]

$$C_S(t) = \exp\left[ -\left( t/\tau_S \right)^{\beta} \right], \tag{24}$$

where $\tau_S = 1/k$ denotes the characteristic survival timescale of the cage state in terms of an effective jump rate $k$ and $\beta$ characterizes the degree of temporal heterogeneity in the jump dynamics.[14,43,44] The corresponding fast-limit timescale $\tau_{fast} = 1/k_{fast}$ is obtained from $C_{fast}(t)$. The extracted parameters are shown in Fig. 7.

Figures 7(a) and 7(b) show the temperature dependence of $\tau_S$, together with $\tau_{fast}$, for particles A and B, respectively, at representative densities as a function of $T_g/T$. The temperature dependence of both $\tau_S$ and $\tau_{fast}$ follows trends analogous to the Angell behavior of $\tau_\alpha$, with more pronounced super-Arrhenius growth at higher densities. At high temperatures, the two timescales coincide for all densities. Upon cooling, $\tau_S$ progressively exceeds $\tau_{fast}$, indicating that rate fluctuations persist on timescales comparable to the cage lifetime.[43–45,54] At the lowest temperatures investigated for each density, this separation is most pronounced at $\rho = 0.82$, intermediate at $\rho = 0.72$, and least pronounced at $\rho = 0.65$. These results show that rate fluctuations become more persistent upon cooling in the fragile regime.

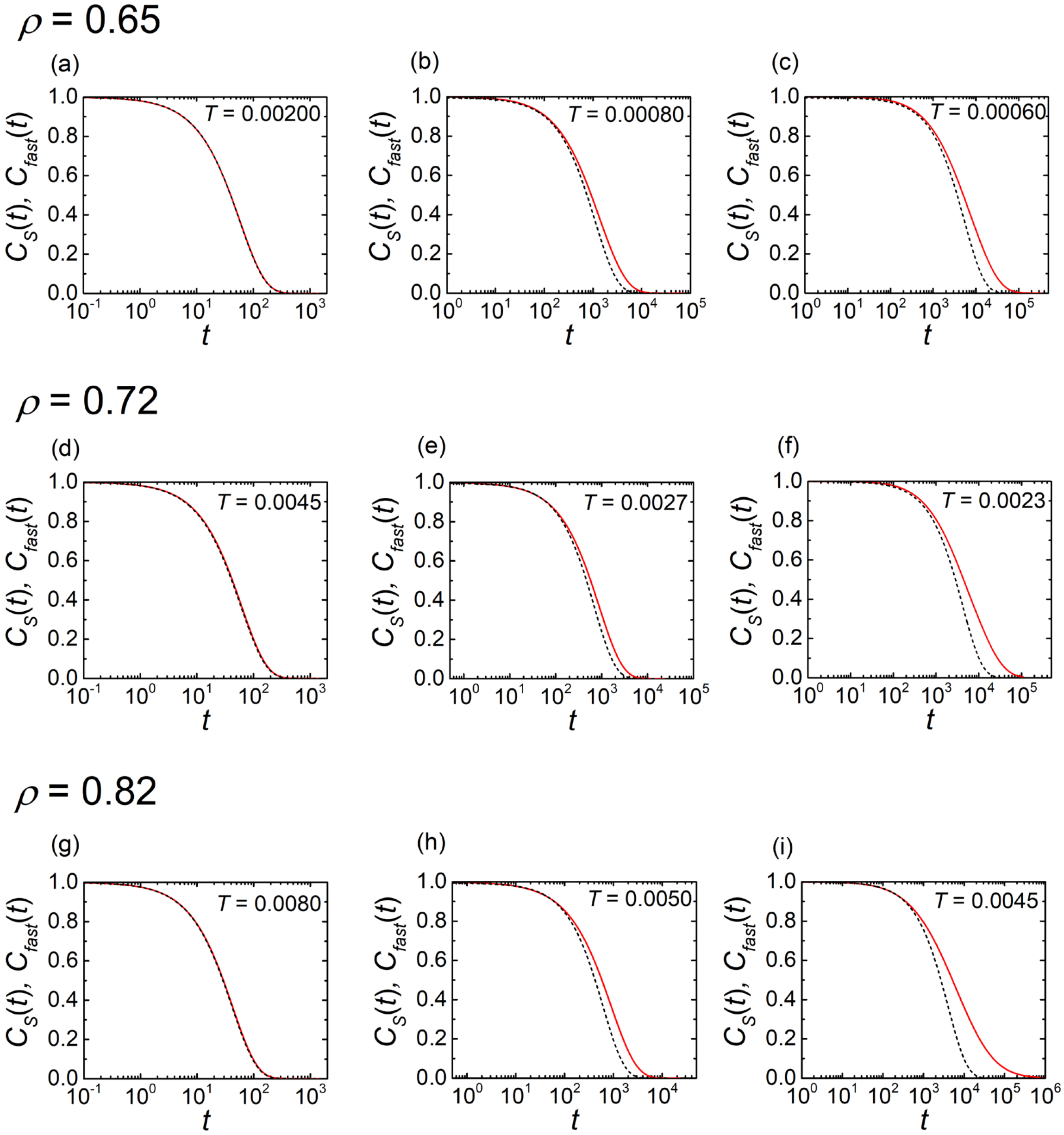

**FIG. 6.** Survival probability of the cage state, $C_S(t)$ (red), and its fast-fluctuation limit, $C_{fast}(t)$ (dashed black), for particle A at representative temperatures. Results are shown for $\rho = 0.65$ [(a)–(c)], $\rho = 0.72$ [(d)–(f)], and $\rho = 0.82$ [(g)–(i)]. At high temperatures, $C_S(t)$ closely follows $C_{fast}(t)$, consistent with Poisson statistics. Upon cooling, systematic deviations emerge, indicating the breakdown of the fast-fluctuation approximation.

Figures 7(c) and 7(d) show the temperature dependence of the stretching exponent $\beta$ for particles A and B, respectively, at representative densities, plotted as a function of $T_g/T$, exhibiting a clear density-dependent separation. At high temperatures, $\beta$ remains close to unity for all densities, indicating a near-exponential decay of $C_S(t)$. Upon cooling, $\beta$ decreases for all densities, reflecting the emergence of temporal heterogeneity in the jump dynamics. At $\rho =$ 0.65, $\beta$ decreases gradually over a broad temperature range, indicating a progressive broadening of the distribution of jump timescales. In contrast, at $\rho = 0.82$, $\beta$ remains close to unity over a wider temperature range and then decreases rapidly at lower temperatures, indicating a more abrupt broadening of the distribution of jump timescales. The intermediate density $\rho = 0.72$ exhibits behavior between these two limits. At the lowest temperatures investigated, $\beta$ attains significantly smaller values at higher densities, reaching $\beta \approx 0.5$ at $\rho =$ 0.82, compared to $\beta \approx 0.7$ at $\rho = 0.65$, indicating that cage-to-jump dynamics become more temporally heterogeneous in the fragile regime.

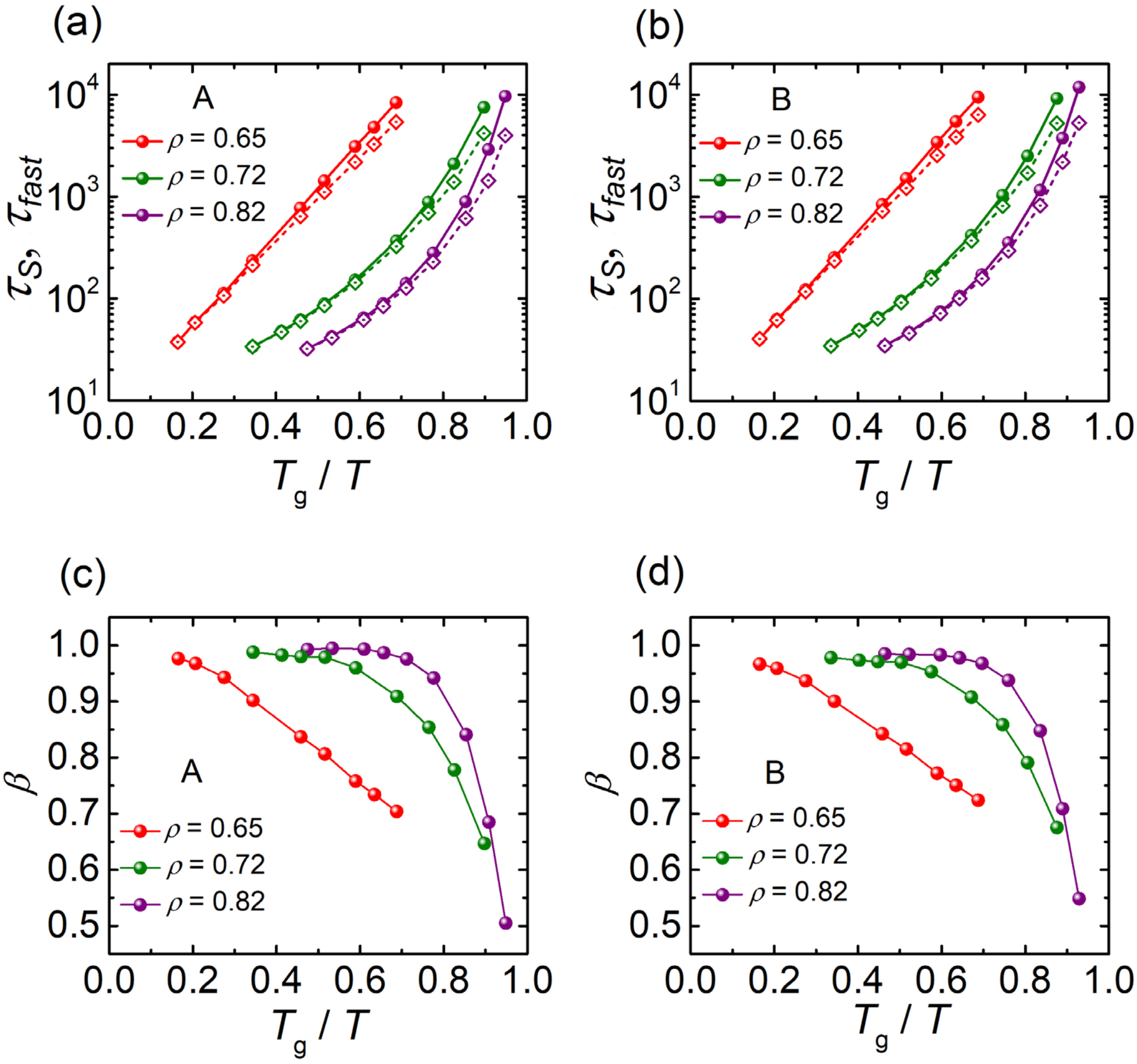

**FIG. 7.** Temperature dependence of the cage-state survival time $\tau_S$, fast-limit timescale $\tau_{fast}$, and stretching exponent $\beta$. (a), (b) $\tau_S$ and $\tau_{fast}$ as functions of $T_g/T$ for particles A and B, respectively, at the representative densities. Filled circles denote $\tau_S$, and open diamonds denote $\tau_{fast}$. (c), (d) Corresponding stretching exponent $\beta$ for particles A and B, respectively. The separation between $\tau_S$ and $\tau_{fast}$, together with the decrease in $\beta$, becomes more pronounced at higher densities, indicating stronger temporal heterogeneity in the fragile regime.

The increasing separation between $\tau_S$ and $\tau_{fast}$, together with the decrease in $\beta$ upon cooling, indicates that the jump dynamics cannot be described using a single characteristic timescale. The effect becomes more pronounced at higher densities, reflecting stronger dynamic disorder in the fragile regime. These results suggest that the hop coordinate $h$ alone is insufficient to fully describe the jump dynamics at lower temperatures across all densities, indicating the presence of additional microscopic variables that govern the rate fluctuations. In the following section, we systematically examine the structural factors associated with dynamic disorder across representative densities corresponding to the strong, intermediate, and fragile regimes.

### D. Microscopic origin of dynamic disorder in jump dynamics

To identify the structural origin of dynamic disorder across fragility regimes, we examined how the local environment evolves in the vicinity of a jump and how these structural changes relate to variations in the jump rate. We first compared the distributions of nearest-neighbor distances of jumping particles at the hop threshold with the corresponding equilibrium distributions. These comparisons show that structural changes involve a broad range of neighbor ranks, making it difficult to directly identify the structural components most relevant to the jump dynamics.

To resolve this, we analyzed the nearest-neighbor distances using the KL divergence $D_{eq}(k)$ and Pearson correlation $\rho_P(k)$, as introduced in Sec. II D. The quantity $D_{eq}(k)$ characterizes the deviation of the distribution of the $k$-th neighbor distance near the hop threshold from its equilibrium counterpart, whereas $\rho_P(k)$ quantifies the coupling between fluctuations in individual neighbor distances and variations in the hop amplitude. These two measures provide complementary information, enabling us to identify the neighbor ranks most relevant to jump motion. Based on the selected neighbor ranks, we constructed structural slowness parameters (SSPs), as described in Sec. II E. The SSP provides a reduced structural coordinate that combines the contributions of relevant neighbors and serves as a candidate slow variable associated with fluctuations in the jump rate.

In the following, we identify the relevant neighbor ranks, construct the corresponding SSPs, and examine the slow-fluctuation description separately for the representative densities corresponding to the strong ($\rho = 0.65$), intermediate ($\rho = 0.72$), and fragile ($\rho = 0.82$) regimes. For clarity, we present results for particle A; the corresponding results for particle B are provided in the Supplementary Material.

### (i) Strong regime ($\rho$ = 0.65)

We first discuss the dynamic disorder in the strong regime ($\rho$ = 0.65). Figure 8(a) shows $D_{eq}(k)$ and $\rho_P(k)$ as a function of neighbor rank $k$ for jumping particles of type A at $T$ = 0.0006. $D_{eq}(k)$ exhibits large values over a limited range of $k$, indicating that the structural rearrangements during the jump of particle A are dominated by these neighbor ranks, whereas $\rho_P(k)$ shows large positive and negative values over the same range, reflecting strong coupling between fluctuations in these neighbor distances and variations in the hop amplitude. In particular, significant contributions are observed for $k$ = 6–10 and 12–20, whereas contributions from other ranks are smaller. These neighbor ranks correspond to the outer part of the first coordination shell and the onset of the second shell, as indicated by the rank-resolved radial distribution profile $g_A\left(\langle r_k \rangle\right)$ shown in the lower subpanel of Fig. 8(a). The sign of $\rho_P(k)$ provides further insight into the nature of these rearrangements: for $k$ = 6–10, $\rho_P(k)$ is predominantly positive, indicating that these neighbors move away from the jumping particle as it approaches the jump state, whereas for $k$ = 12–20, $\rho_P(k)$ is predominantly negative, indicating that these neighbors move closer to the jumping particle. Thus, the jump involves a nonuniform rearrangement of the local environment, in which different neighbor groups are displaced in opposite directions.

Using these identified neighbor ranks and the definition in Eq. [18], we constructed two SSPs, $S^{(6\text{–}10)}$ and $S^{(6\text{–}10,\ 12\text{–}20)}$. Figures 8(b) and 8(c) show the corresponding distributions at equilibrium and $h^*$ for $T$ = 0.0006. The corresponding KL divergences are $D_{eq}(S^{(6\text{–}10)})$ = 0.22 and $D_{eq}(S^{(6\text{–}10,\ 12\text{–}20)})$ = 0.54, both of which are larger than those of the selected individual neighbor ranks, indicating that the combined variables capture the structural changes associated with the jump more effectively. The jump rate is strongly modulated by both SSPs, with a stronger dependence on $S^{(6\text{–}10,\ 12\text{–}20)}$, indicating an additional contribution from ranks 12–20. Both variables are therefore treated as candidate slow variables.

We next assessed the roles of $S^{(6\text{–}10)}$ and $S^{(6\text{–}10,\ 12\text{–}20)}$ as potential slow variables by analyzing the survival probability in the slow-fluctuation limit, where each SSP is treated as an additional coordinate complementing the hop function $h$. At higher temperatures ($T \geq 0.0020$), $C_S(t)$ closely follows $C_{fast}(t)$ [Fig. 9(a)], indicating that the fast-fluctuation description remains sufficient. In this regime, $C_{slow}^{S(6\text{–}10)}(t)$ decays more slowly than $C_S(t)$, indicating that $S^{(6\text{–}10)}$ does not yet control the cage-state survival dynamics. At $T$ = 0.0012, $C_S(t)$ progressively approaches

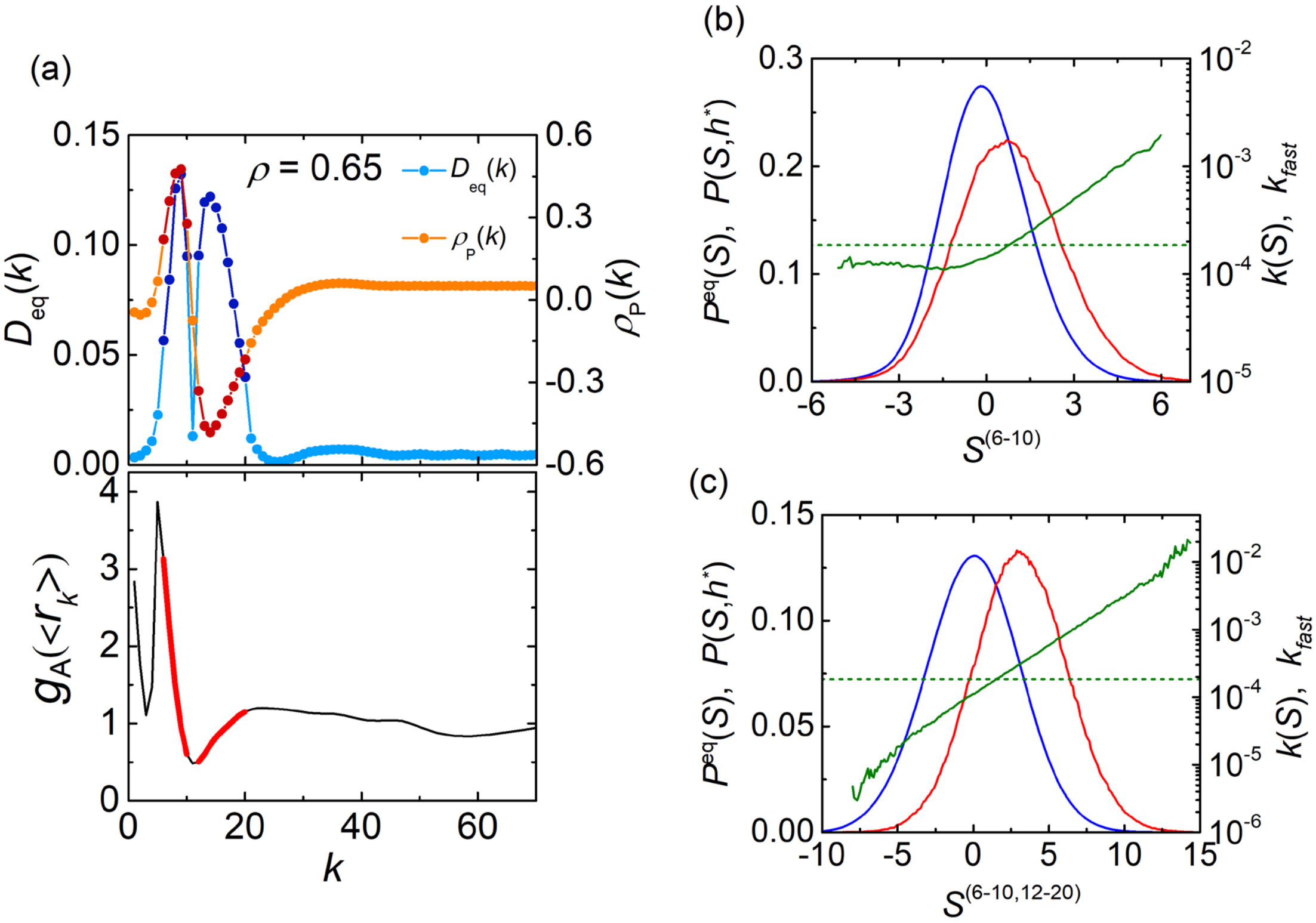

**FIG. 8.** Neighbor-rank analysis for jump motion of particle A at $T = 0.0006$ in the strong regime ($\rho = 0.65$). (a) KL divergence $D_{eq}(k)$ (blue) and Pearson correlation $\rho_P(k)$ (orange) as functions of neighbor rank $k$, where $k$ is defined with respect to the jumping A particle. The selected ranks $k$ = 6–10 and 12–20 are highlighted in dark red for $D_{eq}(k)$ and dark blue for $\rho_P(k)$. The lower subpanel shows the rank-resolved radial distribution profile $g_A(\langle r_k \rangle)$, with the same rank regions highlighted in red to indicate their location in the outer first shell and the onset of the second shell. These ranks are used to construct $S^{(6–10)}$ and $S^{(6–10,\ 12–20)}$. (b) Equilibrium distribution of $S^{(6–10)}$ (blue) and distribution at the hop threshold $h^*$ (red). The jump rate $k[S^{(6–10)}]$ is shown as a green solid line, and the green dashed line denotes $k_{fast}$. (c) Corresponding distributions and jump rate for the extended variable $S^{(6–10,\ 12–20)}$.

$C_{slow}^{S(6–10)}(t)$, indicating that a description based solely on $h$ becomes inadequate, and the selected neighbor ranks contribute to the rate fluctuations [Fig. 9(b)]. At $T = 0.0008$, $C_{slow}^{S(6–10)}(t)$ decays faster than $C_S(t)$, showing that these ranks alone are no longer sufficient to capture the dominant slow contributions [Fig. 9(c)]. We therefore consider the extended SSP $S^{(6–10,\ 12–20)}$. Its corresponding slow-fluctuation survival probability $C_{slow}^{S(6–10,12–20)}(t)$, provides an improved description of $C_S(t)$ for $T \leq 0.0008$. [Figs. 9(c)–9(f)]. At the lowest temperatures studied ($T$ = 0.0006), the two functions approach each other closely [Fig. 9(f)], indicating that the dominant rate fluctuations are captured by $S^{(6-10,\ 12-20)}$. Although the relevant set of neighbor ranks extends from $k = 6$–10 to $k = 6$–10, 12–20 upon cooling, the contributing ranks remain confined to the outer first shell and the onset of the second shell. This indicates that the structural rearrangements governing the jump dynamics remain localized and exhibit only a modest increase in spatial extent upon cooling, consistent with the behavior expected for a strong glass-forming liquid.[16,17,44] A similar trend is also observed for particle B [Fig. S15].

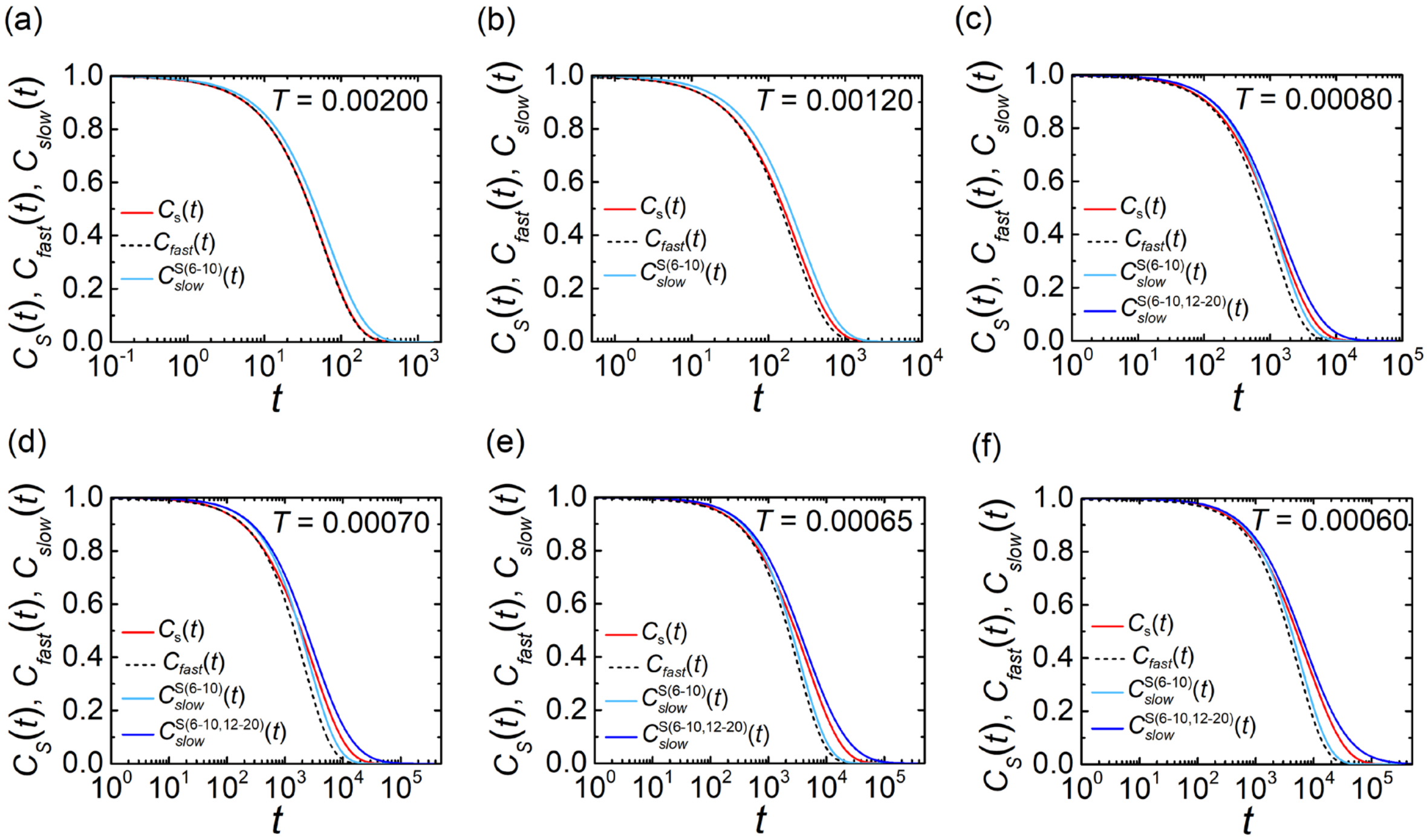

**FIG. 9.** Survival probability for the cage state, $C_S(t)$ (red), its fast-fluctuation limit, $C_{fast}(t)$ (black dashed), and the slow-fluctuation limits $C_{slow}^{S(6-10)}(t)$ (light blue) and $C_{slow}^{S(6-10,12-20)}(t)$ (blue), for particle A in the strong regime ($\rho = 0.65$). Results are shown at (a) $T = 0.0020$, (b) 0.0012, (c) 0.0008, (d) 0.0007, (e) 0.00065, and (f) 0.0006. The slow-fluctuation limit constructed from the extended SSP $S^{(6-10,\ 12-20)}$ is included in panels (c)-(f). Upon cooling, $C_S(t)$ progressively approaches the slow-fluctuation limit, with the extended SSP providing a better description at lower temperatures.

### (ii) Intermediate regime ($\rho = 0.72$)

We next examined the intermediate fragility regime, $\rho = 0.72$. Figure 10(a) depicts $D_{eq}(k)$ and $\rho_P(k)$ as a function of neighbor rank $k$ for jumping particles of type A at $T = 0.0023$. Large values are observed for $k = 6$–10, 12–19, and 29–37, while values at other ranks are finite but relatively smaller. Compared to the strong regime, an additional set of neighbor ranks in the inner part of the second shell becomes relevant, as also indicated by the rank-resolved radial distribution profile $g_A\left(\langle r_k \rangle\right)$, indicating that a broader range of structural rearrangements is associated with jump dynamics. Based on these observations, we constructed two SSPs, $S^{(6\text{–}10,\ 12\text{–}19)}$ and $S^{(6\text{–}10,\ 12\text{–}19,\ 29\text{–}37)}$. Figures 10(b) and 10(c) show the corresponding distributions at equilibrium and $h^*$. The KL divergences are $D_{eq}(S^{(6\text{–}10,\ 12\text{–}19)}) = 0.43$ and $D_{eq}(S^{(6\text{–}10,\ 12\text{–}19,\ 29\text{–}37)}) = 0.66$, indicating that the inclusion of the additional ranks results in a greater structural change along the SSP. Both variables are therefore treated as candidate slow variables in the following survival probability analysis.

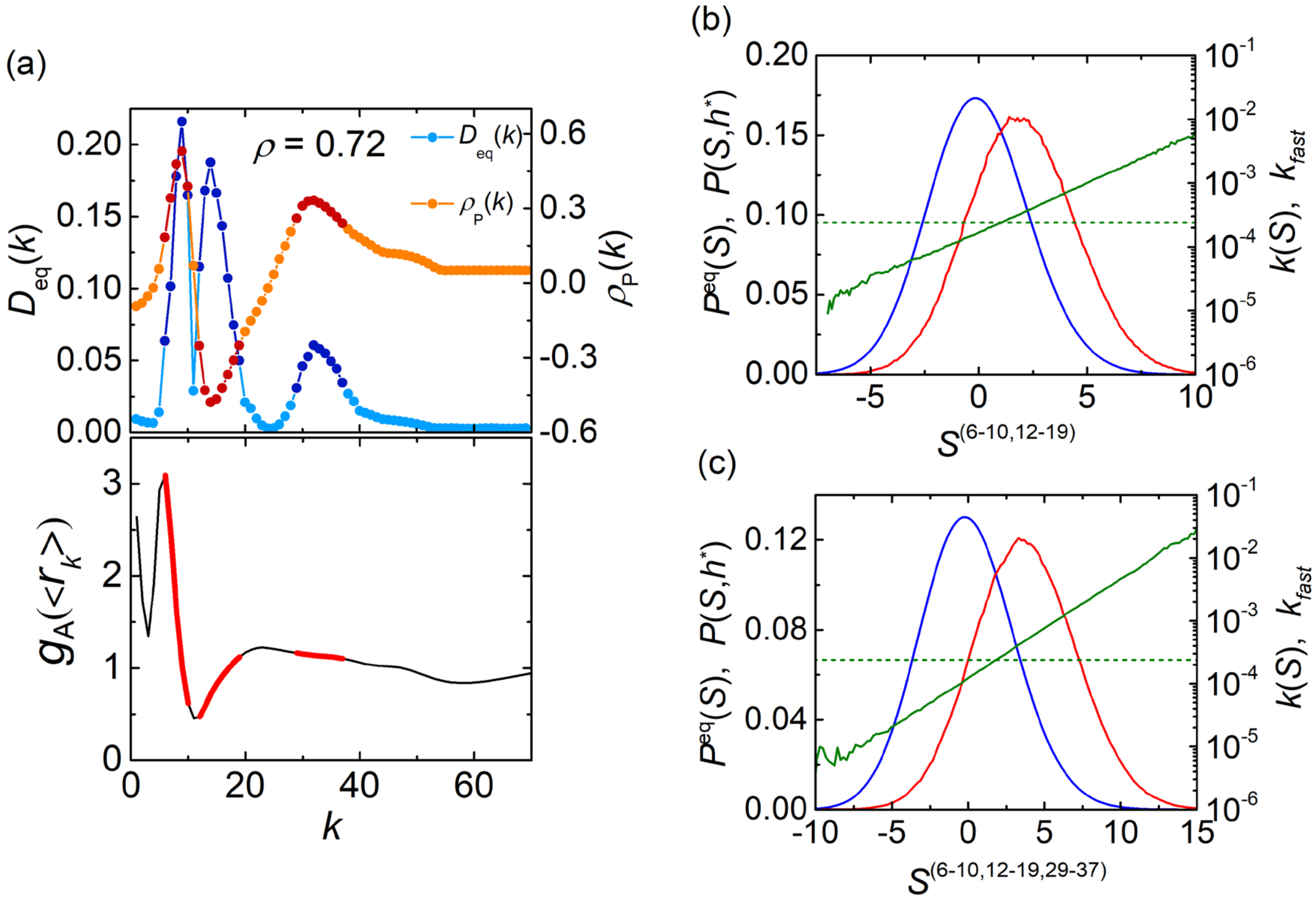

**FIG. 10.** Neighbor-rank analysis for jump motion of particle A at $T = 0.0023$ in the intermediate regime ($\rho = 0.72$). (a) KL divergence $D_{eq}(k)$ (blue) and Pearson correlation $\rho_P(k)$ (orange) as functions of neighbor rank $k$, where $k$ is defined with respect to the jumping A particle. The selected ranks $k = 6$–10, 12–19, and 29–37 are highlighted in dark red for $D_{eq}(k)$ and dark blue for $\rho_P(k)$. The lower subpanel shows the rank-resolved radial distribution profile $g_A\left(\langle r_k \rangle\right)$, with the same rank regions highlighted in red. (b) Equilibrium distribution of $S^{(6\text{–}10,\ 12\text{–}19)}$ (blue) and distribution at the hop threshold $h^*$ (red). The jump rate $k[S^{(6\text{–}10,\ 12\text{–}19)}]$ is shown as a green solid line, and the green dashed line denotes $k_{fast}$. (c) Corresponding distributions and jump rate for the extended variable $S^{(6\text{–}10,\ 12\text{–}19,\ 29\text{–}37)}$.

We next analyzed the survival probability in the slow-fluctuation limit using the SSPs defined above, $S^{(6\text{-}10,\ 12\text{-}19)}$ and $S^{(6\text{-}10,\ 12\text{-}19,\ 29\text{-}37)}$. At higher temperatures ($T \geq 0.0045$), $C_{fast}(t)$ closely approximates $C_S(t)$ [Fig. 11(a)], indicating that the jump dynamics are described by $h$. In this regime, $C_{slow}^{S(6-10,12-19)}(t)$ decays more slowly than $C_S(t)$, indicating that the corresponding structural fluctuations evolve on timescales shorter than the jump dynamics. As the temperature decreases ($T$ = 0.0030 and 0.0027), $C_S(t)$ approaches $C_{slow}^{S(6-10,12-19)}(t)$, indicating an increasing influence of these neighbor ranks on the rate fluctuations [Figs. 11(c)-11(d)]. However, upon further cooling ($T$ = 0.0025), $C_{slow}^{S(6-10,12-19)}(t)$ becomes faster than $C_S(t)$, indicating that this set alone does not capture the dominant slow contributions [Fig. 11(e)]. We therefore consider the extended SSP $S^{(6\text{-}10,\ 12\text{-}19,\ 29\text{-}37)}$. The corresponding slow-fluctuation survival probability, $C_{slow}^{S(6-10,12-19,29-37)}(t)$, decays more slowly than $C_S(t)$ and provides an improved description at lower temperatures. At the lowest temperature studied ($T$ = 0.0023), $C_{slow}^{S(6-10,12-19,29-37)}(t)$ provides a consistent description of the slow decay of $C_S(t)$ [Fig. 11(e)], indicating that the dominant rate fluctuations are captured within this set of ranks. These results indicate that, in contrast to the strong regime, contributions from more distant neighbor ranks become relevant upon cooling, reflecting a more pronounced increase in the spatial extent of the fluctuations that govern the jump dynamics.

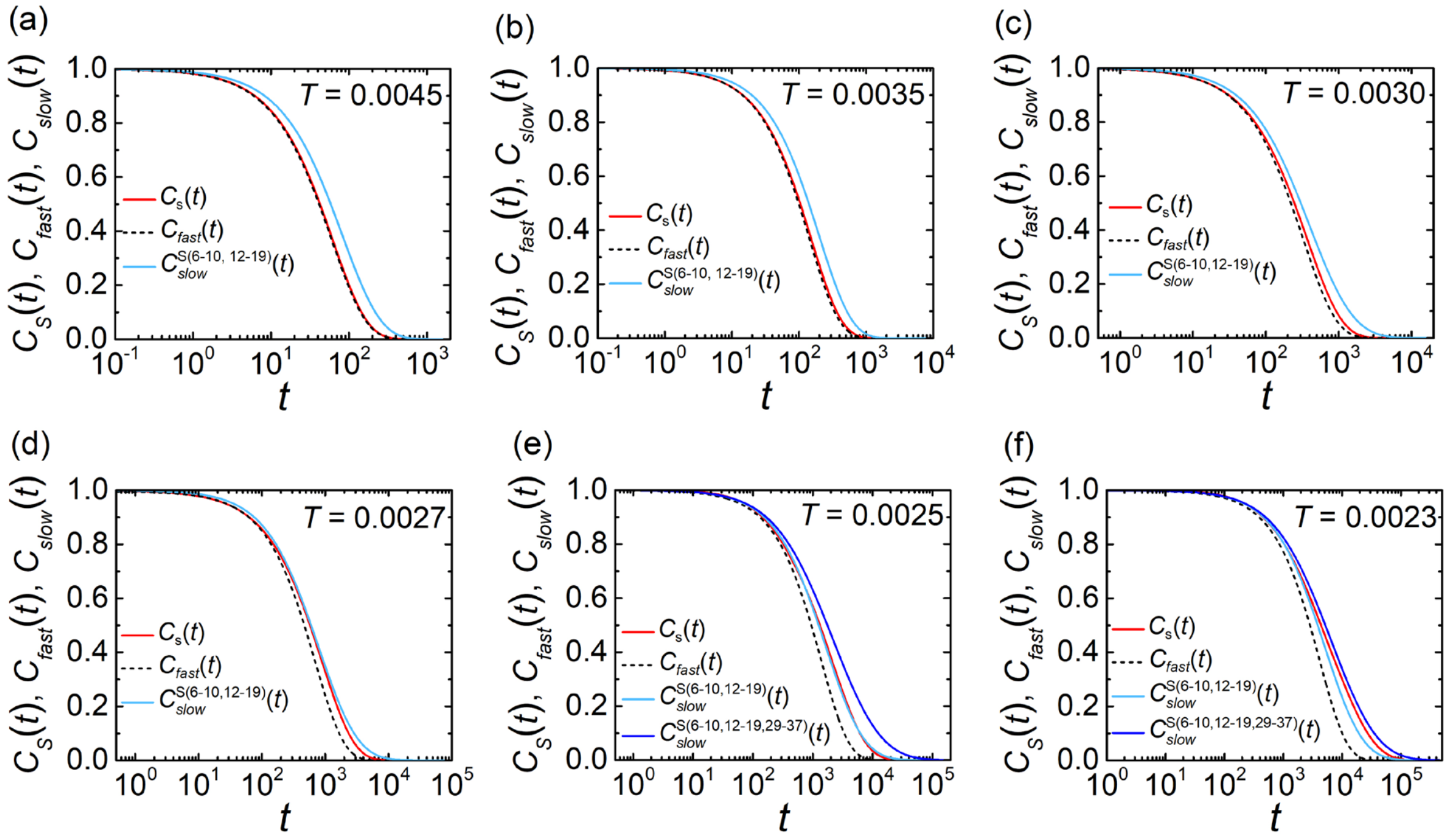

**FIG. 11.** Survival probability for the cage state, $C_S(t)$ (red), its fast-fluctuation limit, $C_{fast}(t)$ (black dashed), and the slow-fluctuation limits $C_{slow}^{S(6-10,12-19)}(t)$ (light blue) and $C_{slow}^{S(6-10,12-20,29-37)}(t)$ (blue) for particle A in the intermediate regime ($\rho = 0.72$). Results are shown at (a) $T = 0.0045$, (b) 0.0035, (c) 0.0030, (d) 0.0027, (e) 0.0025, and (f) 0.0023. The extended slow variable is shown only at the two lowest temperatures, where $S^{(6-10,\ 12-19)}$ no longer provides a sufficient description of $C_S(t)$. Upon cooling, the extended variable provides a more consistent description of the slow decay of $C_S(t)$, indicating the increasing role of more distant neighbor ranks in the intermediate regime.

### (iii) Fragile regime ($\rho$ = 0.82)

We finally examined the fragile regime corresponding to $\rho$ = 0.82. Figure 12(a) shows $D_{eq}(k)$ and $\rho_P(k)$ as a function of neighbor rank $k$ for jumping particles of type A at $T$ = 0.0045. In contrast to the strong and intermediate regimes, significant values of both $D_{eq}(k)$ and $\rho_P(k)$ are observed over a broader range of neighbor ranks, with prominent contributions for $k$ = 6–17, and additional contributions for $k$ = 24–55. These contributions arise from neighbor ranks associated with the outer part of the first coordination shell, the onset of the second shell, and the middle to outer part of the second shell, as indicated by the rank-resolved radial distribution profile $g_A(\langle r_k \rangle)$, [Fig. 12(a), lower subpanel]. Accordingly, we constructed two SSPs, $S^{(6-17)}$ and $S^{(6-17,\,24-55)}$, corresponding to a smaller and a more extended set of contributing neighbor ranks. The distributions of these SSPs at equilibrium and $h^*$ are shown in Figs. 12(b) and 12(c). The corresponding KL divergences are $D_{eq}(S^{(6-17)})$ = 0.51 and $D_{eq}(S^{(6-17,\,24-55)})$ = 0.73, with the extended SSP yielding a larger value, indicating a stronger structural distinction between the equilibrium and $h^*$ states.

We then examined whether the SSPs $S^{(6-17)}$ and $S^{(6-17,\,24-55)}$ can account for the survival probability in the slow-fluctuation limit. At higher temperatures ($T \geq 0.0080$), $C_S(t)$ coincides with $C_{fast}(t)$, whereas $C_{slow}^{S(6-17)}(t)$ decays more slowly than $C_S(t)$ [Fig. 13(a)]. Upon cooling ($T$ = 0.0060 and 0.0055), $C_S(t)$ shifts toward $C_{slow}^{S(6-17)}(t)$, indicating that the contribution of these neighbor ranks increases [Figs. 13(b)-13(c)]. However, at $T$ = 0.0050 $C_{slow}^{S(6-17)}(t)$ becomes faster than $C_S(t)$ at intermediate times, showing that this set does not capture the dominant slow contribution [Fig. 13(d)]. We therefore consider the extended SSP $S^{(6-17,\,24-55)}$. $C_{slow}^{S(6-17,24-55)}(t)$ decays more slowly than $C_S(t)$ down to $T$ = 0.0047 and provides an improved description compared to $C_{slow}^{S(6-17)}(t)$, indicating that these additional neighbor ranks play an important role in modulating the jump dynamics.

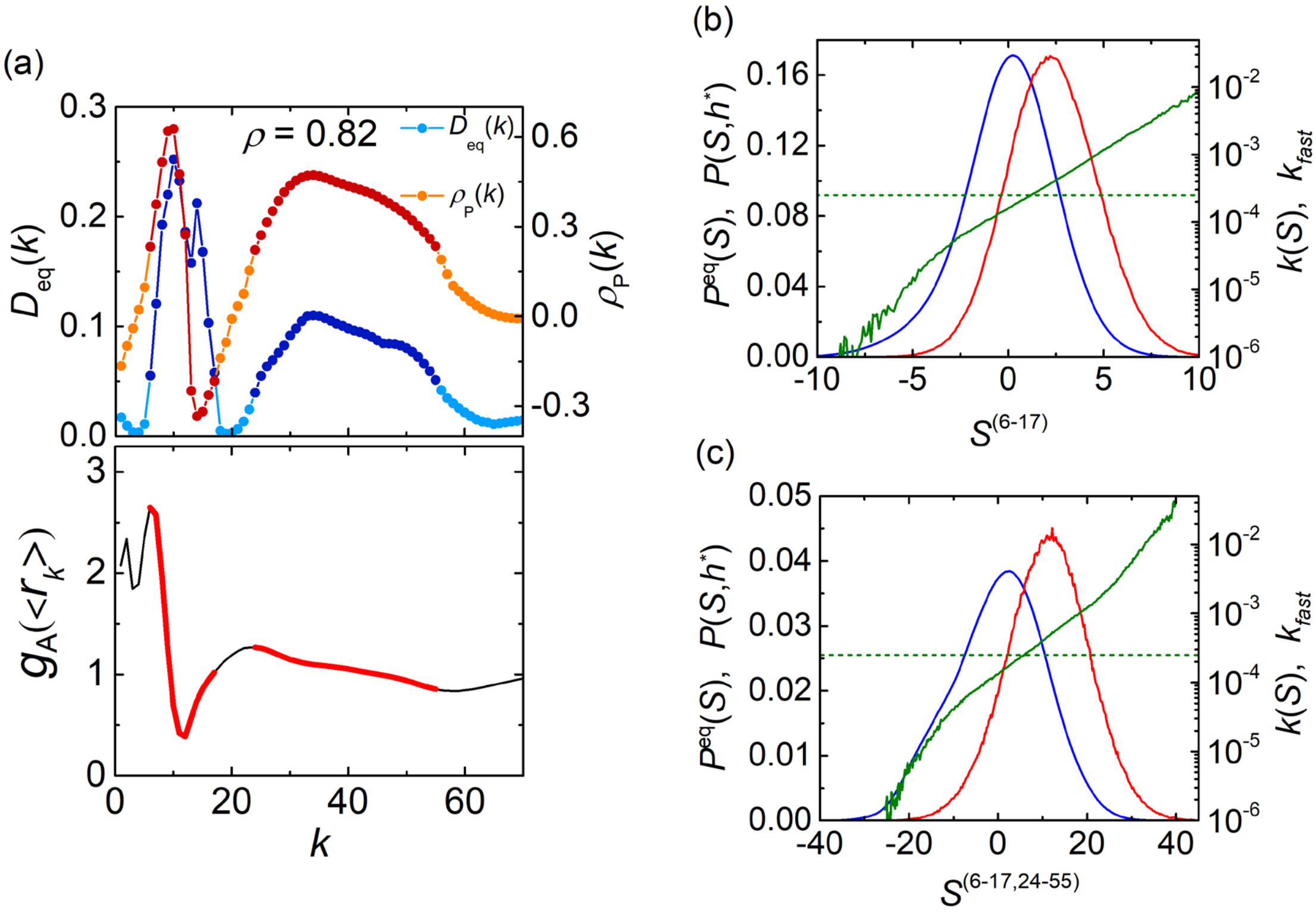

**FIG. 12.** Neighbor-rank analysis for jump motion of particle A at $T$ = 0.0045 in the fragile regime ($\rho$ = 0.82). (a) KL divergence $D_{eq}(k)$ (blue) and Pearson correlation $\rho_P(k)$ (orange) as functions of neighbor rank $k$, where $k$ is defined with respect to the jumping A particle. The selected ranks $k$ = 6–17 and 24–55 are highlighted in dark red for $D_{eq}(k)$ and dark blue for $\rho_P(k)$. The lower subpanel shows the rank-resolved radial distribution profile $g_A(\langle r_k \rangle)$, with the same rank regions highlighted in red. (b) Equilibrium distribution of $S^{(6–17)}$ (blue) and distribution at the hop threshold $h^*$ (red). The jump rate $k[S^{(6–17)}]$ is shown as a green solid line, and the green dashed line denotes $k_{fast}$. (c) Corresponding distributions and jump rate for the extended variable $S^{(6–17,\,24–55)}$.

At $T$ = 0.0045, however, $C_{slow}^{S(6–17,\,24–55)}(t)$ becomes faster than $C_S(t)$ [Fig. 13(f)], indicating that even this extended set of neighbor ranks does not completely capture the dominant slow contribution. Including neighbors from the third and subsequent shells does not improve the description (data not shown), indicating that further increasing the spatial range of pair-distance-based descriptors alone is insufficient. This suggests that the dominant slow fluctuations are not fully captured by pair-distance-based descriptors and may involve additional structural features associated with local packing. We therefore considered the Voronoi free volume as an additional structural variable. The slow-limit survival probability constructed using the Voronoi free volume alone, $C_{slow}^{v}(t)$, decays much faster than $C_S(t)$ (Fig. S18), demonstrating that it does not capture the dominant slow contribution. We therefore construct a two-dimensional slow variable using $S^{(6-17,\ 24-55)}$ and the Voronoi free volume ($v$). The corresponding slow-limit survival probability, $C_{slow}^{S(6-17,\,24-55)\times v}(t)$, provides a significantly improved description of $C_S(t)$ [Fig. 13(f)].

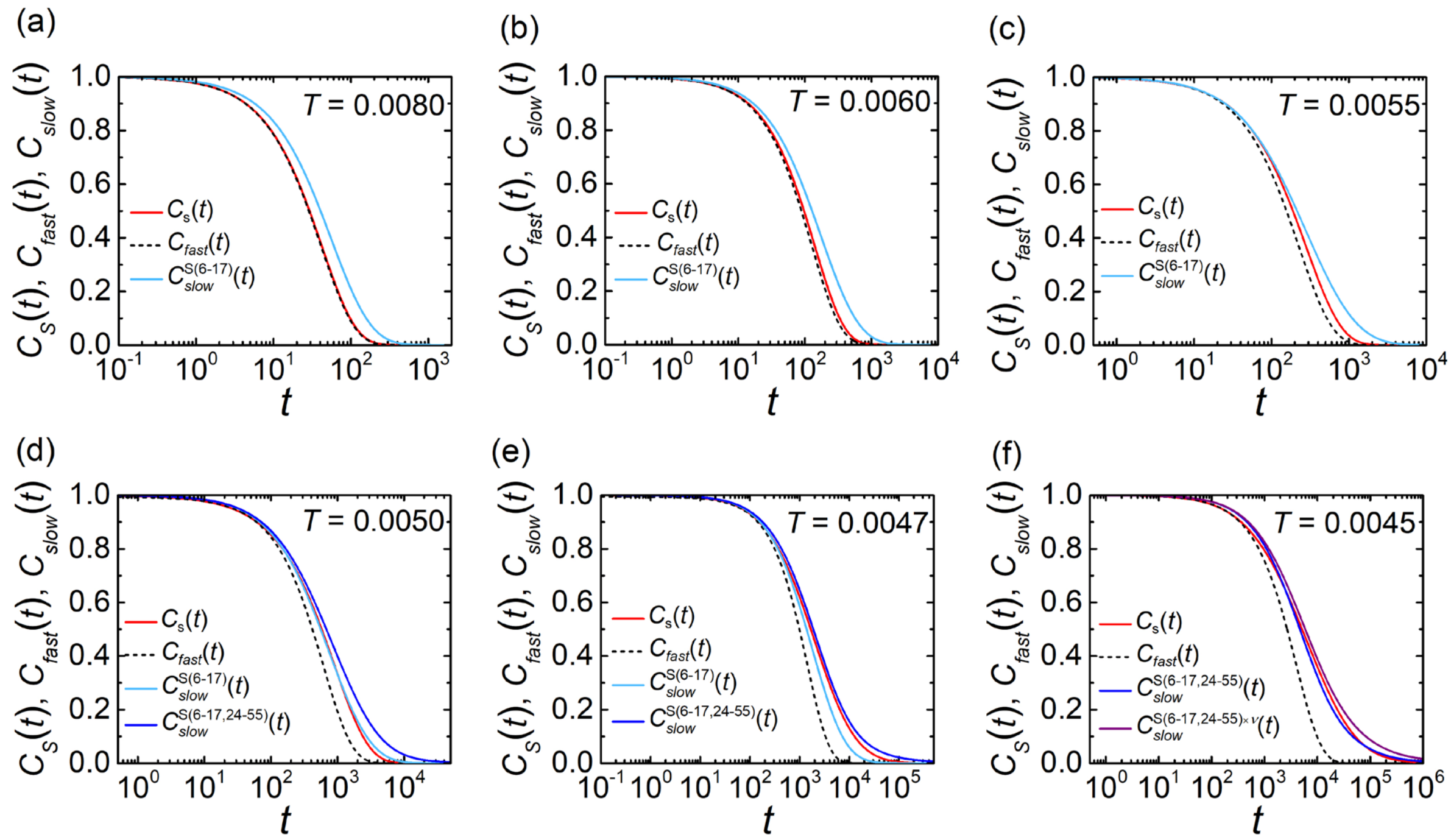

**FIG. 13.** Survival probability for the cage state, $C_S(t)$ (red), its fast-fluctuation limit, $C_{fast}(t)$ (black dashed), and slow-fluctuation limits for particle A in the fragile regime ($\rho$ = 0.82). Results are shown at (a) $T$ = 0.0080, (b) 0.0060, (c) 0.0055, (d) 0.0050, (e) 0.0047, and (f) 0.0045. The light-blue curve represents $C_{slow}^{S(6-17)}(t)$, and the blue curve represents $C_{slow}^{S(6-17,\,24-55)}(t)$. In (f), the magenta curve represents the slow-fluctuation limit $C_{slow}^{S(6-17,\,24-55)\times v}(t)$, constructed from the combined variable $S^{(6-17,\ 24-55)}\times v$, where $v$ denotes the Voronoi free volume. The extended distance-based SSP improves the description upon cooling, while the combined $S^{(6-17,\ 24-55)}\times v$ variable provides a better description at the lowest temperature, indicating that both interparticle distances and local packing contribute to the dominant slow fluctuations in the fragile regime.

To examine whether this improvement arises simply from adding another local structural descriptor, we also tested the bond-orientational order parameter $q_6$ as an alternative additional variable. Unlike the Voronoi free volume, which reflects fluctuations in local packing and available space around a particle, $q_6$ primarily characterizes the angular organization of neighboring particles.[87,88] When the two-dimensional slow variable is constructed using $S^{(6\text{-}17,\,24\text{-}55)}$ together with $q_6$, instead of the Voronoi free volume, the resulting slow-limit survival probability $C_{slow}^{S(6-17,\,24-55)\times q_6}(t)$ still decays faster than $C_S(t)$ (Fig. S18). This indicates that local bond-orientational order does not account for the residual slow dynamics not captured by the distance-based SSP as effectively as the Voronoi free volume. These results indicate that the dominant slow fluctuations arise from a coupled interplay between interparticle distances and local packing.

For particles of type B, however, the corresponding slow-limit is adequately described using distance-based SSPs alone (Fig. S20), and the Voronoi free volume is not required to capture the slow dynamics, consistent with the larger particle size reducing the role of local packing fluctuations.

These results reveal a systematic evolution in the structural origin of dynamic disorder across fragility regimes. While the slow dynamics in the strong regime are governed by localized structural rearrangements, progressively larger sets of neighbor ranks become relevant in the intermediate and fragile regimes, indicating an increasing spatial extent of the fluctuations controlling the jump dynamics.[44,45] In the fragile regime, the failure of distance-based descriptors alone and the need to incorporate local packing information further highlight the complex, many-body nature of these fluctuations.[19,21,89–93] This motivates a quantitative characterization of their spatial extent, which we examine using point-to-set correlations in the next section.

### E. Static length scale from point-to-set correlations

To quantify the spatial extent of the structural correlations associated with the slow dynamics identified above, we evaluated the point-to-set (PTS) correlation length, $\xi_{PTS}$, using the cavity protocol described in Sec. II F. The PTS length provides a static measure of amorphous order by quantifying the extent to which frozen boundary conditions constrain particle configurations within the cavity.[71,75,77,78] The overlap function was evaluated for the total system as well as separately for particles of types A and B. For clarity, we present results for particles of type A in the main text; corresponding results for the total system and particles of type B are provided in the Supplementary Material.

Figures 14(a)-14(c) show the dependence of the excess overlap $\tilde{q}(R)$ on the cavity

radius $R$ at different temperatures for the strong ($\rho = 0.65$), intermediate ($\rho = 0.72$), and fragile ($\rho = 0.82$) regimes, respectively, for particles of type A. The temperature dependence of $\tilde{q}(R)$ is relatively weak in the strong regime, whereas a more systematic separation among temperatures appears at higher densities. This behavior indicates an increasing sensitivity to the frozen boundary upon cooling, particularly in the fragile regime. The PTS length $\xi_{PTS}$ is obtained by fitting the excess overlap as a function of $R$, as described in Sec. II F.

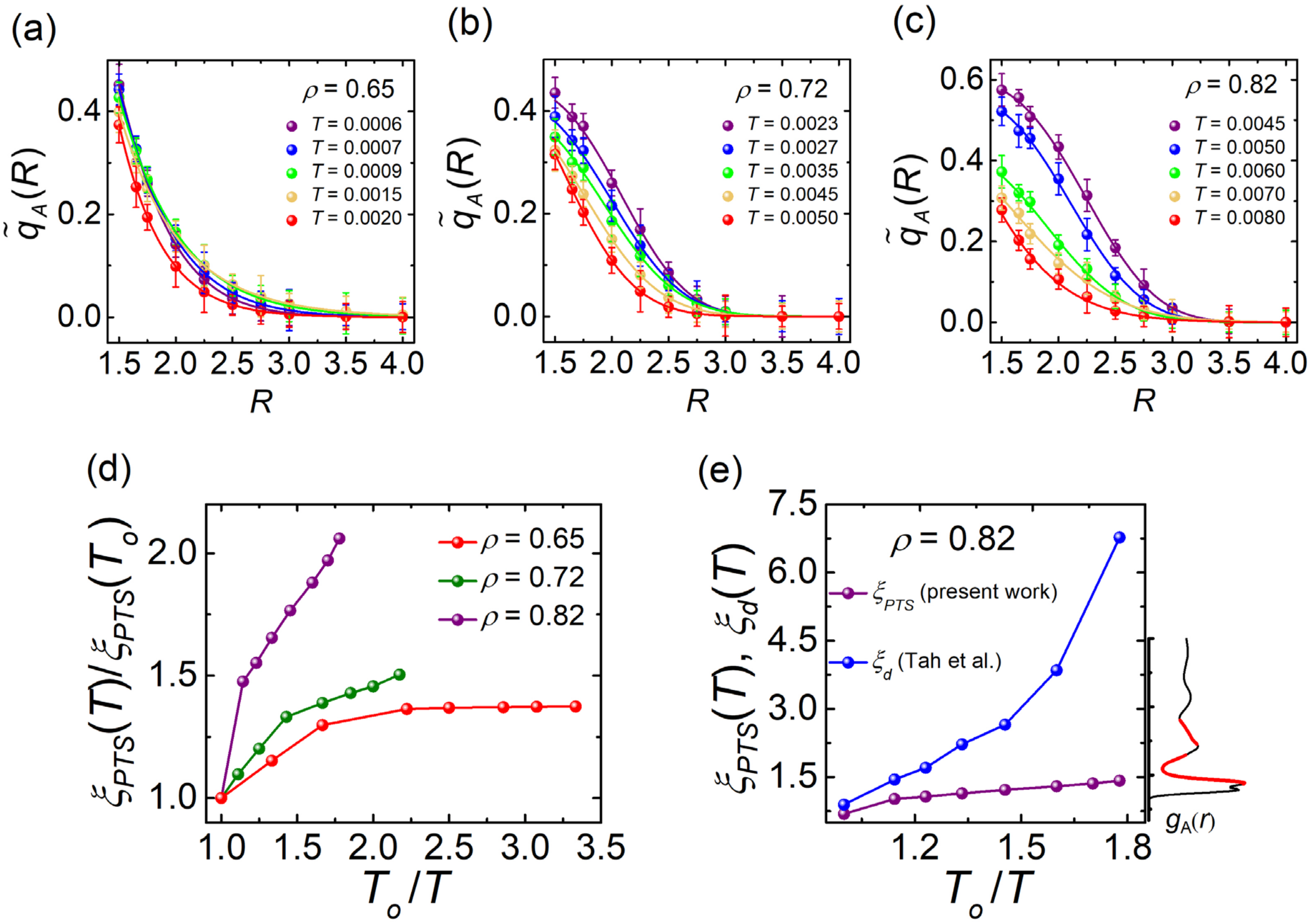

**FIG. 14.** Point-to-set correlations for particles of type A. (a)–(c) Excess overlap $\tilde{q}_A(R)$ as a function of cavity radius $R$ at different temperatures for $\rho = 0.65$, 0.72, and 0.82, respectively. Symbols denote the simulation results, error bars indicate the statistical uncertainty obtained from independent cavity samples, and solid curves show the fits used to extract the PTS correlation length $\xi_{PTS}$. (d) Normalized PTS length, $\xi_{PTS}(T)/\xi_{PTS}(T_0)$, as a function of $T_0/T$ for the three representative densities, where $T_0$ denotes the onset temperature. (e) Temperature dependence of $\xi_{PTS}$ in the fragile regime ($\rho = 0.82$), compared with the collective dynamic correlation length $\xi_d$, whose values were taken from Ref. [33]. The radial distribution profile $g_A(r)$ at $T = 0.0045$ is shown on the right for comparison with the operational spatial extent of slow variables. The red portion of $g_A(r)$ marks the radial range spanned by the neighbor ranks contributing to the slow dynamics.

The temperature dependence of the normalized PTS length, $\xi_{PTS}(T)/\xi_{PTS}(T_0)$, is shown in Fig. 14(d), where $T_0$ is the reference onset temperature reported previously for this model with $T_0$ = 0.0020, 0.0050, and 0.0080 for the strong ($\rho$ = 0.65), intermediate ($\rho$ = 0.72), and fragile ($\rho$ = 0.82) regimes, respectively.[33] This normalization allows comparison of the relative growth of $\xi_{PTS}$ across densities on the same scale as earlier studies. In the strong regime, $\xi_{PTS}$ shows only a modest initial increase upon cooling and tends to saturate at lower temperatures. In contrast, the fragile regime exhibits a rapid and continuous growth of the normalized PTS length over the accessible temperature range. The intermediate density lies between these two limits, indicating that the growth of static amorphous correlations becomes progressively more pronounced with increasing fragility. This fragility-dependent trend is consistent with the results of Tah and Karmakar[33] for this model. The weak growth of $\xi_{PTS}$ in the strong regime is qualitatively similar to that observed in silica melt,[44] whereas the more pronounced growth in the fragile regime resembles that found for the KALJ liquid.[45,78]

To assess how the static length scale relates to the spatial range of the slow dynamics, we compared $\xi_{PTS}$ with the spatial extent inferred from the slow variables identified above. Figure 14(e) shows this comparison for the fragile regime. Although $\xi_{PTS}$ increases upon cooling, its magnitude remains smaller than the spatial extent associated with the slow variables. In particular, the structural descriptors that modulate the jump dynamics involve neighbor ranks extending beyond the length scale indicated by $\xi_{PTS}$, indicating that the static length does not fully capture the spatial range of the rate-modulating structural fluctuations. This comparison reflects the different nature of the two quantities: $\xi_{PTS}$ measures the spatial range of static amorphous order through the overlap response to frozen boundaries, whereas the SSP analysis identifies the structural variables that modulate the cage-to-jump rate. We emphasize that the neighbor-rank range is an operational measure of structural extent and is not intended to represent a conventional correlation length. A similar mismatch between $\xi_{PTS}$ and the spatial extent associated with the slow variables is observed at other densities [Fig. S22]. This separation contrasts with previous observations in silica and KALJ systems, where the PTS length and the spatial extent of the slow variables were more closely correlated.[44,45] The present result suggests that the correspondence between the static amorphous order and spatial extent of dynamic disorder may depend on the interaction potential, consistent with the interaction-potential dependence of nonequilibrium relaxation timescales reported previously.[94] In the present soft-repulsive mixture, local packing and free-volume fluctuations appear to generate slow structural fluctuations that extend beyond the static length scale obtained from PTS correlations.

To place this operational structural extent in the context of a conventional dynamic length scale, we also show in Fig. 14(e) the collective dynamic correlation length $\xi_d$ reported by Tah and Karmakar[33] for the fragile regime of this model. The magnitude of $\xi_d$ is substantially larger than both $\xi_{PTS}$ and the radial range spanned by the relevant neighbor ranks. A nonlinear relation between the static and dynamic lengths, approximately $\xi_d \sim \left(\xi_{PTS}\right)^4$, was also reported.[33] This comparison indicates that the structural variables modulating the jump rate extend beyond the static PTS length but remain localized relative to the broader spatial extent of dynamic heterogeneity characterized by $\xi_d$. Since $\xi_{PTS}$, $\xi_d$, and the neighbor-rank range are obtained from different observables, the comparison is intended to establish their relative spatial scales rather than imply a direct quantitative correspondence.

Taken together, these results show that in the present soft-repulsive mixture, $\xi_{PTS}$ follows the expected fragility-dependent growth of static amorphous order but remains smaller than the spatial extent inferred from the slow variables governing the jump dynamics. At the same time, the substantially larger magnitude of $\xi_d$ indicates that these rate-modulating structural fluctuations remain localized relative to the broader spatial extent of dynamic heterogeneity. Thus, the slow variables governing jump-rate fluctuations cannot be described solely by a single static-overlap-based length scale. The dynamic-disorder bottleneck is instead controlled by multiple local structural features, including selected neighbor-distance fluctuations and, in the fragile regime, packing fluctuations captured by the Voronoi free volume.

## IV. CONCLUDING REMARKS

In this work, we systematically investigated the microscopic origin of dynamic disorder in a density-tunable soft-repulsive binary mixture by analyzing particle jump dynamics across strong, intermediate, and fragile regimes. By characterizing residence-time distributions and survival probabilities, we showed that the slowdown originates from the increasing heterogeneity of cage lifetimes, indicating that the rate-controlling bottleneck lies in the slow structural fluctuations preceding a jump. The residence-time distributions of the jump state remain short and only weakly temperature dependent, whereas the cage-state distributions develop increasingly broad long-time tails upon cooling. The growth of the randomness parameter $R$, together with the increasing deviation of $C_S(t)$ from the fast-fluctuation limit $C_{fast}(t)$ and the decrease of the stretching exponent $\beta$, demonstrates that the cage-to-jump rate becomes increasingly heterogeneous and temporally persistent at lower temperatures. Importantly, when compared as a function of $T_g/T$, $R$ and $\beta$ exhibit a clear density-dependent separation that follows the fragility of the system. In the strong regime, deviations from Poisson statistics and fast-fluctuation behavior develop gradually upon cooling, whereas in the fragile regime, they emerge more sharply at lower temperatures and become significantly stronger at the lowest temperatures studied. Thus, fragility is reflected in both the onset and strength of the dynamic disorder governing cage-to-jump motion.

The microscopic origin of this dynamic disorder was elucidated by identifying the structural variables that modulate the jump rate. The KL divergence $D_{eq}(k)$ and the Pearson correlation $\rho_P(k)$ analyses showed that the neighbor ranks coupled to jump motion change systematically with fragility. Since the relevant structural response involves multiple neighbor ranks, we introduced a structural slowness parameter (SSP) that combines their contributions into a reduced slow coordinate for evaluating the slow-fluctuation survival probability. In the strong regime, the SSP constructed from a compact set of neighbor ranks near the outer part of the first coordination shell and the onset of the second shell describes the slow-fluctuation survival probability over the investigated temperature range, indicating that rate modulation is governed by localized neighbor-distance fluctuations around the jumping particle. In the intermediate regime, additional neighbor ranks become relevant upon cooling, showing that the structural fluctuations controlling the jump rate extend farther into the surrounding environment. In the fragile regime, the relevant ranks span a broader region, extending deeper into the second coordination shell, indicating that the jump probability is controlled by more extended rearrangements of the surrounding particles. More importantly, for particles of type A at the lowest temperature, the extended distance-based SSP alone does not fully reproduce $C_S(t)$. The improved description obtained by combining this descriptor with the Voronoi free volume shows that the slow variable in the fragile regime involves both interparticle-distance fluctuations and local packing fluctuations. Thus, the structural bottleneck responsible for dynamic disorder evolves from localized distance-based constraints in the strong regime to extended and packing-sensitive fluctuations in the fragile regime.

The comparison with PTS correlations further clarifies the relation between the rate-controlling slow variables and static amorphous order. The PTS length shows a fragility-dependent growth, with weak temperature dependence in the strong regime and a more pronounced increase in the fragile regime, consistent with earlier studies of this model.[33] However, in the present soft-repulsive mixture, the spatial extent inferred from the slow variables remains larger than $\xi_{PTS}$. Thus, although PTS correlations capture the expected growth of static amorphous order with fragility, they do not fully characterize the slow fluctuations that regulate the cage-to-jump rate. This separation contrasts with previous observations in silica and KALJ systems[44,45], where the PTS length and the spatial extent of the slow variables were more closely correlated. The present result suggests that the correspondence between static amorphous order and dynamic disorder depends on how the interaction potential shapes the structural fluctuations controlling microscopic rate fluctuations, consistent with earlier observations that relaxation timescales are sensitive to the form of the interaction potential.[94] This distinction becomes most evident in the fragile regime of the present soft-repulsive mixture, where the rate-controlling fluctuations involve extended neighbor rearrangements together with local packing contributions that are not fully represented by the static length scale obtained from PTS correlations.

Overall, this study establishes a microscopic framework for understanding fragility-dependent slowdown by linking dynamic disorder in jump motion to the structural variables that regulate microscopic rate fluctuations. We showed that the microscopic origin of the rate fluctuations evolves systematically across the strong–fragile spectrum, from localized neighbor-distance fluctuations in the strong regime to extended neighbor rearrangements and packing-sensitive fluctuations in the fragile regime. The temperature dependence of $\xi_{PTS}$ also changes with fragility: in the strong regime, $\xi_{PTS}$ tends to saturate rapidly with decreasing temperature, whereas in the fragile regime it continues to grow over the accessible temperature range. These observations provide a microscopic basis for the broad cage-time distributions, enhanced intermittency, and strongly non-exponential survival probabilities observed with increasing fragility. Together with previous studies of water,[43] silica,[44] and KALJ liquids,[45] the present results show that dynamic disorder can emerge through distinct microscopic routes depending on the nature of the liquid: hydrogen-bond-network fluctuations in water, localized network constraints in silica, neighbor-shell rearrangements in KALJ, and extended packing-sensitive fluctuations in the present soft-repulsive mixture. Thus, the combined analysis of survival probabilities, SSPs, and PTS correlations provides a general route to identify the microscopic constraints that govern relaxation in glass-forming liquids. Extending this framework to systems in which relaxation and transport are controlled by rare events that depend on the local environment, such as confined liquids,[95–97] ionic liquids,[98] glassy solid-state electrolytes,[63,99] and biomolecular systems,[53,54,58,100] offers a direct route to identify the microscopic constraints that regulate transition rates and heterogeneous dynamics. In such systems, slowly evolving local structures can modulate relaxation or transport pathways, giving rise to cooperative motion and functional dynamics across complex materials.

## SUPPLEMENTARY MATERIAL

See the supplementary material for additional details referenced in the text.

### ACKNOWLEDGEMENTS

The present study was supported by JSPS KAKENHI (JP21H04676, JP23K17361, and JP26H02273) and DAICEL. The calculations were partially carried out using the supercomputers at the Research Center for Computational Science in Okazaki (Projects: 25-IMS-C223 and 26-IMS-C235).

### AUTHOR DECLARATIONS

#### Conflict of Interest

The authors have no conflicts to disclose.

### DATA AVAILABILITY STATEMENT

The data that support the findings of this study are available within the article and its Supplementary Material.

### REFERENCES

[1] L. Berthier, and G. Biroli, Rev. Mod. Phys. **83**, 587–645 (2011).

[2] V. Lubchenko, and P.G. Wolynes, Annu. Rev. Phys. Chem. **58**, 235–266 (2007).

[3] P.G. Debenedetti, and F.H. Stillinger, Nature **410**, 259–267 (2001).

[4] G. Biroli, and J.P. Garrahan, J. Chem. Phys. **138**, 12A301 (2013).

[5] M.D. Ediger, C.A. Angell, and S.R. Nagel, J. Phys. Chem. **100**, 13200–13212 (1996).

[6] S. Sastry, P.G. Debenedetti, and F.H. Stillinger, Nature **393**, 554–557 (1998).

[7] S. Sastry, Nature **409**(6817), 164–167 (2001).

[8] R. Böhmer, K.L. Ngai, C.A. Angell, and D.J. Plazek, J. Chem. Phys. **99**, 4201–4209 (1993).

[9] C. A. Angell, J. Non. Cryst. Solids **131**, 13–31 (1991).

[10] C.A. Angell, Science **267**, 1924–1935 (1995).

[11] C.A. Angell, K.L. Ngai, G.B. McKenna, P.F. McMillan, and S.W. Martin, J. Appl. Phys. **88**,

3113–3157 (2000).
[12] G. Adam, and J.H. Gibbs, J. Chem. Phys. **43**, 139–146 (1965).
[13] X. Xia, and P.G. Wolynes, Proc. Natl. Acad. Sci. U. S. A. **97**, 2990–2994 (2000).
[14] M.D. Ediger, Annu. Rev. Phys. Chem. **51**, 99–128 (2000).
[15] A.C. Pan, J.P. Garrahan, and D. Chandler, Phys. Rev. E **72**, 041106 (2005).
[16] K. Kim, and S. Saito, J. Chem. Phys. **138**, 12A506 (2013).
[17] H. Staley, E. Flenner, and G. Szamel, J. Chem. Phys. **143**, 244501 (2015).
[18] M. Adhikari, S. Karmakar, and S. Sastry, J. Phys. Chem. B **125**, 10232–10239 (2021).
[19] H. Tanaka, J. Phys. Chem. B **129**, 789–813 (2025).
[20] D. Coslovich, and G. Pastore, J. Chem. Phys. **127**, 124505 (2007).
[21] I. Tah, S.A. Ridout, and A.J. Liu, J. Chem. Phys. **157**, 124501 (2022).
[22] G. Sun, L. Li, and P. Harrowell, J. Non-Crystalline Solids X **13**, 100080 (2022).
[23] M. Wilson, and P.S. Salmon, Phys. Rev. Lett. **103**, 157801 (2009).
[24] D.L. Sidebottom, Phys. Rev. E **92**, 062804 (2015).
[25] C. Yildirim, J.Y. Raty, and M. Micoulaut, Nat. Commun. **7**, 1–6 (2016).
[26] M. Ozawa, K. Kim, and K. Miyazaki, J. Stat. Mech. Theory Exp. **2016**, 074002 (2016).
[27] Z. Zheng, R. Ni, F. Wang, M. Dijkstra, Y. Wang, and Y. Han, Nat. Commun. **5**(May), (2014).
[28] S. Kumar, S. Sarkar, and B. Bagchi, J. Chem. Phys. **160**, 224501 (2024).
[29] J.C. Mauro, P.K. Gupta, and R.J. Loucks, J. Chem. Phys. **130**, 234503 (2009).
[30] J. Krausser, K.H. Samwer, and A. Zaccone, Proc. Natl. Acad. Sci. U. S. A. **112**, 13762–13767 (2015).
[31] P. Lunkenheimer, F. Humann, A. Loidl, and K. Samwer, J. Chem. Phys. **153**, 124507 (2020).
[32] L. Berthier, and T.A. Witten, Epl **86**, 10001 (2009).
[33] I. Tah, and S. Karmakar, Phys. Rev. Mater. **6**, 035601 (2022).
[34] E.R. Weeks, J.C. Crocker, A.C. Levitt, A. Schofield, and D.A. Weitz, Science **287**, 627–631 (2000).
[35] P. Chaudhuri, L. Berthier, and W. Kob, Phys. Rev. Lett. **99**, 060604 (2007).
[36] R. Candelier, O. Dauchot, and G. Biroli, Phys. Rev. Lett. **102**, 1–4 (2009).
[37] R. Candelier, A. Widmer-Cooper, J.K. Kummerfeld, O. Dauchot, G. Biroli, P. Harrowell, and D.R. Reichman, Phys. Rev. Lett. **105**, 135702 (2010).
[38] R. Pastore, A. Coniglio, and M. Pica Ciamarra, Soft Matter **10**, 5724–5728 (2014).
[39] M.P. Ciamarra, R. Pastore, and A. Coniglio, Soft Matter **12**, 358–366 (2016).
[40] S. Indra, and S. Daschakraborty, Chem. Phys. Lett. **685**, 322–327 (2017).
[41] S. Dueby, V. Dubey, and S. Daschakraborty, J. Phys. Chem. B **123**, 7178–7189 (2019).

[42] T. Kikutsuji, K. Kim, and N. Matubayasi, J. Chem. Phys. **150**, 204502 (2019).
[43] S. Saito, J. Chem. Phys. **160**, 194506 (2024).
[44] S. Kumar, Z. Tang, and S. Saito, J. Chem. Phys. **164**, 024503 (2026).
[45] Z. Tang, S. Kumar, and S. Saito, J. Chem. Phys. **164**, 144504 (2026).
[46] O. Rubner, and A. Heuer, Phys. Rev. E **78**, 011504 (2008).
[47] J. Helfferich, F. Ziebert, S. Frey, H. Meyer, J. Farago, A. Blumen, and J. Baschnagel, Phys. Rev. E **89**, 042603 (2014).
[48] M.K. Nandi, and S. Maitra Bhattacharyya, J. Phys. Condens. Matter **32**, 064001 (2020).
[49] J. Helfferich, K. Vollmayr-Lee, F. Ziebert, H. Meyer, and J. Baschnagel, Epl **109**, 36004 (2015).
[50] R. Zwanzig, Acc. Chem. Res. **23**, 148–152 (1990).
[51] J. Wang, and P. Wolynes, Phys. Rev. Lett. **74**, 4317–4320 (1995).
[52] F.L.H. Brown, Phys. Rev. Lett. **90**, 028302 (2003).
[53] B.P. English, W. Min, A.M. Van Oijen, T.L. Kang, G. Luo, H. Sun, B.J. Cherayil, S.C. Kou, and X.S. Xie, Nat. Chem. Biol. **2**, 87–94 (2006).
[54] Y. Matsumura, and S. Saito, J. Chem. Phys. **154**, 224113 (2021).
[55] J. Cao, Chem. Phys. Lett. **327**, 38–44 (2000).
[56] I. Goychuk, J. Chem. Phys. **122**, 164506 (2005).
[57] V. Chernyak, M. Schulz, and S. Mukamel, J. Chem. Phys. **111**, 7416–7425 (1999).
[58] B. Halle, and F. Persson, J. Chem. Theory Comput. **9**, 2838–2848 (2013).
[59] A.P. Thompson, H.M. Aktulga, R. Berger, D.S. Bolintineanu, W.M. Brown, P.S. Crozier, P.J. in 't Veld, A. Kohlmeyer, S.G. Moore, T.D. Nguyen, R. Shan, M.J. Stevens, J. Tranchida, C. Trott, and S.J. Plimpton, Comput. Phys. Commun. **271**, 108171 (2022).
[60] L. Verlet, Phys. Rev. **159**, 98–103 (1967).
[61] D.J. Evans, and B.L. Holian, J. Chem. Phys. **83**, 4069–4074 (1985).
[62] A. Smessaert, and J. Rottler, Phys. Rev. E **88**, 022314 (2013).
[63] B. Kang, J. Yu, S. Saito, J. Jang, and B.J. Sung, Adv. Sci. **e09205**, 1–10 (2025).
[64] X. Ma, Z.S. Davidson, T. Still, R.J.S. Ivancic, S.S. Schoenholz, A.J. Liu, and A.G. Yodh, Phys. Rev. Lett. **122**, 028001 (2019).
[65] M.J. Schnitzer, and S.M. Block, Cold Spring Harb. Symp. Quant. Biol. **60**, 793–802 (1995).
[66] S. Yang, J. Cao, R.J. Silbey, and J. Sung, Biophys. J. **101**, 519–524 (2011).
[67] Solomon. Kullback, and Richard A. Leibler, Ann. Math. Stat. **22**, 79–86 (1951).
[68] K. Pearson, Proc. R. Soc. London **58**, 240–242 (2015).
[69] F.W. Starr, S. Sastry, J.F. Douglas, and S.C. Glotzer, Phys. Rev. Lett. **89**, 125501 (2002).

[70] C.H. Rycroft, Chaos **19**, 041111 (2009).

[71] J.P. Bouchaud, and G. Biroli, J. Chem. Phys. **121**, 7347–7354 (2004).

[72] R.L. Jack, and J.P. Garrahan, J. Chem. Phys. **123**, 164508 (2005).

[73] A. Montanari, and G. Semerjian, J. Stat. Phys. **125**, 23–54 (2006).

[74] A. Cavagna, T.S. Grigera, and P. Verrocchio, Phys. Rev. Lett. **98**, 187801 (2007).

[75] G. Biroli, J.P. Bouchaud, A. Cavagna, T.S. Grigera, and P. Verrocchio, Nat. Phys. **4**(10), 771–775 (2008).

[76] S. Karmakar, C. Dasgupta, and S. Sastry, Proc. Natl. Acad. Sci. U. S. A. **106**, 3675–3679 (2009).

[77] L. Berthier, and W. Kob, Phys. Rev. E **85**, 011102 (2012).

[78] G.M. Hocky, T.E. Markland, and D.R. Reichman, Phys. Rev. Lett. **108**, 225506 (2012).

[79] B. Charbonneau, P. Charbonneau, and G. Tarjus, Phys. Rev. Lett. **108**, 035701 (2012).

[80] H. Vogel, Phys. Z. **22**, 645–646 (1921).

[81] G.S. Fulcher, J. Am. Ceram. Soc. **8**, 339–355 (1925).

[82] G. Tammann, and W. Hesse, Z. Anorg. Allg. Chem. **156**, 245–257 (1926).

[83] L. Berthier, G. Biroli, J.P. Bouchaud, W. Kob, K. Miyazaki, and D.R. Reichman, J. Chem. Phys. **126**, 184503 (2007).

[84] A. Furukawa, and H. Tanaka, Phys. Rev. E **94**, 052607 (2016).

[85] R. Pastore, T. Kikutsuji, F. Rusciano, N. Matubayasi, K. Kim, and F. Greco, J. Chem. Phys. **155**, 114503 (2021).

[86] G. Williams, and D.C. Watts, Trans. Faraday Soc. **66**, 80–85 (1970).

[87] P.J. Steinhardt, D.R. Nelson, and M. Ronchetti, Phys. Rev. B **28**, 784–805 (1983).

[88] H. Tanaka, Eur. Phys. J. E **35**, 113 (2012).

[89] E.D. Cubuk, S.S. Schoenholz, J.M. Rieser, B.D. Malone, J. Rottler, D.J. Durian, E. Kaxiras, and A.J. Liu, Phys. Rev. Lett. **114**, 108001 (2015).

[90] S.S. Schoenholz, E.D. Cubuk, D.M. Sussman, E. Kaxiras, and A.J. Liu, Nat. Phys. **12**, 469–471 (2016).

[91] H. Tong, and H. Tanaka, Nat. Commun. **10**, 5596 (2019).

[92] V. Bapst, T. Keck, A. Grabska-Barwińska, C. Donner, E.D. Cubuk, S.S. Schoenholz, A. Obika, A.W.R. Nelson, T. Back, D. Hassabis, and P. Kohli, Nat. Phys. **16**, 448–454 (2020).

[93] E. Boattini, F. Smallenburg, and L. Filion, Phys. Rev. Lett. **127**, 088007 (2021).

[94] S. Kumar, S. Acharya, and B. Bagchi, Phys. Rev. E **107**(2), 24138 (2023).

[95] M. Neek-Amal, F.M. Peeters, I. V. Grigorieva, and A.K. Geim, ACS Nano **10**, 3685–3692 (2016).

[96] D. Muñoz-Santiburcio, and D. Marx, Chem. Rev. **121**, 6293–6320 (2021).
[97] S. Kumar, and B. Bagchi, J. Chem. Phys. **156**, 224501 (2022).
[98] Z. Hu, and C.J. Margulis, Proc. Natl. Acad. Sci. U. S. A. **103**, 831–836 (2006).
[99] H. Park, S. Saito, and B.J. Sung, J. Am. Chem. Soc. **148**, 24151−24162 (2026).
[100] J. Ono, Y. Matsumura, T. Mori, and S. Saito, J. Phys. Chem. B **128**, 20–32 (2024).

# Supplementary Material

## Unveiling Structural Bottlenecks of Dynamic Disorder in a Density-Tunable Glass Former: From Strong to Fragile Regimes

Shubham Kumar[1] and Shinji Saito[1,2,*]

[1]Institute for Molecular Science, Myodaiji, Okazaki, Aichi 444-8585, Japan

[2]The Graduate University for Advanced Studies (SOKENDAI), Myodaiji, Okazaki, Aichi 444-8585, Japan

*Author for correspondence: shinji@ims.ac.jp

**This PDF file includes:**

Tables S1 to S5

Figure S1 to S22

**Table S1:** Simulation state points for the soft-repulsive binary mixture at each number density $\rho$. Each column corresponds to a fixed density, and the entries list the temperatures studied from the normal liquid regime to deeply supercooled conditions. Temperatures are reported in reduced units.

| | $\rho = 0.65$ | $\rho = 0.69$ | $\rho = 0.72$ | $\rho = 0.78$ | $\rho = 0.82$ |
|---|---|---|---|---|---|
| $T_1$ | 0.00250 | 0.0050 | 0.0060 | 0.0080 | 0.0090 |
| $T_2$ | 0.00200 | 0.0040 | 0.0050 | 0.0070 | 0.0080 |
| $T_3$ | 0.00150 | 0.0035 | 0.0045 | 0.0060 | 0.0070 |
| $T_4$ | 0.00120 | 0.0030 | 0.0040 | 0.0055 | 0.0065 |
| $T_5$ | 0.00090 | 0.0025 | 0.0035 | 0.0050 | 0.0060 |
| $T_6$ | 0.00080 | 0.0020 | 0.0030 | 0.0045 | 0.0055 |
| $T_7$ | 0.00070 | 0.0017 | 0.0027 | 0.0040 | 0.0050 |
| $T_8$ | 0.00065 | 0.0015 | 0.0025 | 0.0037 | 0.0047 |
| $T_9$ | 0.00060 | — | 0.0023 | — | 0.0045 |

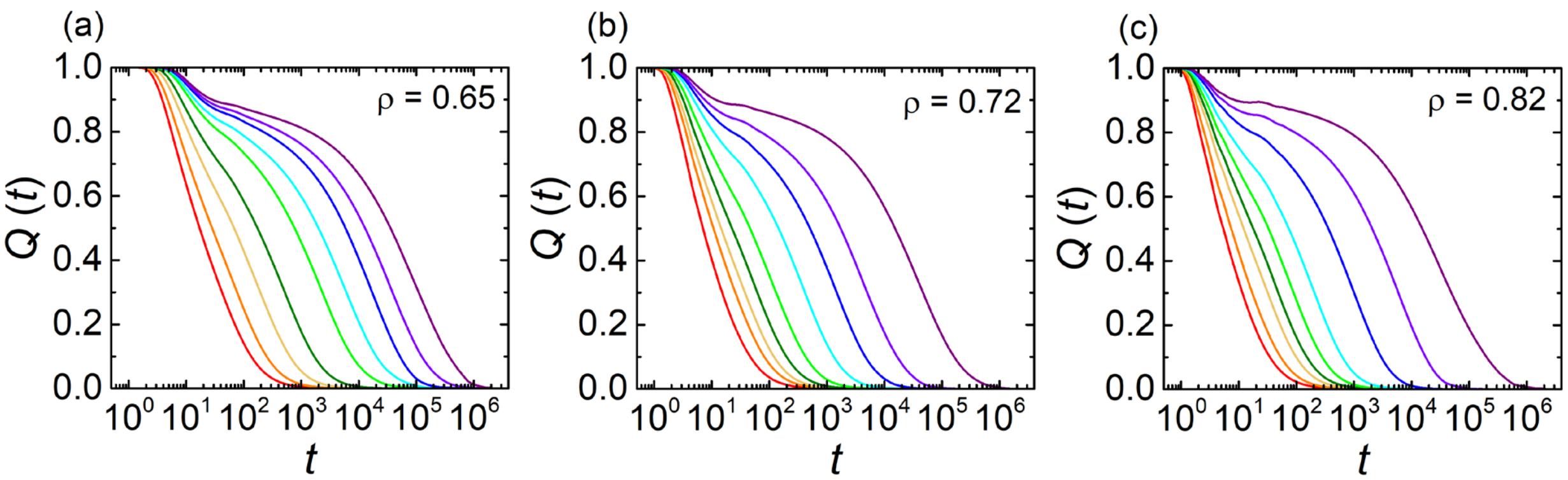

**FIG. S1.** Temperature dependence of the self-overlap function $Q(t)$, calculated over all particles, for the soft-repulsive binary mixture at representative densities: (a) $\rho = 0.65$, (b) $\rho = 0.72$, and (c) $\rho = 0.82$. The structural relaxation time was defined by $Q(\tau_\alpha) = e^{-1}$. In each panel, colors indicate decreasing temperature from red to purple; the corresponding temperature values are listed in Table S1. Cooling leads to a progressively slower decay of $Q(t)$ at all densities, with the strongest temperature dependence observed in the fragile regime.

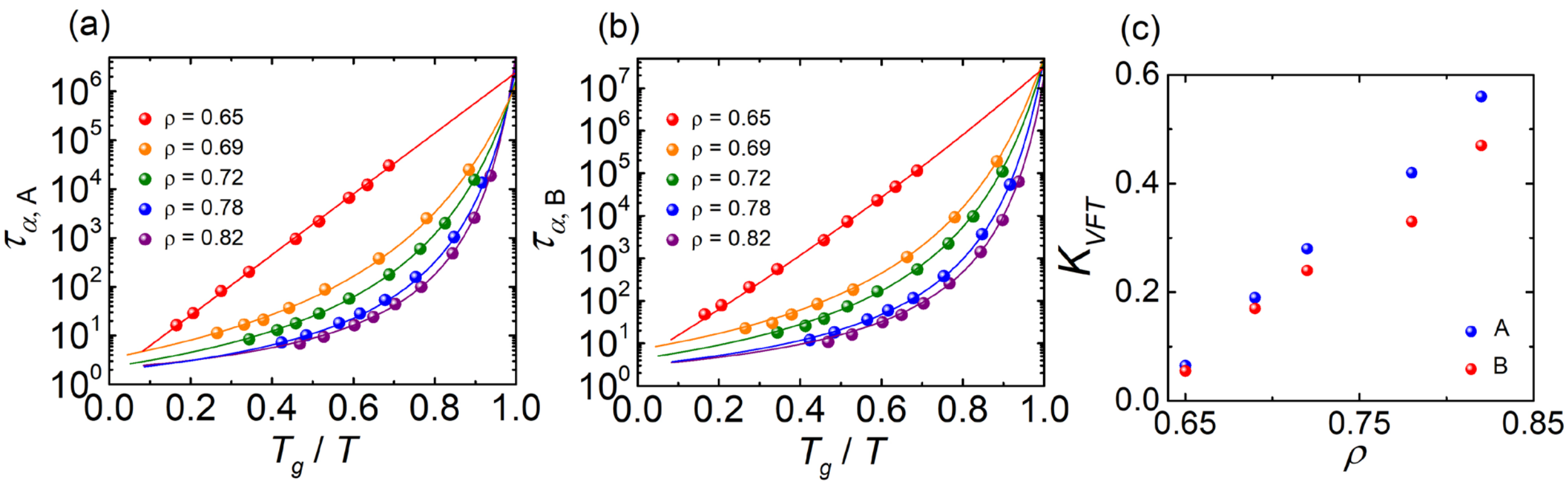

**FIG. S2.** Species-resolved Angell plots of the structural relaxation times for (a) particles A, $\tau_{\alpha,A}$, and (b) particles B, $\tau_{\alpha,B}$, as a function of the rescaled inverse temperature $T_g/T$. The same density-dependent $T_g$, defined from the total relaxation time by $\tau_\alpha\left(T_g\right)=5\times10^6$, is used for both species. Symbols denote simulation data, and solid lines represent VFT fits. (c) Species-resolved kinetic fragility $K_{VFT}$ obtained from the VFT fits as a function of density.

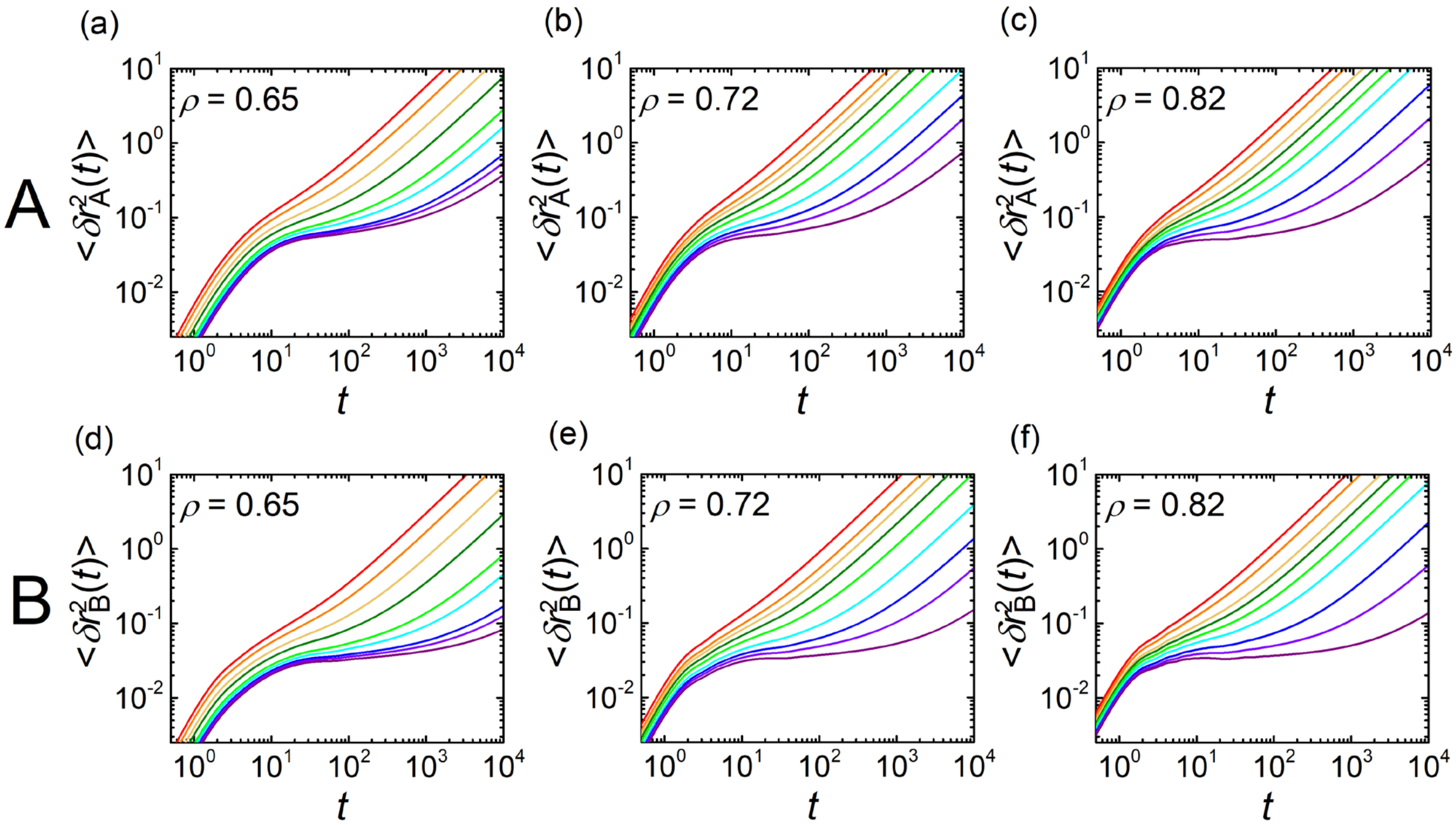

**FIG. S3.** Mean-squared displacement $\left\langle \delta r^2(t)\right\rangle$ for particles A and B at representative densities. Results are shown for particles A [(a)–(c)] and particles B [(d)–(f)] at $\rho = 0.65$, 0.72, and 0.82. Colors indicate decreasing temperature from red to purple; the corresponding temperature values are listed in Table S1.

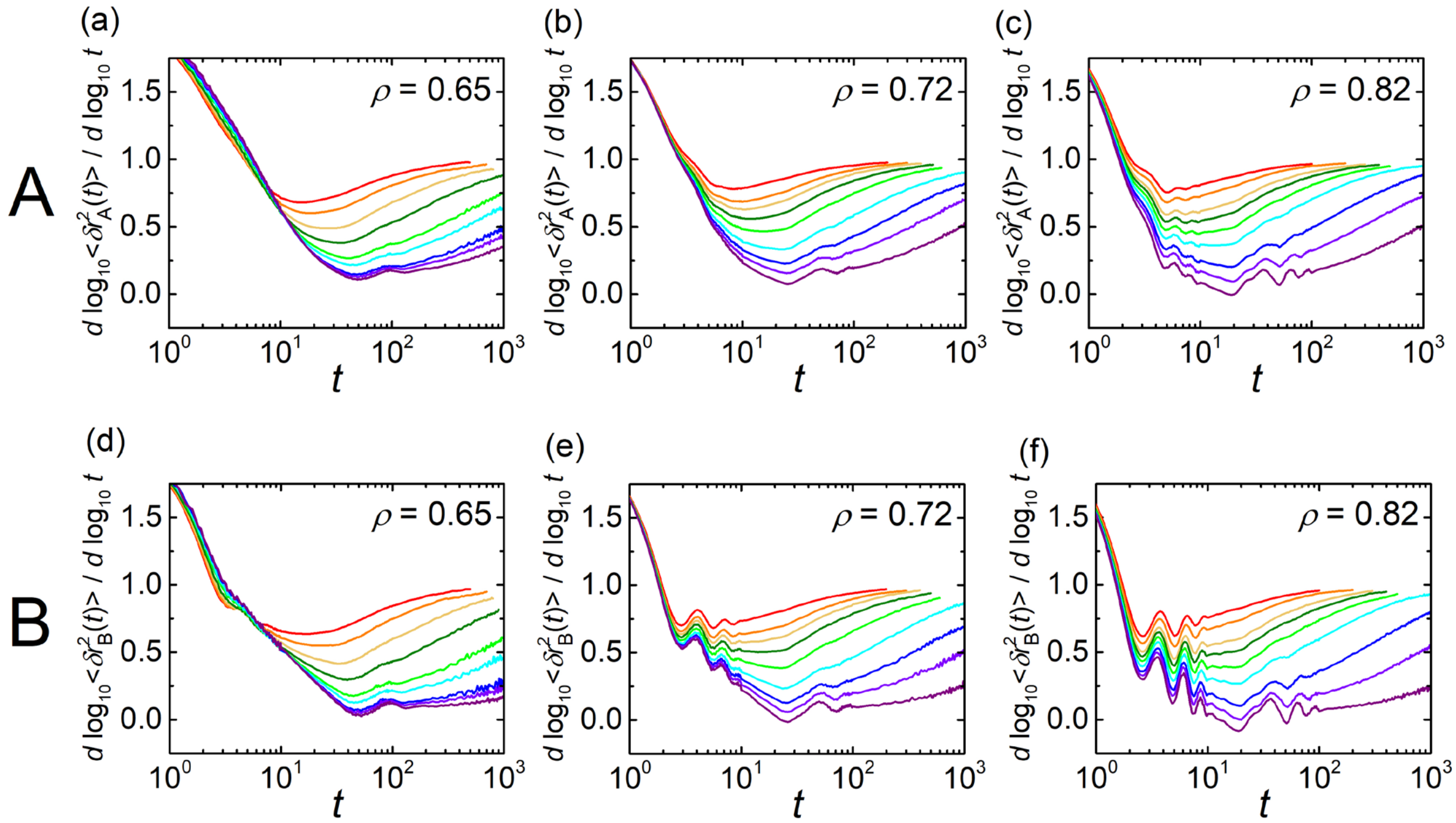

**FIG. S4.** Logarithmic derivative of the mean-squared displacement, $d\log\left\langle \delta r_{\alpha}^{2}(t)\right\rangle / d\log t$, for particles A and B at representative densities. Results are shown for particles A [(a)–(c)] and particles B [(d)–(f)] at $\rho = 0.65$, 0.72, and 0.82. Colors indicate decreasing temperature from red to purple; the corresponding temperature values are listed in Table S1. The minima identify the plateau regime of maximal cage confinement used to select the averaging window $\Delta t$ for the hop function.

**TABLE S2.** Averaging window $\Delta t$ used in the hop-function analysis for particles A and B at $\rho =$ 0.65. The values were selected from the plateau regime of the MSD using the minima of $d\log\left\langle \delta r_{\alpha}^{2}(t)\right\rangle / d\log t$, with $\alpha$ = A, B.

| $T$ | $\Delta t_A$ | $\Delta t_B$ |
|---|---|---|
| 0.00250 | 15.4 | 18.1 |
| 0.00200 | 18.5 | 23.4 |
| 0.00150 | 29.6 | 34.1 |
| 0.00120 | 34.7 | 39.4 |
| 0.00090 | 40.9 | 44.5 |
| 0.00080 | 48.2 | 47.5 |
| 0.00070 | 48.4 | 51.3 |
| 0.00065 | 49.5 | 52.5 |
| 0.00060 | 50.4 | 52.4 |

**TABLE S3.** Averaging window $\Delta t$ used in the hop-function analysis for particles A and B at $\rho =$ 0.72. The values were selected from the plateau regime of the MSD using the minima of $d\log\left\langle \delta r_{\alpha}^{2}(t)\right\rangle / d\log t$, with $\alpha$ = A, B.

| $T$ | $\Delta t_A$ | $\Delta t_B$ |
|---|---|---|
| 0.0060 | 7.8 | 5.6 |
| 0.0050 | 9.9 | 5.6 |
| 0.0045 | 10.3 | 8.6 |
| 0.0040 | 11.2 | 14.6 |
| 0.0035 | 15.0 | 22.8 |
| 0.0030 | 23.0 | 23.8 |
| 0.0027 | 24.9 | 26.3 |
| 0.0025 | 25.2 | 26.3 |
| 0.0023 | 25.4 | 26.3 |

**TABLE S4.** Averaging window $\Delta t$ used in the hop-function analysis for particles A and B at $\rho =$ 0.82. The values were selected from the plateau regime of the MSD using the minima of $d\log\left\langle \delta r_{\alpha}^{2}(t)\right\rangle / d\log t$, with $\alpha$ = A, B.

| $T$ | $\Delta t_A$ | $\Delta t_B$ |
|---|---|---|
| 0.0090 | 4.9 | 5.1 |
| 0.0080 | 5.1 | 4.9 |
| 0.0070 | 5.0 | 4.9 |
| 0.0065 | 7.6 | 5.0 |
| 0.0060 | 9.6 | 7.7 |
| 0.0055 | 13.0 | 17.0 |
| 0.0050 | 19.0 | 20.0 |
| 0.0047 | 20.0 | 20.0 |
| 0.0045 | 20.0 | 19.5 |

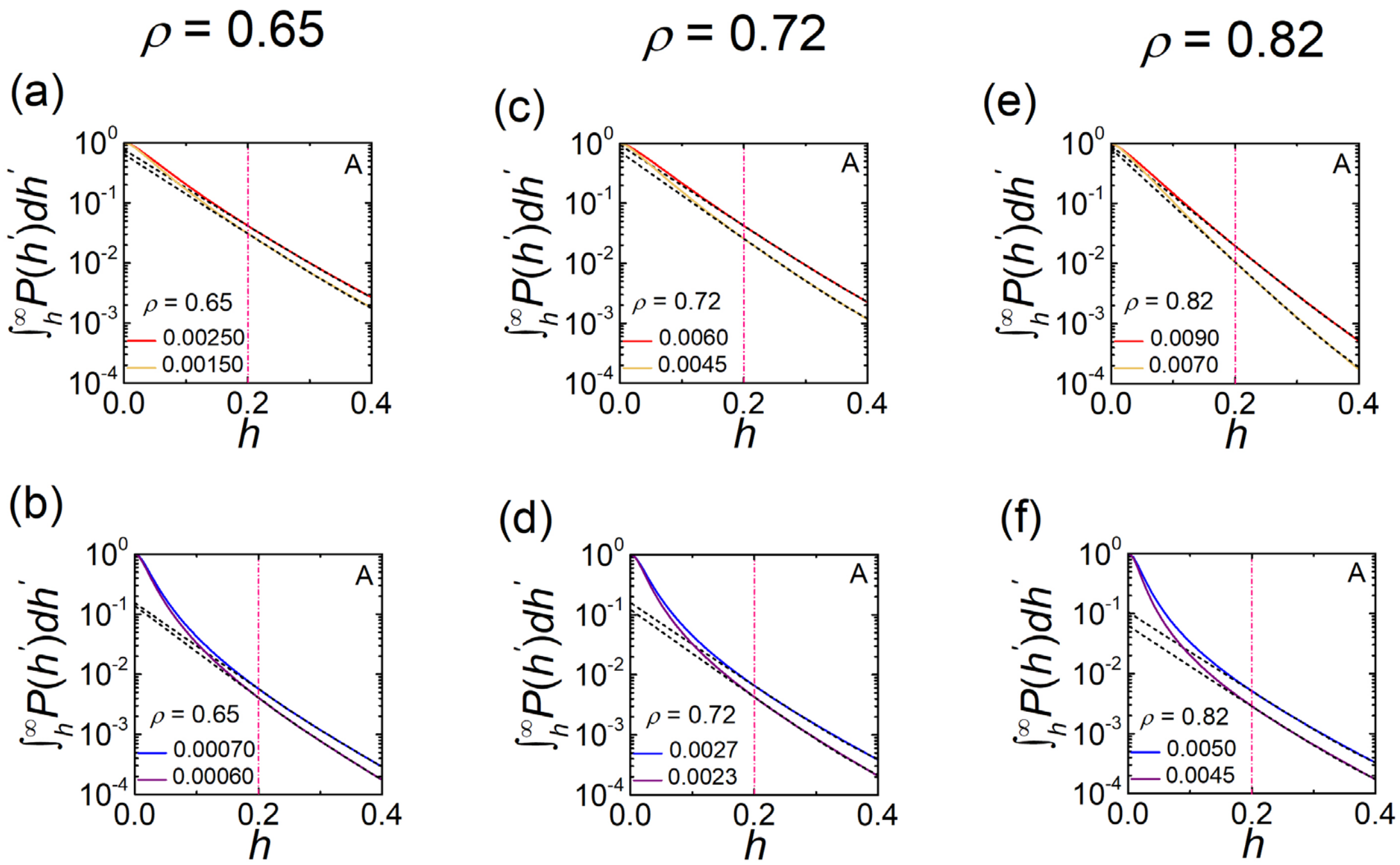

**FIG. S5.** Cumulative probability distributions of the hop function $h$ for particles A at representative densities and temperatures. Results are shown for $\rho = 0.65$ [(a), (b)], $\rho = 0.72$ [(c), (d)], and $\rho = 0.82$ [(e), (f)], corresponding to the strong, intermediate, and fragile regimes, respectively. For each density, the upper panel shows representative high-temperature state points and the lower panel shows representative low-temperature state points. The vertical dashed line indicates the threshold value $h_A^* = 0.2$ used for distinguishing cage and jump states.

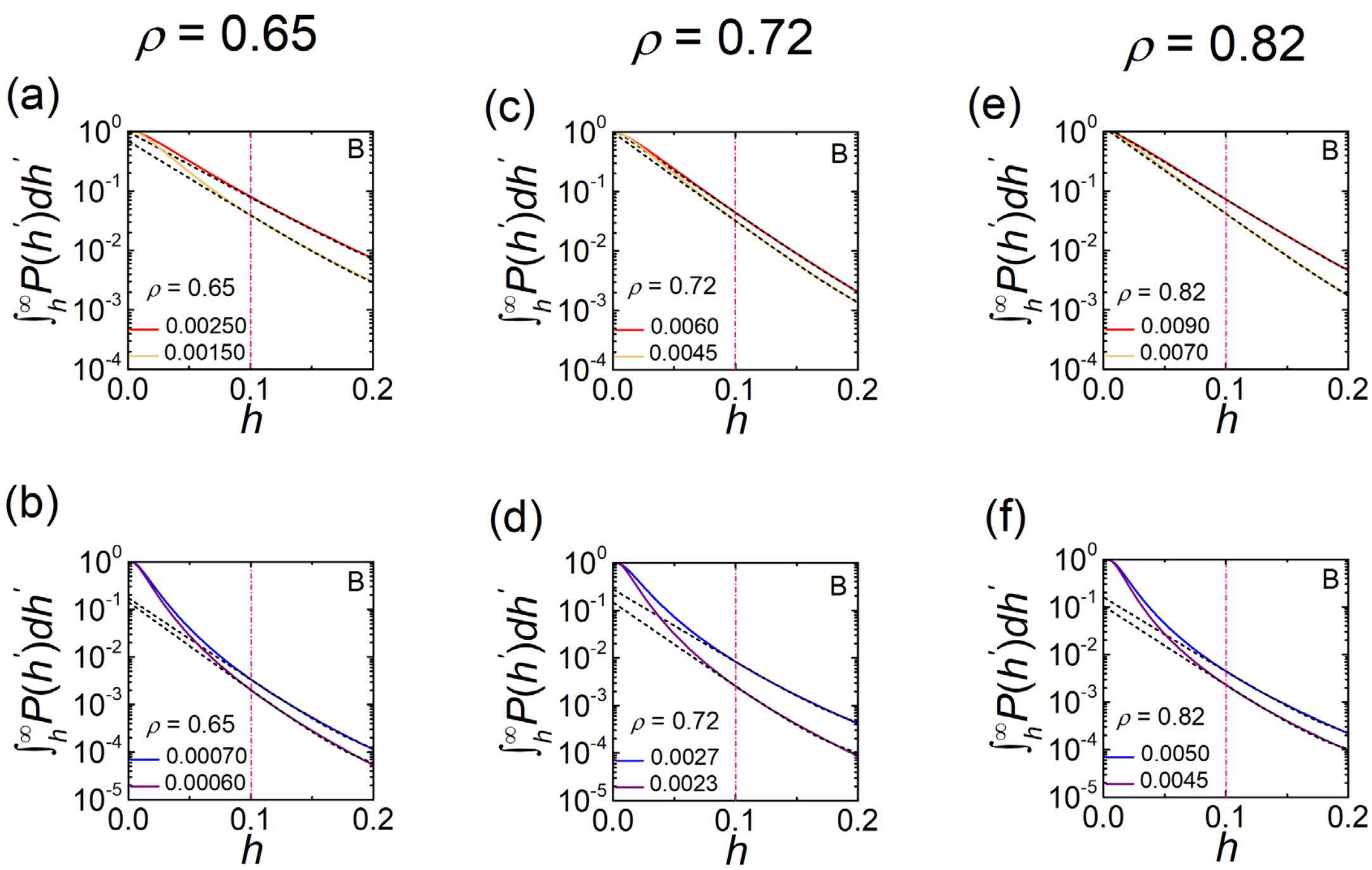

**FIG. S6.** Cumulative probability distributions of the hop function $h$ for particles B at representative densities and temperatures. Results are shown for $\rho = 0.65$ [(a), (b)], $\rho = 0.72$ [(c), (d)], and $\rho = 0.82$ [(e), (f)], corresponding to the strong, intermediate, and fragile regimes, respectively. For each density, the upper panel shows representative high-temperature state points and the lower panel shows representative low-temperature state points. The vertical dashed line indicates the threshold value $h_B^* = 0.1$ used for distinguishing cage and jump states.

## Free-energy profiles along the hop coordinate

The free-energy profile along the hop coordinate is obtained as $\Delta F(h) = -k_B T \ln P(h) + C$, where $P(h)$ is the probability distribution of $h$, and $C$ is chosen such that the minimum of $\Delta F(h)$ is zero, allowing profiles at different state points to be compared.[43–45] Figures 3(a)-3(c) show the free-energy profiles $\Delta F(h)/(k_B T)$ for particle A at different temperatures for the densities $\rho = 0.65$, 0.72, and 0.82, respectively, whereas the corresponding results for particle B are shown in Figs. 3(d)–3(f). For each density, lowering the temperature leads to a systematic increase in the free-energy barrier along the hop coordinate.

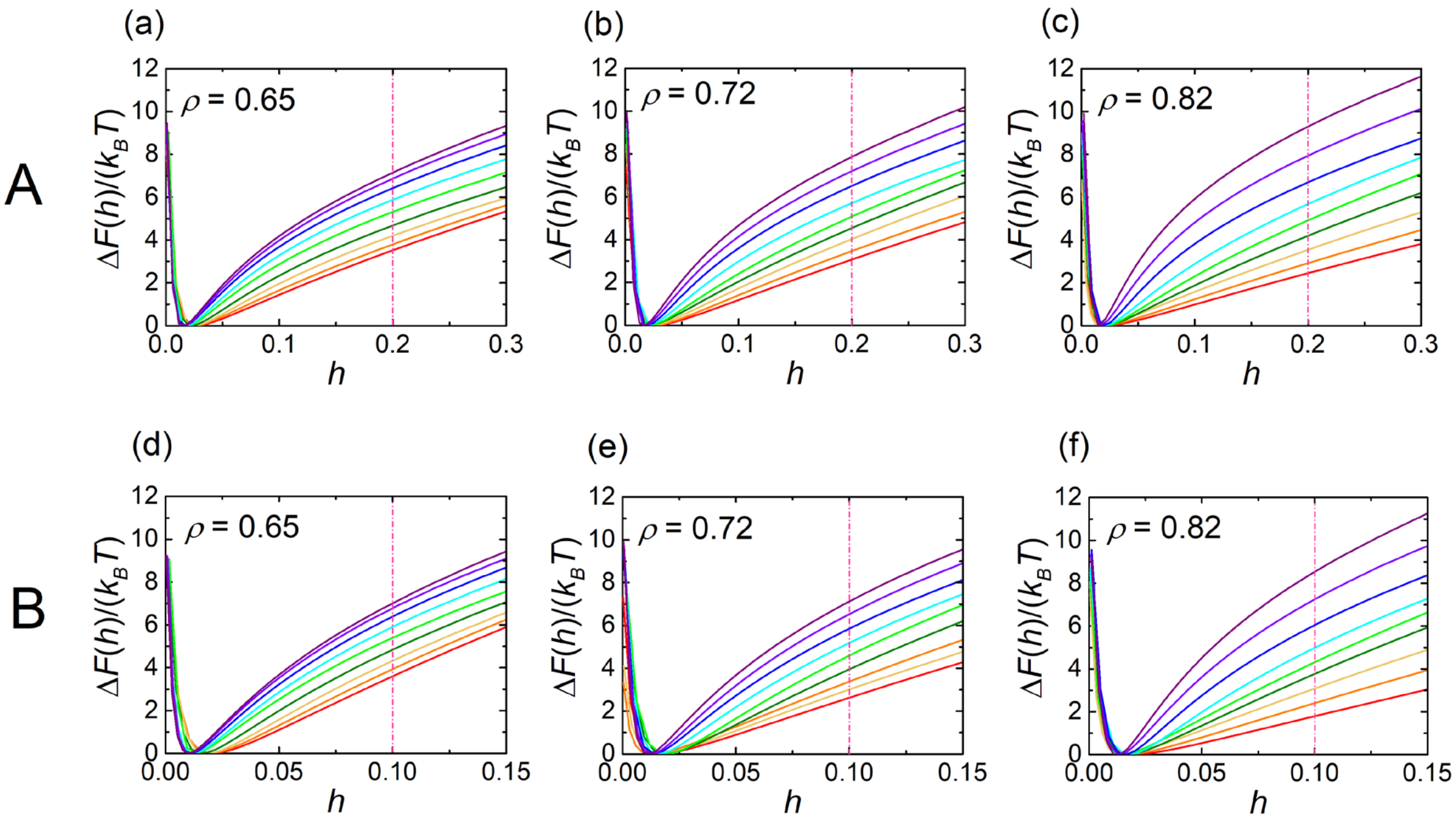

**FIG. S7.** Free-energy profiles $\Delta F(h)/(k_BT)$ along the hop coordinate $h$ for particles A [(a)–(c)] and B [(d)–(f)] at representative densities. The densities are $\rho = 0.65$ [(a), (d)], $\rho = 0.72$ [(b), (e)], and $\rho = 0.82$ [(c), (f)], corresponding to the strong, intermediate, and fragile regimes, respectively. Colors indicate decreasing temperature from red to purple in each panel; the corresponding temperature values are listed in Table S1. Vertical dashed lines denote the hop thresholds $h_A^* = 0.2$ for particle A and $h_B^* = 0.1$ for particle B. Lowering the temperature leads to an increase in the free-energy barrier along the hop coordinate at all densities.

To quantify the temperature dependence of the barrier at the hop threshold, we compared the dimensionless barrier height $\Delta F(h^*)/(k_BT)$ evaluated at the highest and lowest temperatures studied for each density. The corresponding values for particles A and B are summarized in Table I. In the strong regime ($\rho = 0.65$), the barrier at the hop threshold increases moderately upon cooling for both particle types. At the intermediate density ($\rho = 0.72$), the increase in $\Delta F(h^*)/(k_BT)$ upon cooling becomes more pronounced. In contrast, in the fragile regime ($\rho = 0.82$), the change in the barrier between the highest and lowest temperatures becomes substantially larger. Thus, the increase in the free-energy barrier at the hop threshold becomes progressively more pronounced upon cooling as the system evolves from the strong to the fragile regime. This trend reflects the enhanced temperature sensitivity of jump events in the fragile regime, leading to a stronger suppression of large-hop events at low temperatures. We note that the temperature dependence of $\Delta F(h^*)/(k_B T)$ reflects changes along the specific microscopic coordinate $h$ and should not be directly identified with the activation barrier governing structural relaxation, particularly in the strong regime where the relaxation remains nearly Arrhenius.

**TABLE S5.** Dimensionless free-energy barrier at the hop threshold, $\Delta F(h^*)/(k_B T)$, evaluated from the free-energy profiles along the hop coordinate for particles A and B at the representative densities. $T_{\mathrm{high}}$ and $T_{\mathrm{low}}$ denote the highest and lowest temperatures used for the comparison at each density. The last column reports the increase in $\Delta F(h^*)/(k_B T)$ upon cooling, calculated as the difference between the values at $T_{\mathrm{low}}$ and $T_{\mathrm{high}}$.

| $\rho$ | Particle | $T_{\mathrm{high}}$ | $\Delta F(h^*)/(k_B T)$ at $T_{\mathrm{high}}$ | $T_{\mathrm{low}}$ | $\Delta F(h^*)/(k_B T)$ at $T_{\mathrm{low}}$ | Increase in barrier |
|---|---|---|---|---|---|---|
| **0.65** | A | 0.0025 | 3.54 | 0.0006 | 7.20 | 3.66 |
| **0.65** | B | 0.0025 | 3.74 | 0.0006 | 7.01 | 3.27 |
| **0.72** | A | 0.0060 | 3.10 | 0.0023 | 7.86 | 4.76 |
| **0.72** | B | 0.0060 | 2.72 | 0.0023 | 7.12 | 4.40 |
| **0.82** | A | 0.0090 | 2.50 | 0.0045 | 9.42 | 6.92 |
| **0.82** | B | 0.0090 | 1.80 | 0.0045 | 8.62 | 6.82 |

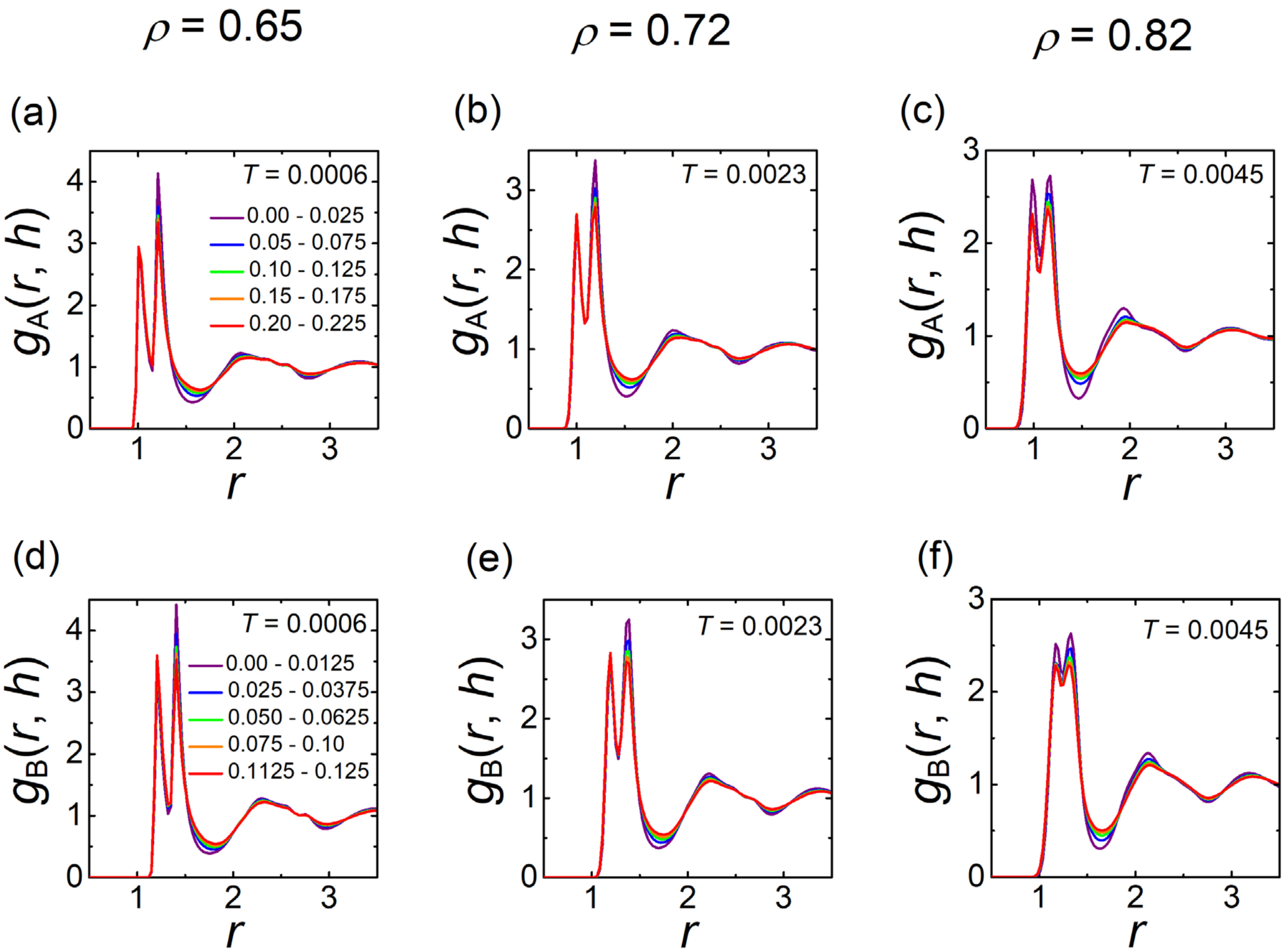

**FIG. S8.** Hop-resolved radial distribution functions $g_A(r, h)$ and $g_B(r, h)$ between a jumping particle and its surrounding neighbors at the lowest temperatures for the representative densities. Panels (a)–(c) show $g_A(r, h)$ for jumping particle A at $\rho = 0.65$, 0.72, and 0.82, respectively, whereas panels (d)–(f) show the corresponding $g_B(r, h)$ for jumping particle B. Different colors represent different ranges of the hop amplitude $h$. The hop-amplitude ranges are indicated in the legends shown in panels (a) and (d) for particles A and B, respectively.

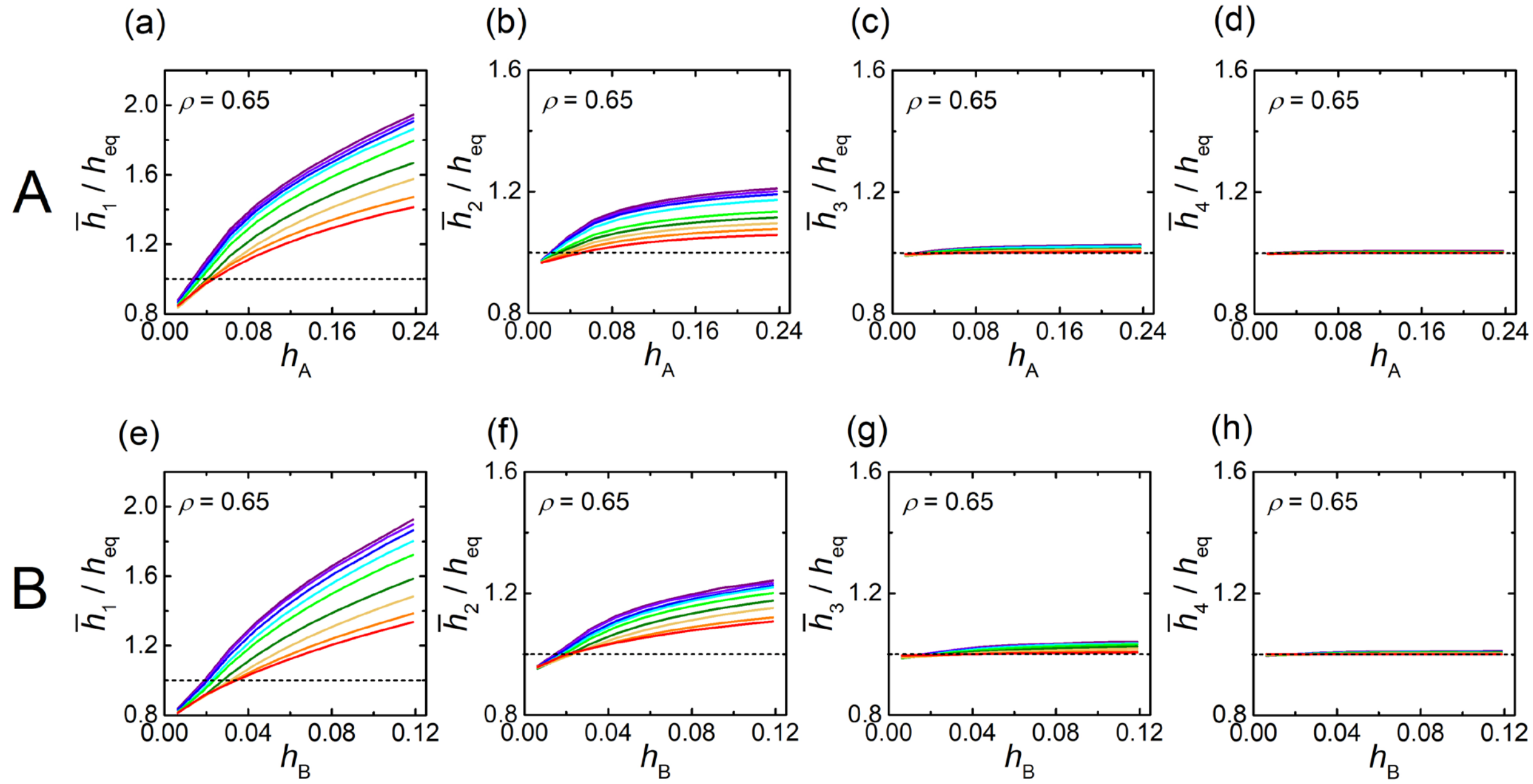

**FIG. S9.** Temperature dependence of the normalized average hop amplitude $\bar{h}_n / h_{eq}$ of particles in the $n$th coordination shell around a jumping particle at $\rho = 0.65$ (strong regime). Results are shown for jumping particles A [(a)–(d)] and B [(e)–(h)]. The subscripts $n$ = 1-4 correspond to the first to fourth coordination shells. Colors indicate decreasing temperature from red to purple; the corresponding temperature values are listed in Table S1. Dashed horizontal lines denote $\bar{h}_n / h_{eq} = 1$, corresponding to the equilibrium average hop amplitude.

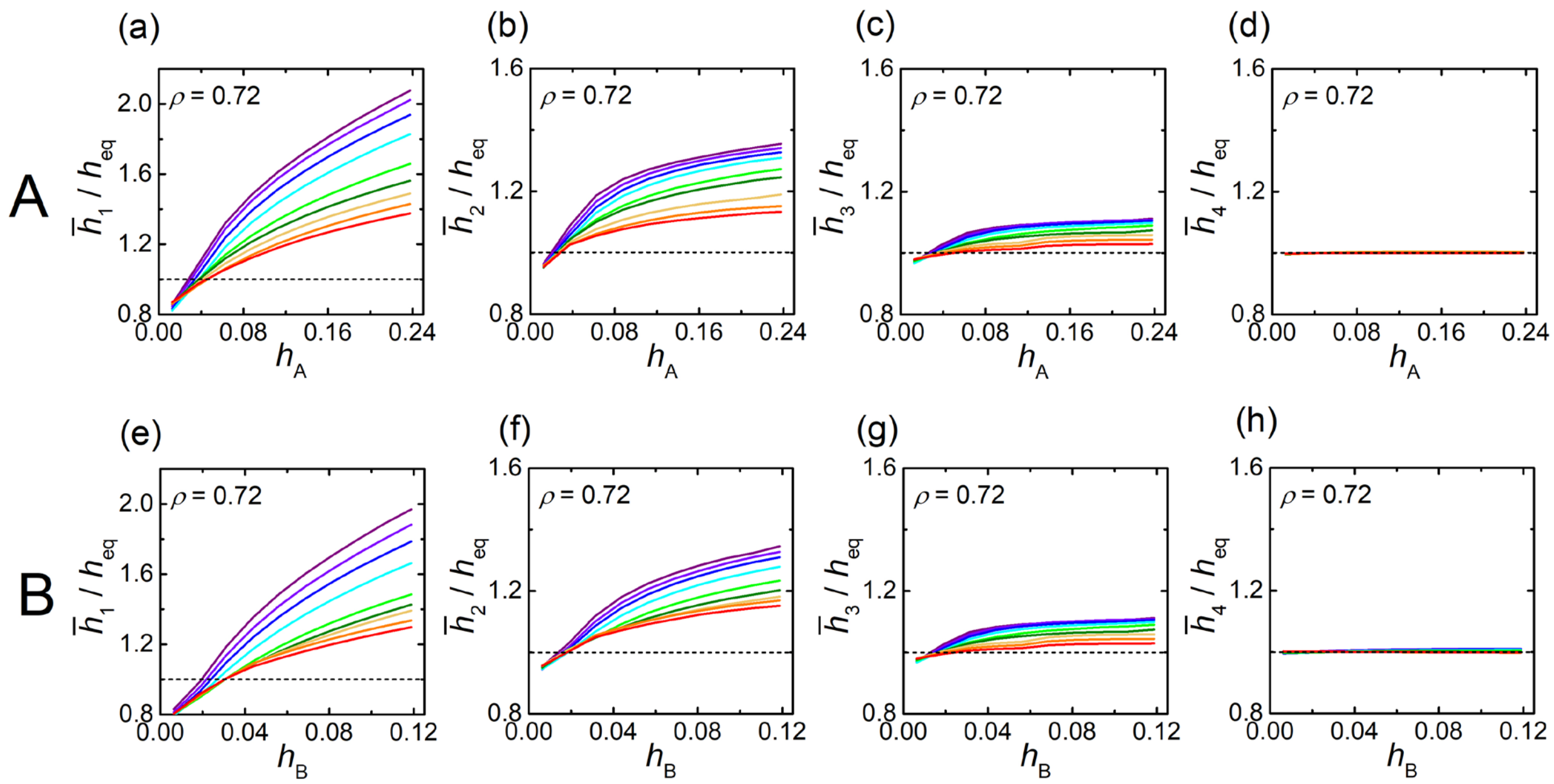

**FIG. S10.** Temperature dependence of the normalized average hop amplitude $\bar{h}_n / h_{eq}$ of particles in the *n*th coordination shell around a jumping particle at $\rho = 0.72$ (intermediate regime). Results are shown for jumping particles A [(a)–(d)] and B [(e)–(h)]. The subscripts $n$ = 1-4 correspond to the first to fourth coordination shells. Colors indicate decreasing temperature from red to purple; the corresponding temperature values are listed in Table S1. Dashed horizontal lines denote $\bar{h}_n / h_{eq} = 1$, corresponding to the equilibrium average hop amplitude.

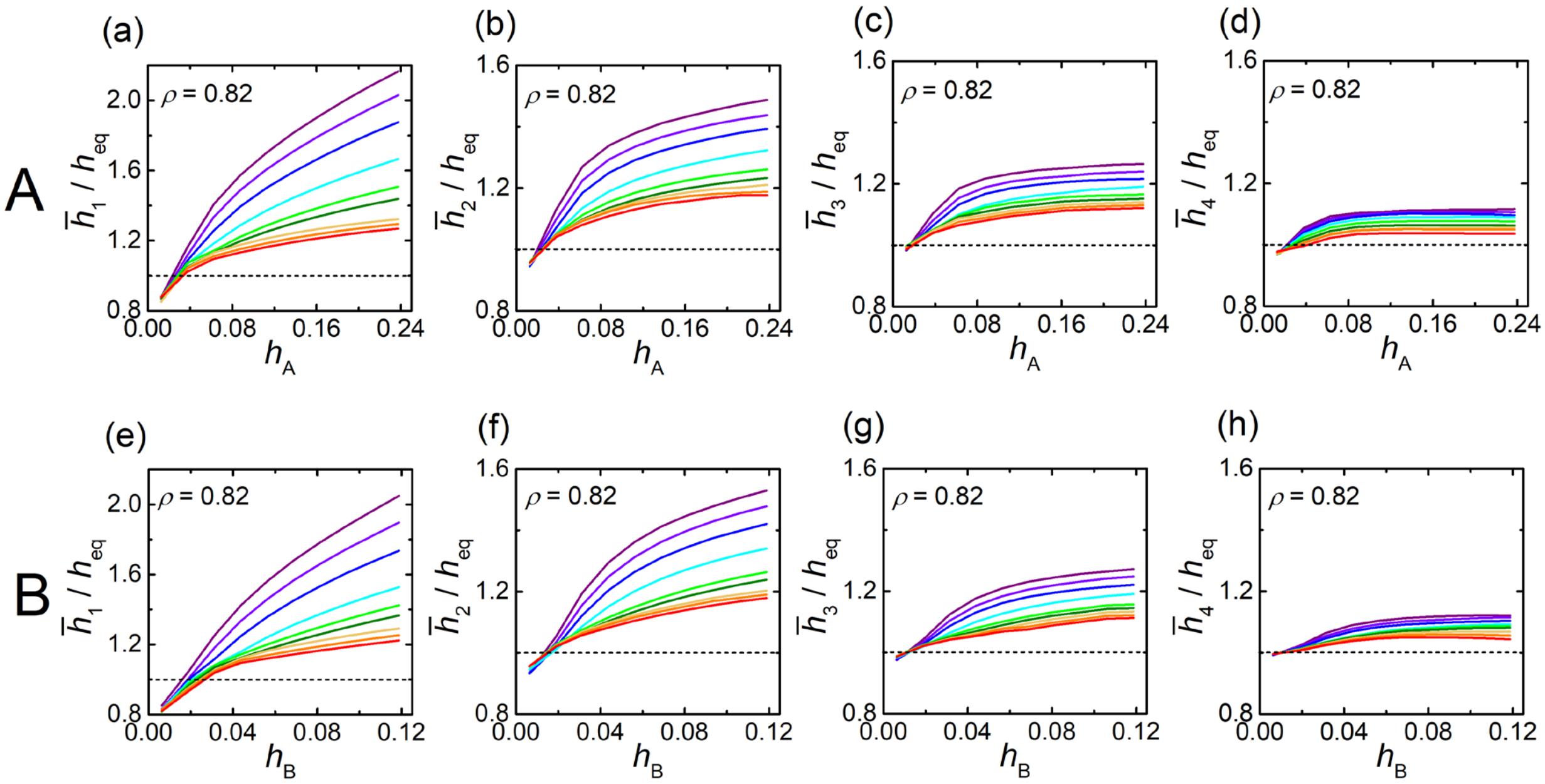

**FIG. S11.** Temperature dependence of the normalized average hop amplitude $\bar{h}_n / h_{eq}$ of particles in the $n$th coordination shell around a jumping particle at $\rho = 0.82$ (fragile regime). Results are shown for jumping particles A [(a)–(d)] and B [(e)–(h)]. The subscripts $n$ = 1-4 correspond to the first to fourth coordination shells. Colors indicate decreasing temperature from red to purple; the corresponding temperature values are listed in Table S1. Dashed horizontal lines denote $\bar{h}_n / h_{eq} = 1$, corresponding to the equilibrium average hop amplitude.

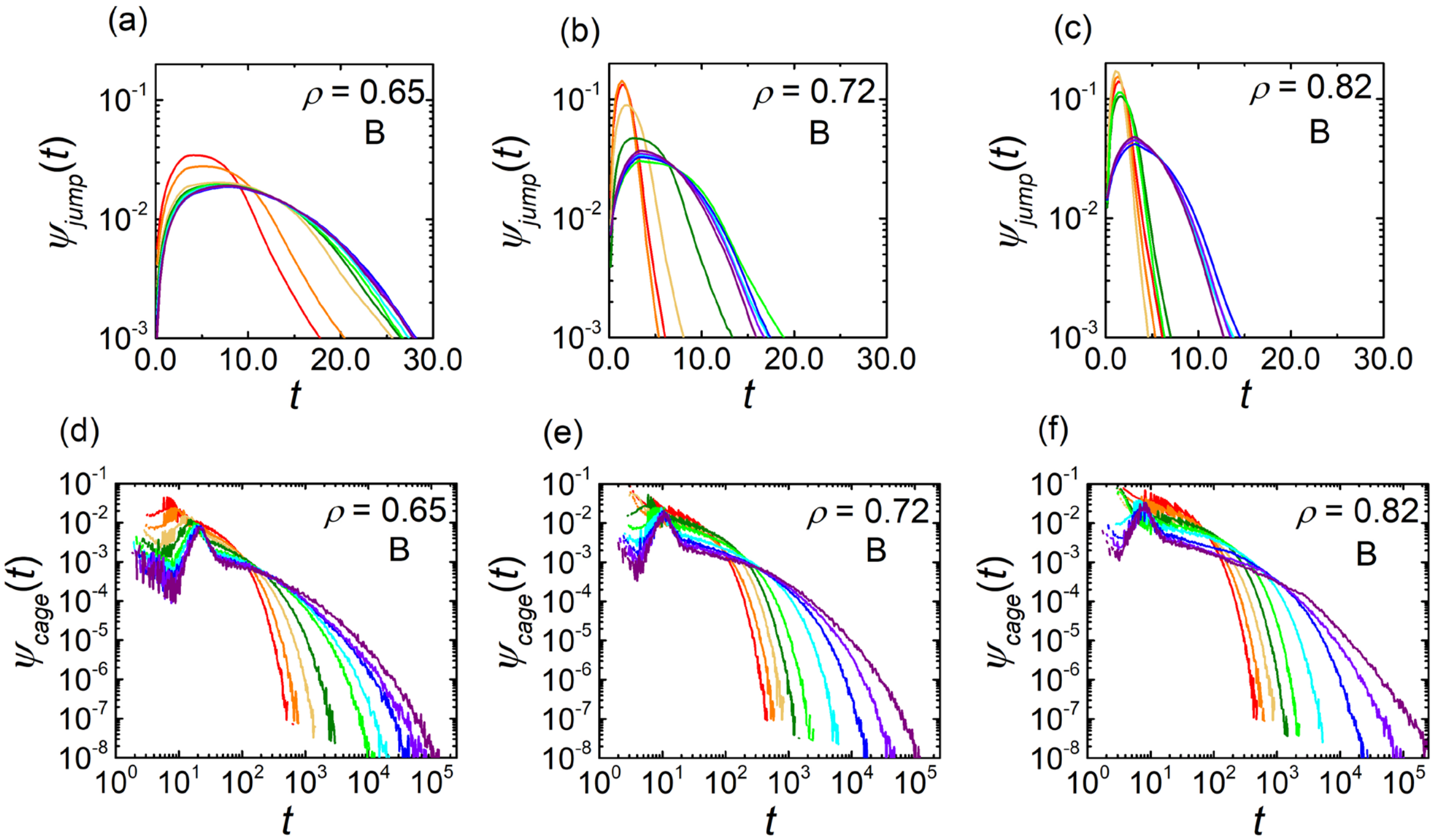

**FIG. S12.** Residence-time distributions of the jump and cage states for particle B at representative densities. Panels (a)–(c) show the jump-state residence-time distributions $\psi_{jump}(t)$, and panels (d)–(f) show the cage-state residence-time distributions $\psi_{cage}(t)$, for $\rho$ = 0.65, 0.72, and 0.82, respectively. Colors indicate decreasing temperature from red to purple in each panel; the corresponding temperature values are listed in Table S1. The jump-state distributions remain short and weakly temperature dependent, whereas the cage-state distributions broaden strongly and develop long-time tails upon cooling.

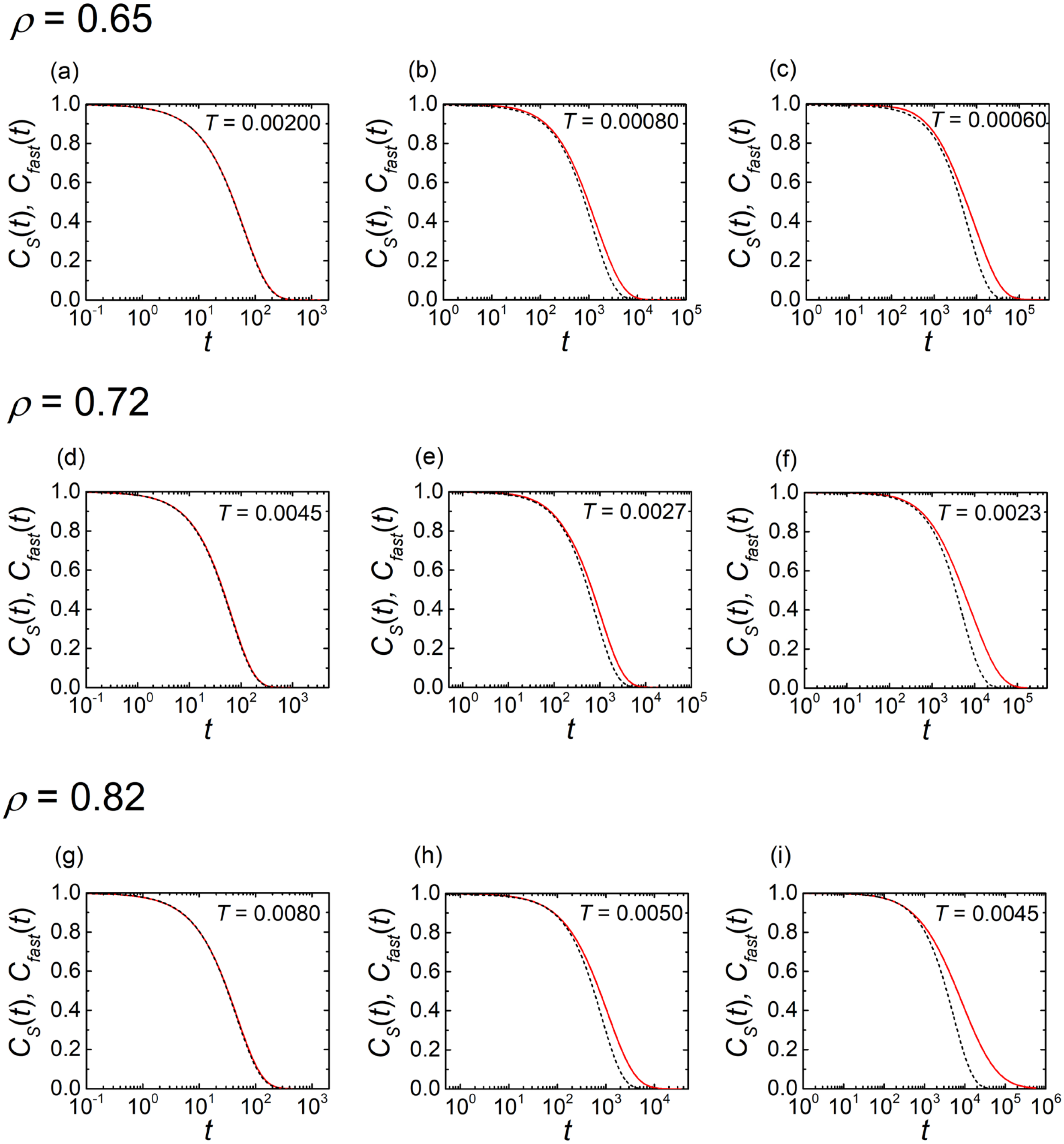

**FIG. S13.** Survival probability of the cage state, $C_S(t)$ (red), and its fast-fluctuation limit, $C_{fast}(t)$ (dashed black), for particle B at representative temperatures. Results are shown for $\rho = 0.65$ [(a)–(c)], $\rho = 0.72$ [(d)–(f)], and $\rho = 0.82$ [(g)–(i)]. At high temperatures, $C_S(t)$ closely follows $C_{fast}(t)$, consistent with Poisson statistics. Upon cooling, systematic deviations emerge, indicating the breakdown of the fast-fluctuation approximation.

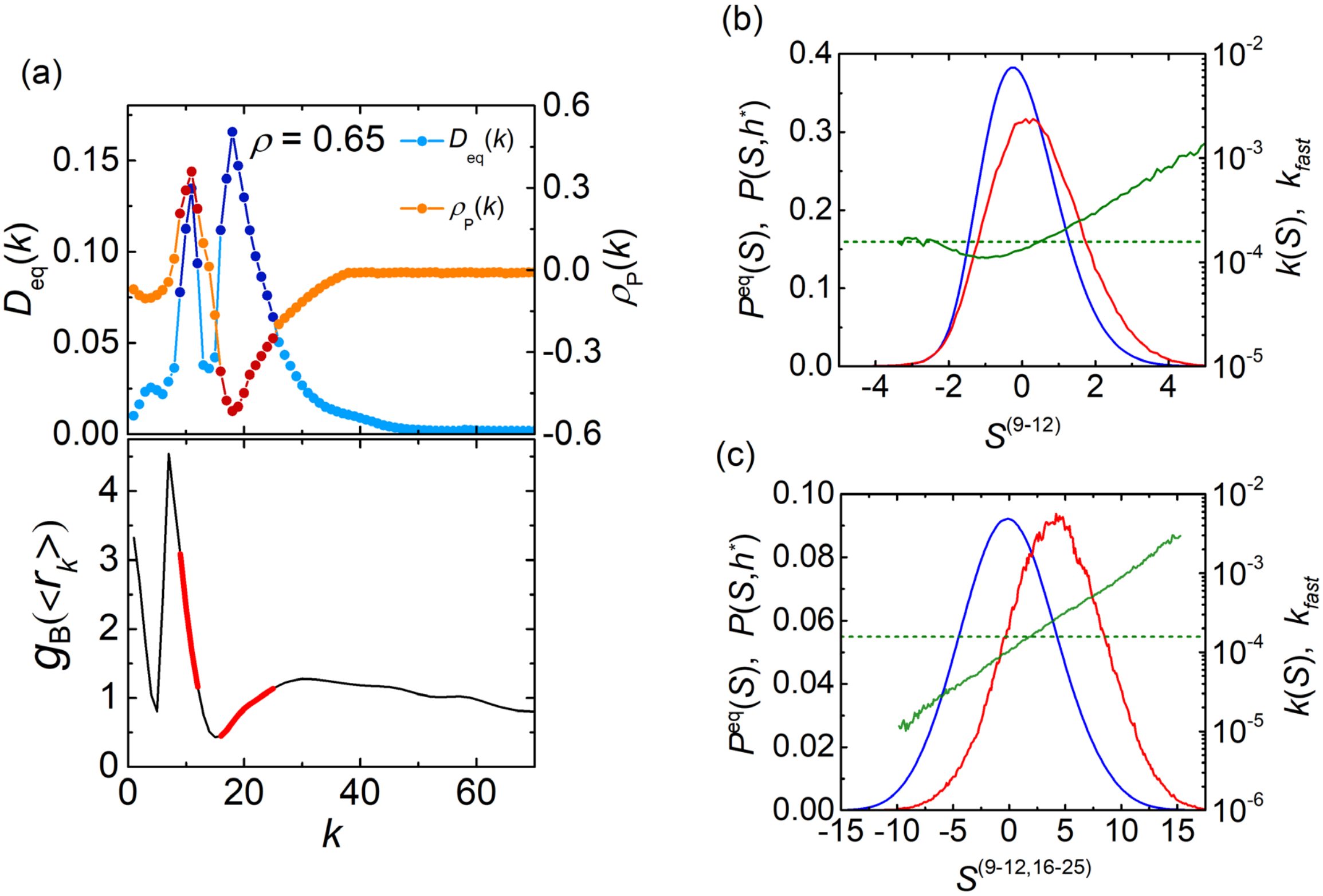

**FIG. S14.** Neighbor-rank analysis for jump motion of particle B at $T = 0.0006$ in the strong regime ($\rho = 0.65$). (a) KL divergence $D_{eq}(k)$ (blue) and Pearson correlation $\rho_P(k)$ (orange) as functions of neighbor rank $k$, where $k$ is defined with respect to the jumping B particle. The selected ranks $k$ = 9-12 and 16-25 are highlighted in dark red for $D_{eq}(k)$ and dark blue for $\rho_P(k)$. The lower subpanel shows the rank-resolved radial distribution profile $g_B\left(\langle r_k \rangle\right)$, with the same rank regions highlighted in red to indicate their location in the outer first shell and the onset of the second shell. (b) Equilibrium distribution of $S^{(9\text{-}12)}$ (blue) and distribution at the hop threshold $h^*$ (red). The jump rate $k[S^{(9\text{-}12)}]$ is shown as a green solid line, and the green dashed line denotes $k_{fast}$. (c) Corresponding distributions and jump rate for the extended variable $S^{(9\text{-}12,\,16\text{-}25)}$.

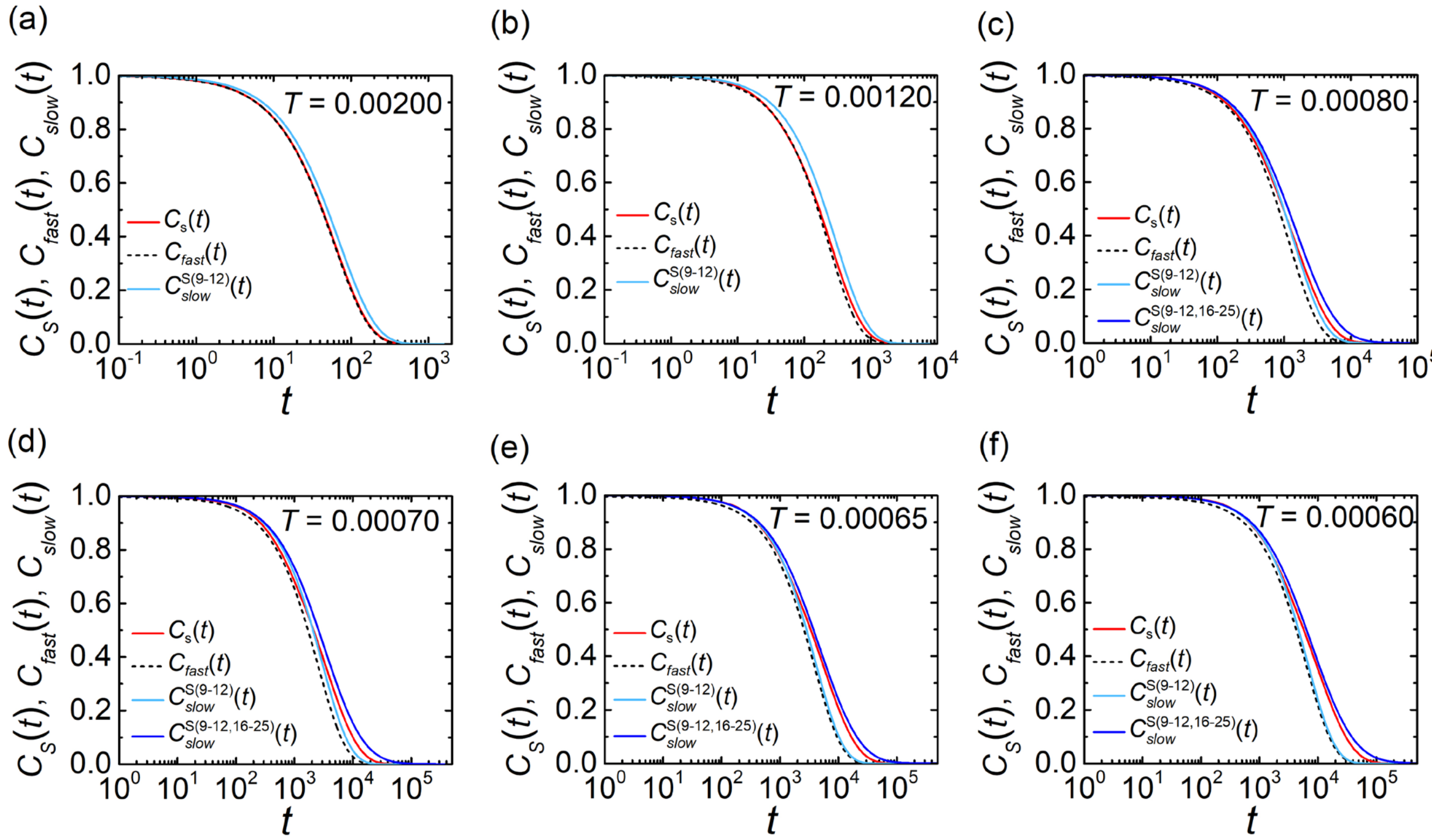

**FIG. S15.** Survival probability for the cage state, $C_S(t)$ (red), its fast-fluctuation limit, $C_{fast}(t)$ (black dashed), and the slow-fluctuation limits $C_{slow}^{S(9-12)}(t)$ (light blue) and $C_{slow}^{S(9-12,16-25)}(t)$ (blue), for particle A in the strong regime ($\rho = 0.65$). Results are shown at (a) $T = 0.0020$, (b) 0.0012, (c) 0.0008, (d) 0.0007, (e) 0.00065, and (f) 0.0006. The slow-fluctuation limit constructed from the extended SSP $S^{(9\text{-}12,\ 16\text{-}25)}$ is included in panels (c)-(f). Upon cooling, $C_S(t)$ progressively approaches the slow-fluctuation limits, with the extended SSP providing a better description at lower temperatures.

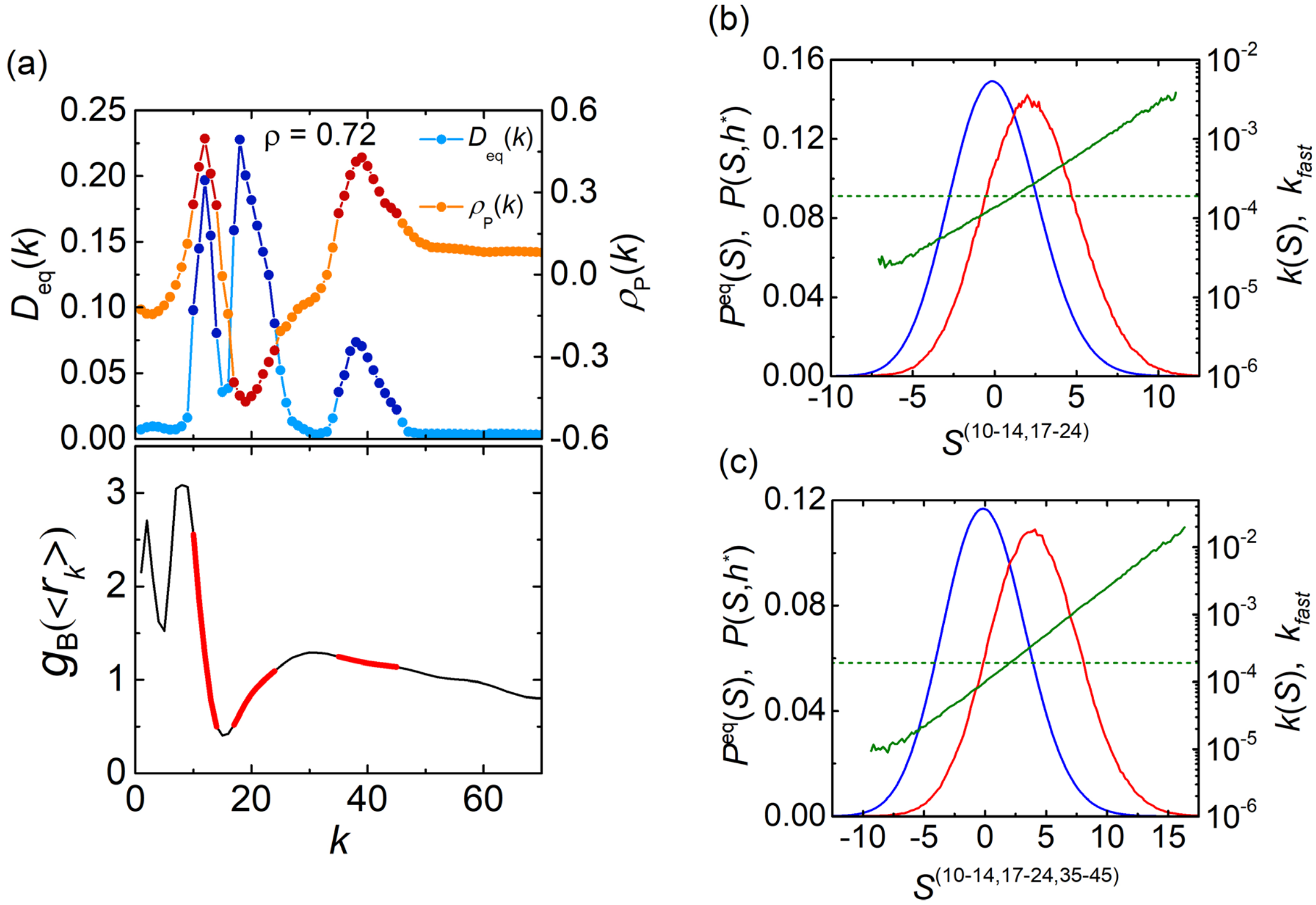

**FIG. S16.** Neighbor-rank analysis for jump motion of particle B at $T = 0.0023$ in the intermediate regime ($\rho = 0.72$). (a) KL divergence $D_{eq}(k)$ (blue) and Pearson correlation $\rho_P(k)$ (orange) as functions of neighbor rank $k$, where $k$ is defined with respect to the jumping B particle. The selected ranks $k$ = 10-14, 17-24, and 35-45 are highlighted in dark red for $D_{eq}(k)$ and dark blue for $\rho_P(k)$. The lower subpanel shows the rank-resolved radial distribution profile $g_B\left(\left\langle r_k \right\rangle\right)$, with the same rank regions highlighted in red. (b) Equilibrium distribution of $S^{(10\text{-}14,\,17\text{-}24)}$ (blue) and distribution at the hop threshold $h^*$ (red). The jump rate $k[S^{(10\text{-}14,\,17\text{-}24)}]$ is shown as the green solid line, and the green dashed line denotes $k_{fast}$. (c) Corresponding distributions and jump rate for the extended variable $S^{(10\text{-}14,\,17\text{-}24,\,35\text{-}45)}$.

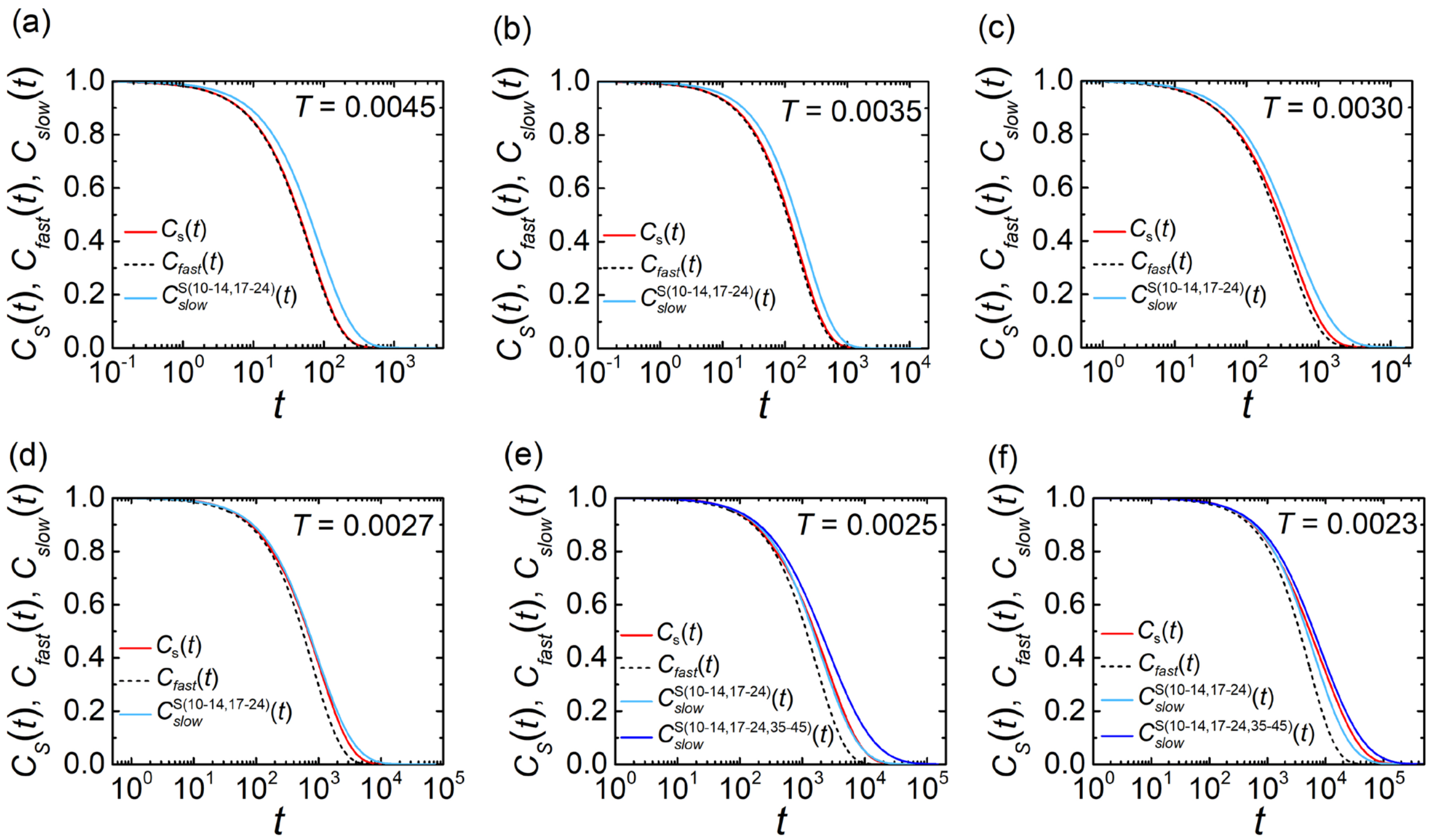

**FIG. S17.** Survival probability for the cage state, $C_S(t)$ (red), its fast-fluctuation limit, $C_{fast}(t)$ (black dashed), and the slow-fluctuation limits $C_{slow}^{S(10-14,17-24)}(t)$ (light blue) and $C_{slow}^{S(10-14,17-24,35-45)}(t)$ (blue) for particle B in the intermediate regime ($\rho = 0.72$). Results are shown at (a) $T = 0.0045$, (b) 0.0035, (c) 0.0030, (d) 0.0027, (e) 0.0025, and (f) 0.0023. The extended slow variable is shown only at the two lowest temperatures, where $S^{(10\text{-}14,\ 17\text{-}24)}$ no longer provides a sufficient description of $C_S(t)$. Upon cooling, the extended variable provides a more consistent description of the slow decay of $C_S(t)$, indicating the increasing role of more distant neighbor ranks in the intermediate regime.

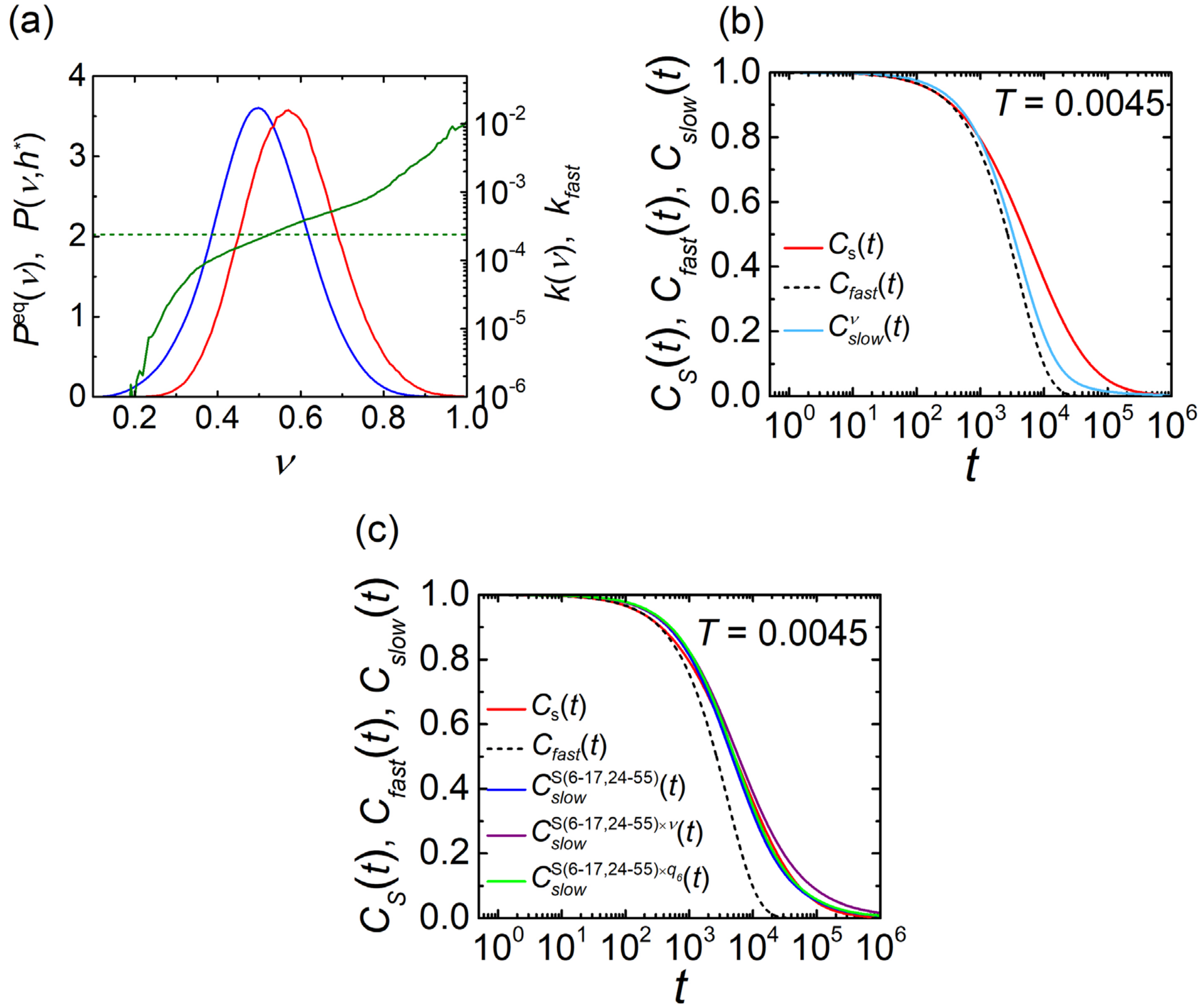

**FIG. S18.** Role of the Voronoi free volume ($\nu$) in the jump dynamics of particle A at $T = 0.0045$ in the fragile regime ($\rho = 0.82$). (a) Distributions of $\nu$ at equilibrium, $P^{eq}(\nu)$ (blue), and at the hop threshold, $P^{eq}(\nu, h^*)$ (red), together with the jump rate $k(\nu)$ (green solid line). The green dashed line denotes the fast-fluctuation rate $k_{fast}$. (b) Cage-state survival probability $C_S(t)$ (red), its fast-fluctuation limit $C_{fast}(t)$ (black dashed), and the slow-fluctuation limit constructed using the Voronoi free volume alone, $C^{\nu}_{slow}(t)$ (light blue). (c) Comparison of the slow-fluctuation limits constructed using the distance-based SSP $S^{(6\text{-}17,\ 24\text{-}55)}$ alone, $C^{S(6-17,\,24-55)}_{slow}(t)$ (blue), together with two-dimensional slow variables combining $S^{(6\text{-}17,\ 24\text{-}55)}$ with the Voronoi free volume, $C^{S(6-17,\,24-55)\times\nu}_{slow}(t)$ (purple), or with the bond-orientational order parameter, $C^{S(6-17,\,24-55)\times q_6}_{slow}(t)$ (green). The comparison shows that including $\nu$ improves the description of $C_S(t)$ more effectively than including $q_6$, indicating the importance of local packing fluctuations.

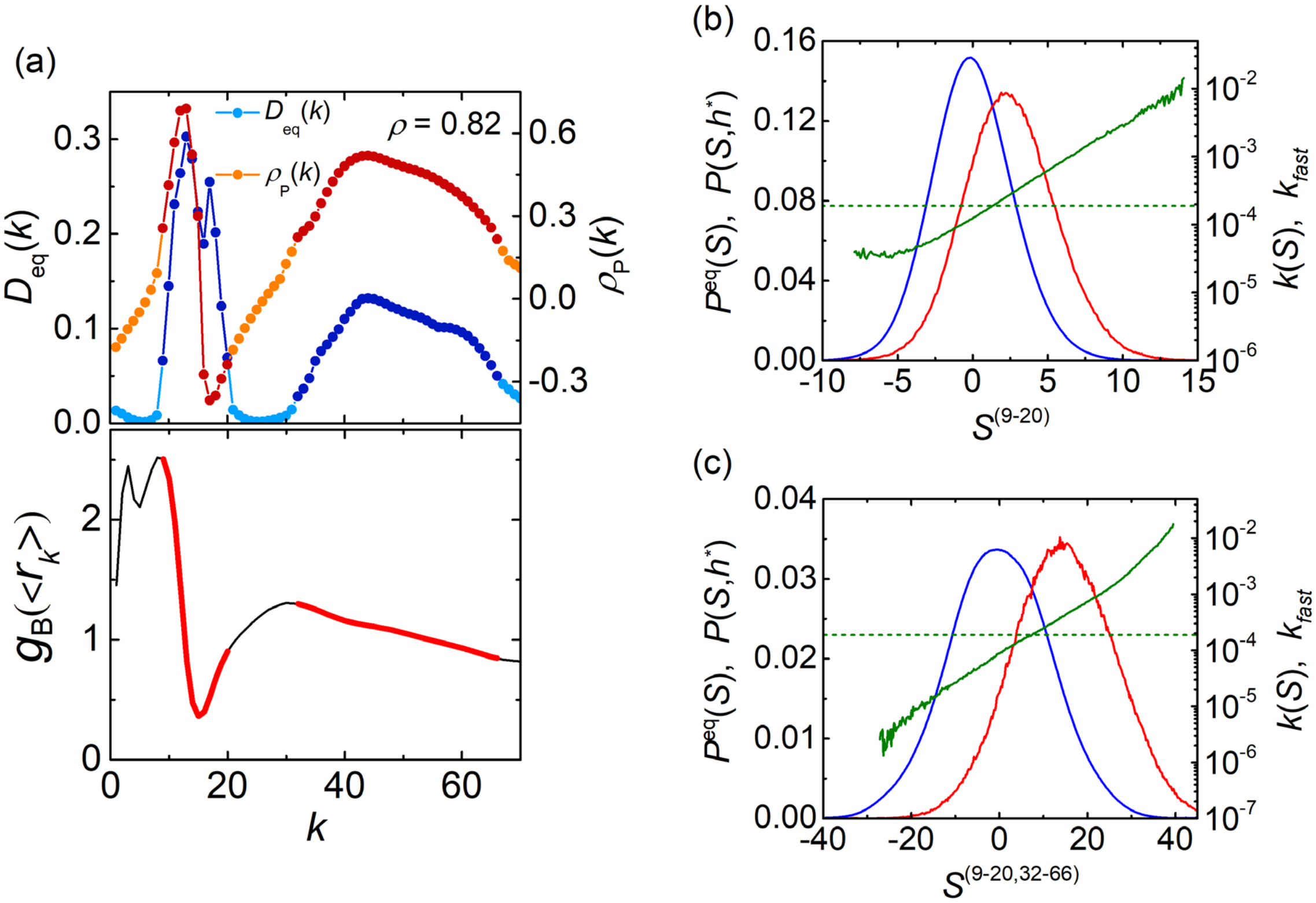

**FIG. S19.** Neighbor-rank analysis for jump motion of particle B at $T = 0.0045$ in the fragile regime ($\rho = 0.82$). (a) KL divergence $D_{eq}(k)$ (blue) and Pearson correlation $\rho_P(k)$ (orange) as functions of neighbor rank $k$, where $k$ is defined with respect to the jumping B particle. The selected ranks $k$ = 9-20 and 32–66 are highlighted in dark red for $D_{eq}(k)$ and dark blue for $\rho_P(k)$. The lower subpanel shows the rank-resolved radial distribution profile $g_A\left(\langle r_k \rangle\right)$, with the same rank regions highlighted in red. (b) Equilibrium distribution of $S^{(9\text{-}20)}$ (blue) and distribution at the hop threshold $h^*$ (red). The jump rate $k[S^{(9\text{-}20)}]$ is shown as a green solid line, and the green dashed line denotes $k_{fast}$. (c) Corresponding distributions and jump rate for the extended variable $S^{(9\text{-}20,\,32\text{-}66)}$.

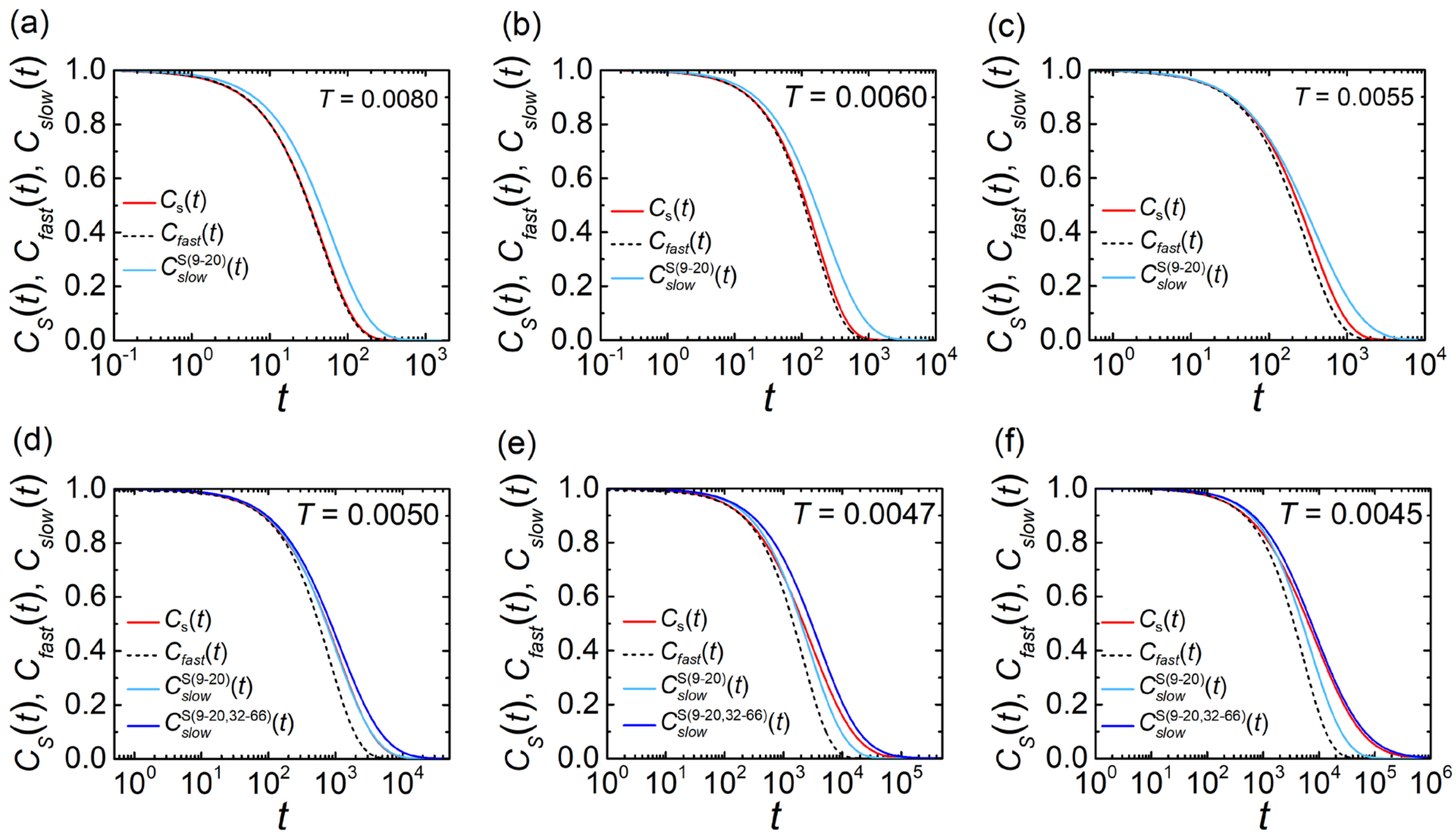

**FIG. S20.** Survival probability for the cage state, $C_S(t)$ (red), its fast-fluctuation limit, $C_{fast}(t)$ (black dashed), and slow-fluctuation limits for particle B in the fragile regime ($\rho = 0.82$). Results are shown at (a) $T = 0.0080$, (b) 0.0060, (c) 0.0055, (d) 0.0050, (e) 0.0047, and (f) 0.0045. The light-blue curve represents $C_{slow}^{S(9-20)}(t)$, and the blue curve represents $C_{slow}^{S(9-20,\,32-66)}(t)$. The extended distance-based SSP provides a consistent slow-fluctuation description at low temperatures, indicating that the dominant rate fluctuations for particle B are captured without including the Voronoi free volume.

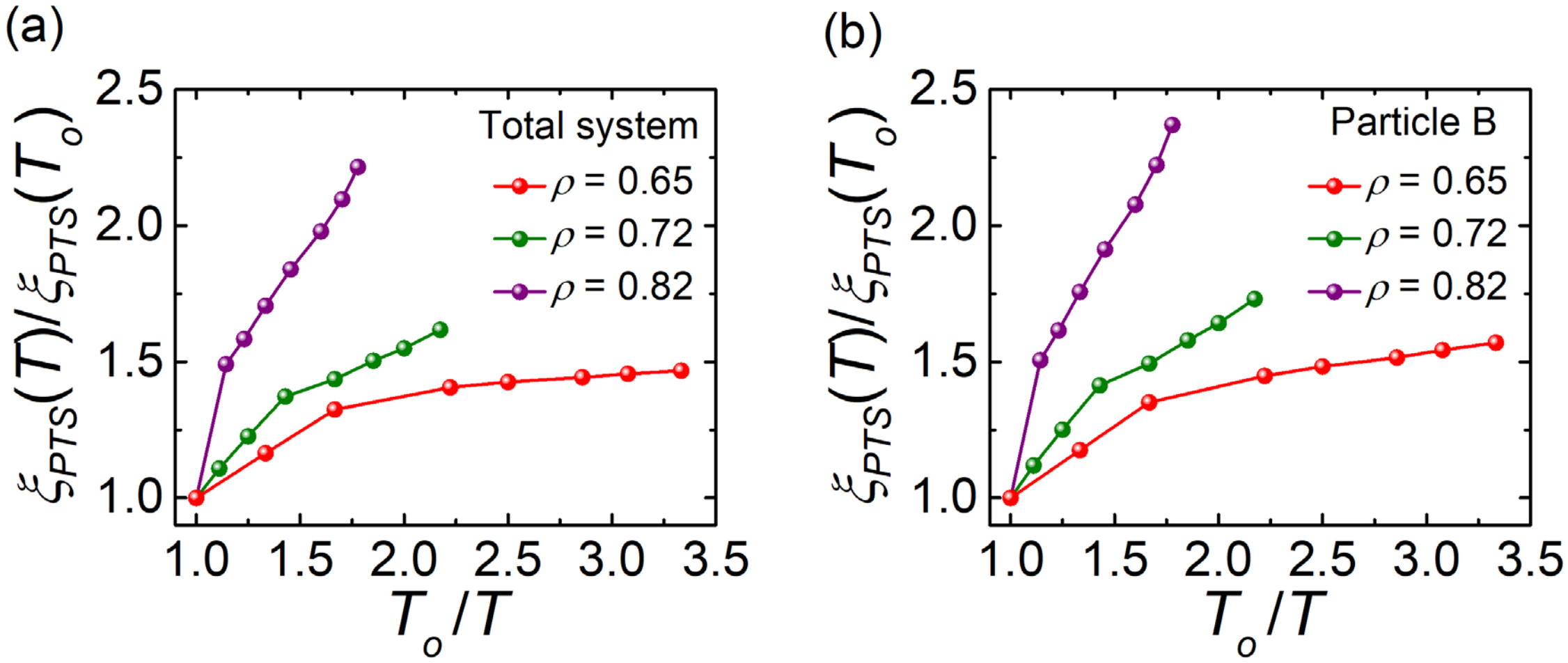

**FIG. S21.** Temperature dependence of the normalized point-to-set correlation length $\xi_{PTS}(T)/\xi_{PTS}(T_0)$ for (a) the total system and (b) particles of type B at the representative densities $\rho = 0.65$, 0.72, and 0.82. The PTS correlation lengths are normalized by their corresponding values at the onset temperature $T_0$. In both cases, the relative growth of $\xi_{PTS}$ becomes progressively more pronounced with increasing density, consistent with the results for particles of type A shown in Fig. 14(d).

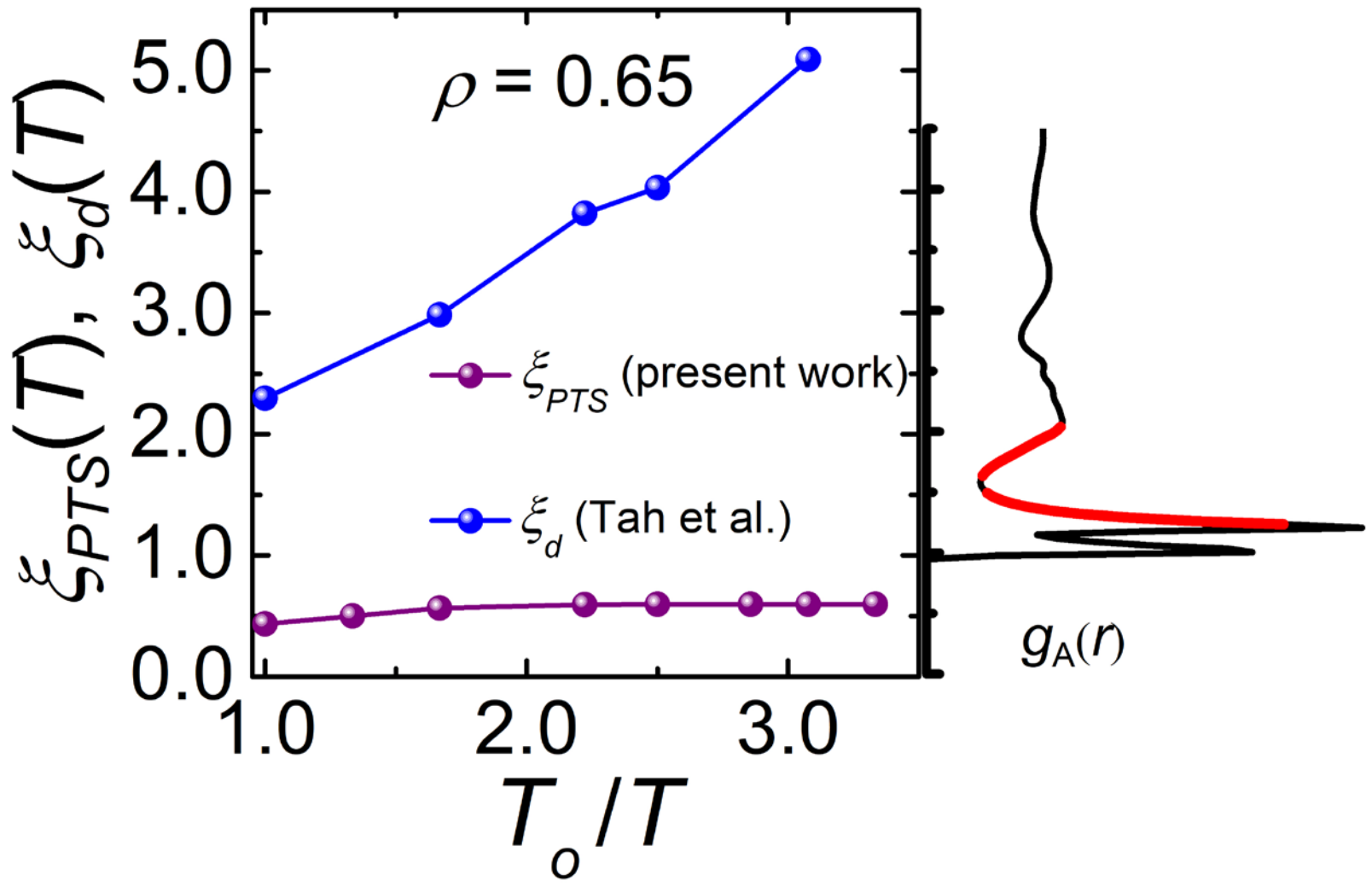

**FIG. S22.** Temperature dependence of $\xi_{PTS}$ in the strong regime ($\rho = 0.65$), compared with the collective dynamic correlation length $\xi_d$, whose values were taken from Ref. [33]. The radial distribution profile $g_A(r)$ at $T = 0.0006$ is shown on the right for comparison with the operational spatial extent of the slow variables. The red portion of $g_A(r)$ marks the radial range spanned by the neighbor ranks contributing to the slow dynamics.